\documentclass[fleqn,usenatbib]{mnras}
\usepackage{newtxtext,newtxmath}
\usepackage{comment}
\usepackage{multirow}
\usepackage{mathtools}
\usepackage[T1]{fontenc}
\usepackage{ae,aecompl}
\usepackage{graphicx}	% Including figure files
\usepackage{amsmath}	% Advanced maths commands
\usepackage{enumitem} % items tools
\usepackage{xfrac} % fractions
\usepackage{algorithm}
\usepackage{algpseudocode}

\setlist[itemize]{leftmargin=*}
\newcommand{\bigO}[1]{\mathcal{O}(#1)}

\newcommand{\norm}[1]{\|#1\|}

\newcommand{\ud}{\mathrm{d}}

\newcommand{\vect}[1]{\boldsymbol{#1}}
\newcommand{\matr}[1]{\boldsymbol{\mathrm{#1}}}

\newcommand{\derfrac}[2]{\frac{\ud #1}{\ud #2}}

\newcommand{\Msun}{M_\odot}

\newcommand{\nbody}{\texttt{NBODY}}
\newcommand{\nbodysix}{\texttt{NBODY6++GPU}}
\newcommand{\mstar}{\texttt{MSTAR}}
\newcommand{\frost}{\texttt{FROST}}
\newcommand{\bifrost}{\texttt{BIFROST}}
\newcommand{\ketju}{\texttt{KETJU}}
\newcommand{\scic}{\texttt{SCIC}}
\newcommand{\mcluster}{\texttt{McLuster}}
\newcommand{\sse}{\texttt{SSE}}
\newcommand{\bse}{\texttt{BSE}}
\newcommand{\mse}{\texttt{MSE}}
\newcommand{\fsi}{\texttt{FSI}}
\newcommand{\hhsfsi}{\texttt{HHS-FSI}}
\newcommand{\logh}{\texttt{LogH}}

\newcommand{\op}[1]{{\vect{\mathrm{#1}}}}
\DeclareMathOperator{\myarsinh}{arsinh}

\title[Simulating subsystems in star clusters]{{\texttt{BIFROST}: simulating compact subsystems in star clusters using a hierarchical fourth-order forward symplectic integrator code}}  
\author[A. Rantala et al.]{Antti Rantala$^{1}$\thanks{E-mail: anttiran@mpa-garching.mpg.de}, Thorsten Naab$^{1}$, Francesco Paolo Rizzuto$^{2}$, Matias Mannerkoski$^{2}$, \newauthor Christian Partmann$^{1}$, Kristina Lautenschütz$^{1}$\vspace{1mm}\\
$^{1}$Max-Planck-Institut f\"ur Astrophysik, Karl-Schwarzschild-Str. 1, 
D-85748, Garching, Germany\\
$^{2}$Department of Physics, University of Helsinki, P.O. Box 64 (Gustaf Hällströmin katu 2), FI-00014, University of Helsinki, Finland\\
}
\date{Accepted XXX. Received YYY; in original form ZZZ}
\pubyear{2022}

\begin{document}
\label{firstpage}
\pagerange{\pageref{firstpage}--\pageref{lastpage}}
\maketitle

\begin{abstract}
We present \bifrost{}, an extended version of the GPU-accelerated hierarchical fourth-order forward symplectic integrator code \frost. \bifrost{} (\texttt{BI}naries in \frost{}) can efficiently evolve collisional stellar systems with arbitrary binary fractions up to $f_\mathrm{bin}=100\%$ by using secular and regularised integration for binaries, triples, multiple systems or small clusters around black holes within the fourth-order forward integrator framework. Post-Newtonian (PN) terms up to order PN3.5 are included in the equations of motion of compact subsystems with optional three-body and spin-dependent terms. PN1.0 terms for interactions with black holes are computed everywhere in the simulation domain. The code has several merger criteria (gravitational-wave inspirals, tidal disruption events and stellar and compact object collisions) with the addition of relativistic recoil kicks for compact object mergers. We show that for systems with $N$ particles the scaling of the code remains good up to $N_\mathrm{GPU} \sim 40\times N / 10^6$ GPUs and that the increasing binary fractions up to 100 per cent hardly increase the code running time (less than a factor $\sim 1.5$). We also validate the numerical accuracy of \bifrost{} by presenting a number of star clusters simulations the most extreme ones including a core collapse and a merger of two intermediate mass black holes with a relativistic recoil kick.
\end{abstract}


%%%%%%%%%%%%%%%%%%%%%%%%%%%%%%%%%%%%%%%%%%%%%%%%%%
%%%%%%%%%%%%%%%%% BODY OF PAPER %%%%%%%%%%%%%%%%%%
%%%%%%%%%%%%%%%%%%%%%%%%%%%%%%%%%%%%%%%%%%%%%%%%%%

\section{Introduction}

A majority of stars form in star clusters \citep{Lada2003}. A large fraction of these stars are members of binary or multiple systems \citep{Raghavan2010,Goodwin2010,Tokovinin2014a,Tokovinin2014b} with the multiplicity fraction increasing with the stellar mass \citep{Duchene2013}. For the most massive stars the number of stars in triple and quadruple systems can even exceed the single and binary stars \citep{Moe2017}.

In addition to the importance for stellar evolution (e.g. \citealt{Kouwenhoven2008,Sana2012}) close binaries strongly affect the global dynamical evolution of the star clusters in which they reside \citep{Heggie2006,Hurley2007,Mackey2008,Wang2016}. The extreme example is their effect on the catastrophic runaway core collapse of a star cluster. Due to the negative specific heat capacity of gravitating systems \citep{LyndenBell1968,LyndenBell1980} hard binaries release energy to other stars in close few-body encounters by further shrinking to smaller separations. This has far-reaching consequences: a relatively small number of binary systems can halt the runaway contraction of the core of a star cluster (\citealt{Sugimoto1983, Bettwieser1984, Heggie1993, Kamlah2022}).

Sufficiently hard compact binary systems of white dwarfs, neutron stars and black holes (BHs) are a source of gravitational radiation (\citealt{Peters1963,Peters1964,Benacquista2013}), originally observed indirectly from the shrinking orbits of pulsars in binary systems \citep{Hulse1975}. Recently, gravitational waves have been directly observed from mergers of binary black holes and neutron stars (e.g. \citealt{Abbott2016,Abbott2017}). In addition, future space-borne gravitational wave detectors are expected to detect a large number of radiating white dwarf binary systems in the local Universe (e.g. \citealt{Korol2017}). Another interesting prospect is the search for intermediate-mass black holes (IMBHs) from extreme and intermediate mass-ratio inspirals, EMRIs and IMRIs (e.g. \citealt{AmaroSeoane2018}), in which a supermassive black hole (SMBH) merges with a lower-mass black hole.

The vast range of physically important timescales in star clusters containing binary systems presents a formidable challenge to N-body simulation codes (e.g. \citealt{Heggie2003}). For example, the orbital period
\begin{equation}
P = 2 \pi \left( \frac{a^3}{G M}\right)^{1/2}
\end{equation}
of a compact binary system with a semi-major axis of $a=1 R_\odot$ and total mass of $M = 3 M_\odot$ is less than two hours. However, the time for the binary system to traverse its host star cluster, known as the crossing time defined using the half-mass radius $r_\mathrm{h}$ and its velocity dispersion $\sigma$ as
\begin{equation}
t_\mathrm{cross} = \frac{r_\mathrm{h}}{\sigma}
\end{equation}
is significantly longer. For an example star cluster with $r_\mathrm{h} = 1$ pc and $\sigma = 10$ km/s the crossing time is $t_\mathrm{cross} \sim 0.1$ Myr, more than $10^8$ orbital periods of our example compact binary. The star cluster itself evolves (even in isolation) on even longer timescales due to relaxation effects. The driving two-body relaxation timescale for a star cluster is typically calculated as
\begin{equation}
t_\mathrm{relax} \sim 0.138 \left( \frac{N r_\mathrm{h}^3}{G \tilde{m}} \right)^{1/2} \frac{1}{\ln{\Lambda}}
\end{equation}
in which $\tilde{m} = M/N \sim 0.5 M_\odot$ is the mean stellar mass for a typical stellar population and $\ln{\Lambda}$ is the Coulomb logarithm \citep{Spitzer1987,Aarseth2003} which has a value of $\ln{\Lambda}\sim 10$ (e.g. \citealt{Konstantinidis2010}). For our example star cluster the half-mass relaxation time is $t_\mathrm{relax} \sim 64$ Myr, over $10^{11}$ periods of our example binary.

For over 50 years, simulation codes (especially the \nbody{} series) have incorporated specialised techniques to integrate binary and multiple systems \citep{Aarseth1971,Aarseth1974}. Widely-used methods include Kepler solvers for binaries (e.g. \citealt{Danby1992,Rein2015,Wisdom2015,Dehnen2017a}) and secular methods for hierarchical multiplets (\citealt{Marchal1990,Naoz2013,Hamers2016,Hamers2021}). A successful framework for integrating arbitrary few-body is regularisation. The techniques based on the KS regularisation \citep{Kustaanheimo1965,Aarseth1974,Heggie1986,Mikkola1993} use both coordinate and time transformations. More recently developed regularised codes use a more simple yet powerful algorithmic regularisation (\citealt{Mikkola1999,Preto1999,Mikkola2002,Mikkola2006,Mikkola2008,Hellstrom2010,Trani2019,Wang2020,Rantala2020,Wang_Yi_Han2021}) which only requires the time transformation.

Despite the large number of specialised integration techniques available in the literature the most commonly used numerical simulation codes have not been able to efficiently integrate massive stellar systems ($N>10^5$) with binary fractions higher than $f_\mathrm{bin}\gtrsim10\%$ until recently \citep{Wang2020b}. This limitation was mostly caused by the serial implementation of the regularised integration methods used in N-body codes: typically only the $\bigO{N^2}$ force loops required by direct summation codes were parallelised.

Fourth-order forward symplectic integrators, hereafter \fsi{} (e.g. \citealt{Chin1997,Chin2005,Chin2007a}), have began to gain attention as viable alternatives for Hermite integrators in N-body simulations \citep{Dehnen2011,Dehnen2017a,Rantala2021}. Hierarchical implementations of the integrator (\hhsfsi) have been developed \citep{Rantala2021}, using the technique of hierarchical Hamiltonian splitting (HHS).

In this work we present \bifrost, an updated version of the hierarchical fourth-order forward symplectic integrator code \frost{} \citep{Rantala2021}. The most important feature of the new \bifrost{} code is the efficient secular and regularised integration of subsystems in the simulations with arbitrary binary fractions up to $f_{\mathrm{bin}} = 100\%$. Other code updates include a more robust time-step assignment for simulation particles as well as stellar and compact object mergers. Relativistic dynamical effects in the code are incorporated by using post-Newtonian equations of motion in compact subsystems. Our implementation also allows for PN1.0 level effects such as the relativistic periapsis advance outside subsystems in the entire simulation domain, a feature which typical N-body simulation codes lack.

The article is structured as follows. After the introduction we review the hierarchical fourth-order forward integrator in Section \ref{section: 2}. The new features of the \bifrost{} code including the secular and regularised subsystem integrators are described in Section \ref{section: 3}. The scaling and timing tests of the code on a supercomputer are presented in Section \ref{section: 4}. The applications demonstrating the accuracy and performance of the code are presented in Section \ref{section: 5}. We briefly discuss the future prospects of the code in Section \ref{section: 6} and finally summarised and conclude in Section \ref{section: 7}. Furthermore, Appendix \ref{section: SCIC} describes our new and fast star cluster initial conditions generator used in this study.

%%%%%%%%%%%%%%%%%%%%%%%%%%%%%%%%%%%%%%%%%%%%%%%%%%%%%%%%%%%

\section{Fourth-order forward symplectic integrators}\label{section: 2}

The novel simulation code \bifrost{} is based on our previous N-body code \frost{} \citep{Rantala2021} with several new features added. In this Section, we briefly review the integrator of the \frost{} code, the fourth-order forward symplectic integrator as the integrator is not yet widely well-known in the literature. For a more in-depth introduction to fourth-order forward symplectic integrator and the details of the numerical implementation in CUDA C we refer the reader to \cite{Rantala2021}. The modifications to the direct summation force calculation parts of the code are minimal from \frost{} to \bifrost.

\subsection{The basic integration algorithm \fsi}\label{section: new-fsi}

We consider a system of $N$ point particles $\mathcal{P}$ interacting gravitationally. The common N-body Hamiltonian $H$ of Newtonian gravity is defined as 
\begin{equation}
H = \sum_\mathrm{i} \frac{1}{2} m_\mathrm{i} \norm{\vect{v}_\mathrm{i}}^2 - \sum_\mathrm{j>i} \frac{G m_\mathrm{i} m_\mathrm{j}}{ \norm{\vect{r}_\mathrm{ji} } },
\end{equation}
in which $m_\mathrm{i}$ are the particle masses, $\vect{v}_\mathrm{i}$ their velocities and $\vect{r}_\mathrm{ji} = \vect{r}_\mathrm{j} - \vect{r}_\mathrm{i}$ their mutual separations. The fourth-order accurate symplectic integrator with strictly positive time-step can be derived from this Hamiltonian (\citealt{Chin1997,Chin2005,Chin2007}, see also \citealt{Xu2010}). The derivation is itself based on the earlier work of on symplectic integrators (e.g. \citealt{Takahashi1984,Sheng1989,Yoshida1990,Suzuki1991,Goldman1996}, see also \citealt{Dehnen2017a}).

In \fsi, the integration cycle over a time-step $\epsilon$ proceeds in five steps (kick, drift, gradient kick, drift, kick) for as follows. First, the particle velocities are advanced for one-sixth of the time-step $\epsilon$ in a kick operation as
\begin{equation}\label{eq: kick-new}
\vect{v}_\mathrm{i} \leftarrow \vect{v}_\mathrm{i} + \frac{\epsilon}{6} \vect{a}_\mathrm{i}. 
\end{equation}
Here the common Newtonian gravitational accelerations $\vect{a}_\mathrm{i}$ are calculated as 
\begin{equation}\label{eq: new-newton-acc}
\vect{a}_\mathrm{i} = \sum_\mathrm{j \neq i} \frac{G m_\mathrm{j}}{\norm{\vect{r}_\mathrm{ji}}^3} \vect{r}_\mathrm{ji}.
\end{equation}
Next, the particle positions are advanced for a time interval of $\epsilon/2$ using their velocities in a drift operation as
\begin{equation}\label{eq: drift-new}
\vect{r}_\mathrm{i} \leftarrow \vect{r}_\mathrm{i} + \frac{\epsilon}{2} \vect{v}_\mathrm{i}. 
\end{equation}
The next step, the $2\epsilon/3$ gradient kick
\begin{equation}\label{eq: kick-gradient-new}
\vect{v}_\mathrm{i} \leftarrow \vect{v}_\mathrm{i} + \frac{2\epsilon}{3} \tilde{\vect{a}}_\mathrm{i}
\end{equation}
updates the particle velocities using the so-called gradient accelerations
$\tilde{\vect{a}}_\mathrm{i}$, which are defined as
\begin{equation}\label{eq: new-grad-acc}
\tilde{\vect{a}}_\mathrm{i} = \vect{a}_\mathrm{i} + \frac{\epsilon^2}{24} \sum_\mathrm{j \neq i} \frac{G m_\mathrm{j}}{ \norm{\vect{r}_\mathrm{ji}}^5 } \Big( \norm{\vect{r}_\mathrm{ji}}^2 \vect{a}_\mathrm{ji} - 3 ( \vect{a}_\mathrm{ji} \cdot \vect{r}_\mathrm{ji}) \vect{r}_\mathrm{ji}  \Big),
\end{equation}
in which $\vect{a}_\mathrm{ji} = \vect{a}_\mathrm{j} - \vect{a}_\mathrm{i}$ are the relative accelerations of the particles. The gradient kick is the key operation of the \fsi{} integration cycle as the final acceleration term proportional to $\epsilon^2$ in Eq. \eqref{eq: new-grad-acc} cancels the leading second-order error terms making the integrator fourth-order accurate. The integration cycle is closed by repeating the one-half drift operation of Eq. \eqref{eq: drift-new} and the one-sixth kick operation of Eq. \eqref{eq: kick-new} again. Using the drift ($e^{\epsilon \op{T}}$), kick ($e^{\epsilon \op{U}}$), and gradient kick operators ($e^{\epsilon \op{\tilde{U}}}$) the \fsi{} integration cycle can be summarised as the \fsi{} time evolution operator as
\begin{equation}\label{eq: fsi-cycle}
e^{\epsilon \op{H}} = e^{\frac{1}{6} \epsilon \op{U}} e^{\frac{1}{2} \epsilon \op{T}} e^{\frac{2}{3} \epsilon \op{\tilde{U}}} e^{\frac{1}{2} \epsilon \op{T}} e^{\frac{1}{6} \epsilon \op{U}}.
\end{equation}

Note that compared to symplectic integrators of \cite{Yoshida1990} type beyond the second order the \fsi{} has strictly positive integrator sub-step lengths. While negative integrator sub-steps are not a problem per se, they prohibit the integration of irreversible systems (e.g. with gravitational-wave radiation reaction losses) and can make hierarchical recursive integration computationally less efficient \citep{Pelupessy2012,Springel2021}. In addition, it has been shown that for the Kepler problem fourth-order symplectic integrators with strictly positive sub-steps outperform the ones including negative sub-steps \citep{Chin2007}.

Directly summing the Newtonian gravitational accelerations for $N$ particles is an $\bigO{N^2}$ operation, and \fsi{} requires two of these summation loops during a single integration cycle. Calculating the gradient accelerations for the particles requires two $\bigO{N^2}$ operations as up-to-date Newtonian accelerations are required to compute the gradient terms. In total a single \fsi{} integration cycle has four $\bigO{N^2}$ operations.

\subsection{The hierarchical integration algorithm \hhsfsi}\label{section: new-hhs-fsi}

In the \fsi{} integrator described above all the particle time-steps are equal, which prohibits the integration of very large N-body systems. Typically in N-body simulations of stellar systems the individual particle time-steps span a range of several orders of magnitude \citep{Dehnen2011}. The individual particle time-steps are most often ordered in a power-of-two block time-step hierarchy \citep{McMillan1986, Aarseth2003}. The so-called hierarchical integration (e.g. \citealt{Pelupessy2012}) is a momentum-conserving alternative to the block time-step scheme. 

Using hierarchical Hamiltonian splitting (HHS), the N-body system is recursively subdivided into smaller systems according to their time-steps. Starting from all the particles $\mathcal{P}$ and an integration interval duration $\tau$ the particles are divided into a set of rapidly evolving fast simulation particles $\mathcal{F}$ (with $\epsilon_\mathrm{i} < \tau$) and slow particles $\mathcal{S}$ (with $\epsilon_\mathrm{i} \geq \tau$). The slow particles are then propagated for an interval of $\tau/2$ while the fast set is further subdivided by using a smaller pivot time-step $\tau/2$. The process is repeated until no fast particles remain.

Our hierarchical version of the \fsi{} algorithm, \hhsfsi, was developed in order to simulate N-body systems with a large dynamical range with fourth-order accuracy \citep{Rantala2021}. The integration cycle of \hhsfsi{} on with a slow and a fast level is summarised in Eq. \eqref{eq: hhsfsi-cycle} as
\begin{equation}\label{eq: hhsfsi-cycle}
e^{\epsilon \op{H}} = e^{\frac{1}{6} \epsilon \op{U_\mathrm{SF}}} e^{\frac{1}{2} \epsilon \op{H_\mathrm{S}}} e^{\frac{1}{2} \epsilon \op{H_\mathrm{F}}} e^{\frac{2}{3} \epsilon \op{\tilde{U}_\mathrm{SF}}} e^{\frac{1}{2} \epsilon \op{H_\mathrm{F}}} e^{\frac{1}{2} \epsilon \op{H_\mathrm{S}}} e^{\frac{1}{6} \epsilon \op{U_\mathrm{SF}}}.
\end{equation}
Each time-step hierarchy level begins with assigning time-steps to the particles and dividing them into slow and fast particles using a pivot sorting algorithm. The integration cycle is opened by an inter-level slow-fast kick $e^{\frac{1}{6} \epsilon \op{U_\mathrm{SF}}}$ in which the Newtonian inter-level particle accelerations 
\begin{equation}
\begin{split}
\vect{a}_\mathrm{i} &= \sum_\mathrm{j \in \mathcal{F}} \frac{G m_\mathrm{j}}{\norm{\vect{r}_\mathrm{ji}}^3} \vect{r}_\mathrm{ji}, & \mathcal{P}_\mathrm{i} \in \mathcal{S}\\
\vect{a}_\mathrm{j} &= \sum_\mathrm{i \in \mathcal{S}} \frac{G m_\mathrm{i}}{\norm{\vect{r}_\mathrm{ij}}^3} \vect{r}_\mathrm{ij}, & \mathcal{P}_\mathrm{j} \in \mathcal{F}\\
\end{split}
\end{equation}
are required. Then, the set of slow particles is integrated for $\epsilon/2$ using the \fsi{} of Eq. \eqref{eq: fsi-cycle} in $e^{\frac{1}{2} \epsilon \op{H_\mathrm{S}}}$. For the fast particles, the full integrator is called again in $e^{\frac{1}{2} \epsilon \op{H_\mathrm{F}}}$ for $\epsilon/2$. Analogously to \fsi{} the middle slow-fast inter-level kick $e^{\frac{2}{3} \epsilon \op{\tilde{U}_\mathrm{SF}}}$ has a length of $2\epsilon/3$ and uses gradient accelerations, which are calculated as
\begin{equation}
\begin{split}
\tilde{\vect{a}}_\mathrm{i} &= \vect{a}_\mathrm{i} + \frac{\epsilon^2}{24} \sum_\mathrm{j \in \mathcal{F}} \frac{G m_\mathrm{j}}{\norm{\vect{r}_\mathrm{ji}}^5} \Big( \norm{\vect{r}_\mathrm{ji}}^2 \vect{a}_\mathrm{ji} - 3(\vect{a}_\mathrm{ji} \cdot \vect{r}_\mathrm{ji}) \vect{r}_\mathrm{ji}\Big), & \mathcal{P}_\mathrm{i} \in \mathcal{S}\\
\tilde{\vect{a}}_\mathrm{j} &= \vect{a}_\mathrm{j} + \frac{\epsilon^2}{24} \sum_\mathrm{i \in \mathcal{S}} \frac{G m_\mathrm{i}}{\norm{\vect{r}_\mathrm{ij}}^5} \Big( \norm{\vect{r}_\mathrm{ij}}^2 \vect{a}_\mathrm{ij} - 3(\vect{a}_\mathrm{ij} \cdot \vect{r}_\mathrm{ij}) \vect{r}_\mathrm{ij}\Big), &\mathcal{P}_\mathrm{j} \in \mathcal{F}.
\end{split}
\end{equation}
Again, the function of the gradient term is to cancel leading-order error terms of the integrator making \hhsfsi{} fourth-order accurate. After the middle slow-fast inter-level gradient kick the fast and slow particles are integrated again for $\epsilon/2$ with the \hhsfsi{} and \fsi{} integrators, respectively. Finally, the integration cycle of the level is closed by an one-sixth Newtonian slow-fast inter-level kick.

The \hhsfsi{} integrator has two important properties. First, as all the kick operations are always synchronous, i.e. pair-wise, the integration algorithm is manifestly momentum-conserving. Second, as the inter-level interactions for a certain particle only include accelerations from faster hierarchy levels rapidly evolving subsystems effectively decouple from slowly evolving particles allowing for very efficient integration of small, dense subsystems.

We note that even though the formal symplecticity of the \fsi{} is lost in \hhsfsi{} due to individual variable time-steps, the relative numerical errors during a single integration interval are small enough ($\sim 10^{-10}$) so that the secularly accumulating total error does not become prohibitively large during simulation time-scales of interest. 

\section{Updated simulation code}\label{section: 3}

\subsection{Aim of the new code}

Our previous simulation code \frost{} is an accurate and efficient GPU-accelerated direct-summation N-body code capable of running million-body simulations of star clusters. However, \frost{} lacks any specialised integration techniques for integrating binaries, close hyperbolic fly-bys, triples, multiple systems or small clusters around black holes, and accurate integration of such systems using \frost{} may become prohibitively expensive.

The novel \bifrost{} (from binaries in \frost{}) codes solves the issue of subsystems by integrating their dynamics using specialised few-body solvers. We use both regularised and secular integration methods depending on the type and properties of the subsystems. Close and bound two-body systems are integrated using either a secular technique or regularised integration, while close unbound fly-bys of two particles are always treated with regularised integration. For stable hierarchical three-body systems, which consist of an inner binary system orbiting an outer third companion, we use a special secular integration algorithm. Non-hierarchical three-body systems, such as strongly interacting three-body systems, and fly-bys of single stars with binaries, are short-lived and integrated using regularised integrators. In this study any subsystem configurations with more than three bodies are integrated using regularisation techniques as well. 

We note that for certain types of systems potentially useful approximate integration techniques (usually perturbative) exist in the literature to replace regularisation and they might be beneficial for \bifrost{} as well. Examples of such systems include distant single-binary and binary-binary fly-bys and hierarchical multiple systems containing four or more bodies.

\subsection{Overview of the new code features}

Novel features of the \bifrost{} code are listed in Table \ref{tab: overview}. The most important new features are the regularised and secular integrators for the subsystems. We use two variants of algorithmically regularised integrators with different parallelization properties: a \logh{} \citep{Mikkola1999} implementation to integrate a large number of few-body systems, and \mstar{} \citep{Rantala2021} for integrating a few large ($N\sim1000$) systems. We use our implementations of secular integrators for close low-period binaries and hierarchical triples.

A second class of code updates is related to interfacing the \fsi{} to the few-body subsystem integrators. A new variant of the \fsi{} integrator is presented in Section \ref{section: subsys-fsi}. Routines for finding and classifying the types of subsystems along with the time-step assignment before integration in \bifrost{} are described in Sections \ref{section: subsys-find}. The updated time-step criteria themselves are presented in Section \ref{section: dt}.

The third set of code updates is the incorporation of relativistic post-Newtonian (PN) dynamics in the code. The user can opt to use post-Newtonian equations of motion in the regularised integrators and their corresponding orbit-averaged versions in the secular integrators. The post-Newtonian effects are mostly discussed with the integrator descriptions in Sections \ref{section: logh} and \ref{section: secular}. An important feature of the \bifrost{} code is the possibility to include the first post-Newtonian term PN1.0 also outside regularised regions in the \fsi. Without this global PN1.0 term discussed in Section \ref{section: global-pn1} relativistic precession effects would be ignored in dense systems harbouring massive black holes, such as nuclear star clusters.

Finally \bifrost{} includes a number of models for astrophysical phenomena not included in \frost. We include e.g. single stellar evolution tracks, mergers of stars and compact objects and relativistic gravitational-wave recoil for merging black holes. A number of numerical techniques are also discussed, such as accurate book-keeping for studying energy conservation and automatic simulation restarts if too much numerical error accumulated during an integration interval.

\begin{table}
	\caption{The most important new features of the \bifrost{} code.}
	\label{tab: overview}
	\begin{tabular}{ll} % four columns, alignment for each
		\hline
		Code feature & Section\\
		\hline
		subsystems in the 4th order integrator& \ref{section: subsys-fsi} \& \ref{section: subsys-find}\\
		improved time-steps & \ref{section: dt}\\
		regularised integration of small subsystems & \ref{section: logh}\\
		PN equations of motion in reg. regions & \ref{section: pnacc}\\
		PN equations of motion outside reg. regions & \ref{section: global-pn1}\\
		regularised integration of large regions & \ref{section: mstar}\\
		secular integration of binaries & \ref{section: secular}\\
		Kepler solver & \ref{section: keplersolver}\\
		secular integration of hierarchical triples & \ref{section: secular-triple}\\
		single stellar evolution & \ref{section: sse}\\
		mergers of stars and compact objects & \ref{section: mergers} \& \ref{section: merger-remnant}\\
		gravitational-wave recoil & \ref{section: gwrecoil}\\
		energy book-keeping & \ref{section: bookkeeping}\\
		adaptive error restarts & \ref{section: error-restart}\\
		\hline
	\end{tabular}
\end{table}

\subsection{Fourth-order forward integration with subsystems}\label{section: subsys-fsi}

We now describe a modification of the \fsi{} which incorporates few-body integrators for subsystems, again using Hamiltonian splitting techniques. We recommend reviewing the \fsi{} integrator derivation from \cite{Rantala2021} as the derivation of the novel forward integrator with subsystems in this Section is very brief and sketch-like in nature.

First we consider a single few-body subsystem $\mathcal{S}$ within our N-body system. The remaining particles not in the subsystem $\mathcal{S}$ belong to the set of single particles we label here $\mathcal{R}$. The Hamiltonian of such an N-body system can be written as
\begin{equation}
\begin{split}
    H &= T+U=T_\mathcal{S} + T_\mathcal{R} + U_\mathcal{SS} + U_\mathcal{RR} +U_\mathcal{SR} \\
    &= \sum_\mathrm{i \in \mathcal{S}} \frac{1}{2} m_\mathrm{i} \norm{\vect{v}_\mathrm{i}}^2 + \sum_\mathrm{i \in \mathcal{R}} \frac{1}{2} m_\mathrm{i} \norm{\vect{v}_\mathrm{i}}^2\\
    & - \sum_{\substack{\mathrm{i,j\in \mathcal{S}}\\ \mathrm{j>i}}} \frac{G m_\mathrm{i} m_\mathrm{j}}{ \norm{\vect{r}_\mathrm{ji} } }
     -\sum_{\substack{\mathrm{i,j\in \mathcal{R}}\\ \mathrm{j>i}}} \frac{G m_\mathrm{i} m_\mathrm{j}}{ \norm{\vect{r}_\mathrm{ji} } }
      -\sum_{\substack{\mathrm{i\in \mathcal{S}}\\ \mathrm{j\in \mathcal{R}}}} \frac{G m_\mathrm{i} m_\mathrm{j}}{ \norm{\vect{r}_\mathrm{ji} } }.
\end{split}    
\end{equation}
Here $T_\mathcal{S}$ and $T_\mathcal{R}$ are the kinetic energy terms of the subsystem and the rest of the particles while $U_\mathcal{SS}$ and $U_\mathcal{RR}$ are the internal potential energies of the two particle sets. The remaining potential energy term $U_\mathcal{SR}$ is an interaction energy term between the subsystems and the rest of the particles. 

A very useful form of this Hamiltonian for fourth-order splitting is obtained by re-arranging the terms as
\begin{equation}\label{eq: splitting-subsys}
\begin{split}
H &= T + U\\
&= T_\mathcal{R} + T_\mathcal{S} + U\\
&= T_\mathcal{R} + T_\mathcal{S} + U + (U_\mathrm{SS} - U_\mathrm{SS})\\
&= T_\mathcal{R} + (U - U_\mathrm{SS}) + (T_\mathcal{S} + U_\mathrm{SS})\\
&\equiv T' + U' + H_\mathcal{S}.
\end{split}
\end{equation}
Specifically, $T'$ is the total kinetic energy excluding the internal kinetic energy of the subsystem and $U'$ is the total potential energy ignoring the contribution of potential energy between the particles in the subsystem. The $H_\mathcal{S}$ is the Hamiltonian of the isolated subsystem.

The corresponding fourth-order forward integrator corresponding the Hamiltonian splitting in Eq. \eqref{eq: splitting-subsys} is
\begin{equation}\label{eq: new-integrator-map}
e^{\epsilon \op{H}} = e^{\frac{1}{6} \epsilon \op{U'}} e^{\frac{1}{2} \epsilon \op{T'}} e^{\frac{1}{2} \epsilon \op{H}_\mathcal{S}} e^{\frac{2}{3} \epsilon \op{\tilde{U'}}} e^{\frac{1}{2} \epsilon \op{H}_\mathcal{S}} e^{\frac{1}{2} \epsilon \op{T'}} e^{\frac{1}{6} \epsilon \op{U^\prime}}.
\end{equation}
In the integrator the drift operator generated by $T'$ moves the particles not included in the subsystem while the kick operator generated by the term $U'$ alters the velocities of the particles ignoring the intra-subsystem forces. Finally, the time evolution operator generated by $H_\mathcal{S}$ governs the dynamics of the subsystem. Note that the internal dynamics of the subsystem is decoupled from the rest of the system so it can be integrated with any accurate few-body solver. In addition, the perturbation of the subsystem motion by the rest of the particles is performed outside the subsystems, i.e. not within regularised or secular integration. Thus the subsystem integration methods do not need to take the local tidal field into account (e.g. by using of perturber particles) making the subsystem integration in \bifrost{} particularly efficient. However, this does not mean that the effect of external perturbations on the subsystem dynamics is ignored: it is taken into account in the terms including $U'$ and $\tilde{U}'$ in Eq. \eqref{eq: new-integrator-map}.

The updated \fsi{} integration algorithm with subsystems is presented in Algorithm \ref{alg: fsi-subsys}. \bifrost{} uses the standard \fsi{} of Eq. \eqref{eq: fsi-cycle} instead of the subsystem algorithm on time-step hierarchy levels containing no subsystems. The approach can be generalised for arbitrary number of subsystems within the main N-body system as
\begin{equation}\label{eq: splitting-subsys-2}
\begin{split}
H &= T + U\\
&= T_\mathcal{R} + \sum_\mathrm{i} T_\mathcal{S}^\mathrm{i} + U\\
&= T_\mathcal{R} + \sum_\mathrm{i} T_\mathcal{S}^\mathrm{i} + U + \sum_\mathrm{i}(U_\mathrm{SS}^\mathrm{i} - U_\mathrm{SS}^\mathrm{i})\\
&= T_\mathcal{R} + (U - \sum_\mathrm{i} U_\mathrm{SS}^\mathrm{i}) + \sum_\mathrm{i} (T_\mathcal{S}^\mathrm{i} + U_\mathrm{SS}^\mathrm{i})\\
&\equiv T' + U' + \sum_\mathrm{i} H_\mathcal{S}^\mathrm{i}.
\end{split}
\end{equation}
We note that the individual subsystem Hamiltonians $H_\mathcal{S}^\mathrm{i}$ are independent of each other. Thus even a large collection of subsystems can be simultaneously integrated in parallel in an efficient manner. This allows for integration of N-body systems with up to $100\%$ of particles in subsystems. For this study we have tested simultaneous parallel integration in \bifrost{} up to a few million binary systems.

\begin{algorithm}
\caption{\texttt{FSI}($\mathcal{P}$, $\epsilon$) with subsystems $\mathcal{S}$}\label{alg: fsi-subsys}
\text{Acceleration calculations ignore intra-subsystem contributions.}\vspace{0.1cm}\\
$\vect{a}_\mathrm{i} \gets$ \texttt{accelerations}($\mathcal{P}$)\vspace{0.1cm}\\
$\vect{v}_\mathrm{i} \gets \vect{v}_\mathrm{i} + \frac{\epsilon}{6} \vect{a}_\mathrm{i}$\Comment{kick}\vspace{0.1cm}\\
\texttt{subsystem\_integration}($\mathcal{S},\epsilon/2$)\vspace{0.1cm}\\
$\vect{r}_\mathrm{i} \gets \vect{r}_\mathrm{i} + \frac{\epsilon}{2} \vect{v}_\mathrm{i}$\Comment{drift single particles}\vspace{0.1cm}\\
$\tilde{\vect{a}_\mathrm{i}} \gets$ \texttt{gradient\_accelerations}($\mathcal{P},\epsilon$)\vspace{0.1cm}\\
$\vect{v}_\mathrm{i} \gets \vect{v}_\mathrm{i} + \frac{2\epsilon}{3} \tilde{\vect{a}}_\mathrm{i}$\Comment{gradient kick}\vspace{0.1cm}\\
\texttt{subsystem\_integration}($\mathcal{S},\epsilon/2$)\vspace{0.1cm}\\
$\vect{r}_\mathrm{i} \gets \vect{r}_\mathrm{i} + \frac{\epsilon}{2} \vect{v}_\mathrm{i}$\Comment{drift single particles}\vspace{0.1cm}\\
$\vect{a}_\mathrm{i} \gets$ \texttt{accelerations}($\mathcal{P}$)\vspace{0.1cm}\\
$\vect{v}_\mathrm{i} \gets \vect{v}_\mathrm{i} + \frac{\epsilon}{6} \vect{a}_\mathrm{i}$\Comment{kick}
\end{algorithm}

\subsection{Finding subsystems}\label{section: subsys-find}

\begin{figure}
\includegraphics[width=\columnwidth]{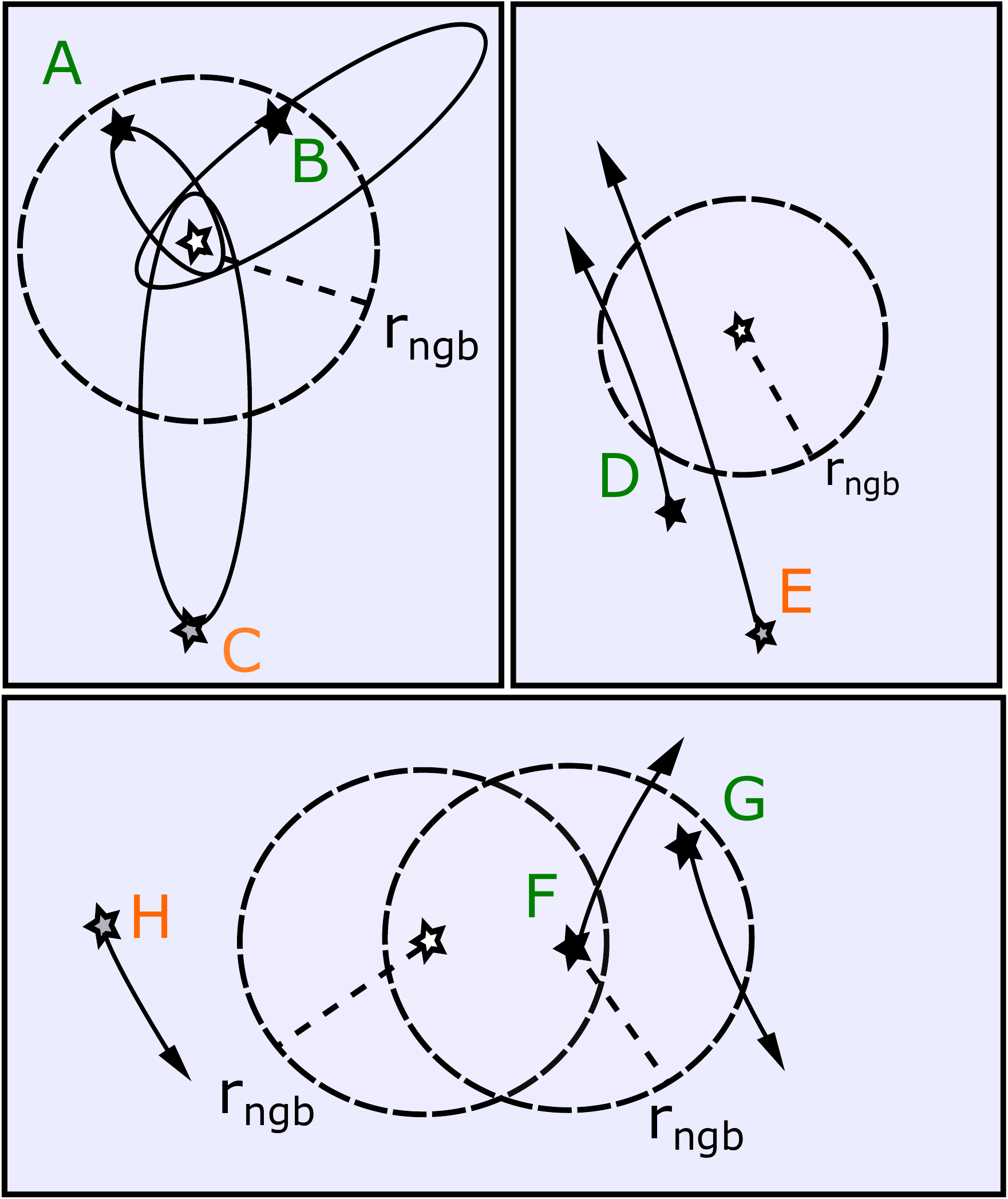}
\centering
\caption{Searching the neighbours of a target star (an open star symbol) in \bifrost. Top left panel illustrates the neighbour criteria for bound particles from Eq. \eqref{eq: ngb-bound}. Particle A fulfils the close binary criteria, B the distant binary near periapsis criteria while particle C is not a neighbour as it is too distant near apoapsis. Top right panel shows the criteria for unbound particles from Eq. \eqref{eq: ngb-unbound}. Particle D is a neighbour as its periapsis distance is smaller than $r_\mathrm{ngb}$ and it will enter inside $r_\mathrm{ngb}$ during the next time-step. Particle E is not a neighbour as it is yet too distant to enter within $r_\mathrm{ngb}$ during the next time-step. The bottom panel illustrates constructing a subsystem of three particles. The target star ends up in a subsystem also containing the particle G even though it is far beyond $r_\mathrm{ngb}$ as they share a common neighbour, particle F. Particle H is not included in the subsystem even though it is closer to the target star than the particle G.}
\label{fig: ngb}
\end{figure}

Before integration, the updated \fsi{} integrator requires information of which particles belong to a subsystem. We perform a neighbour search at the beginning of each integration hierarchy level along with the time-step assignment of the particles. Depending on the number of particles at a time-step hierarchy level we perform the neighbour search either using serial CPU code, MPI-parallelised CPU code or CUDA code on GPUs. The neighbour finding recipe is similar for all the three configurations.

The neighbour search proceeds as follows. First, we compute the orbital elements $a$ (semi-major axis) and $e$ (eccentricity) for each particle pair $i,j$ for which $r_\mathrm{ij} = \norm{\vect{r}_\mathrm{j} - \vect{r}_\mathrm{i} }<r_\mathrm{ngb,max}$. In this study we typically use $r_\mathrm{ngb} = 10^{-3}$ pc and $r_\mathrm{ngb,max} = 0.1$ pc. The neighbour criterion is different for gravitationally bound and unbound particles. Two bound particles $i$ and $j$ are neighbours if either of the two following criteria are fulfilled:
\begin{equation}\label{eq: ngb-bound}
\begin{split}
&a<r_\mathrm{ngb} \hspace{27mm} \text{close binary \hspace{2mm} or}\\
&a \geq r_\mathrm{ngb} \hspace{3mm}\& \hspace{3mm} r_\mathrm{ij} < r_\mathrm{ngb} \hspace{7mm} \text{distant binary near periapsis.}
\end{split}
\end{equation}
For unbound particle pairs, i.e. hyperbolic fly-bys, we consider the periapsis distance $r_\mathrm{peri} = a\:(1-e)$. We reject all neighbour candidates for which $r_\mathrm{peri} \geq r_\mathrm{ngb}$. For pairs with $r_\mathrm{peri} < r_\mathrm{ngb}$ we then estimate whether the particles are close enough to be considered neighbours using their fly-by timescale. The full neighbour criterion for unbound particle pairs is
\begin{equation}\label{eq: ngb-unbound}
\begin{split}
&r < r_\mathrm{ngb} \hspace{52mm} \text{close fly-by}\\
&r \geq r_\mathrm{ngb} \hspace{2mm} \&\; r_\mathrm{peri} < r_\mathrm{ngb} \hspace{2mm} \& \hspace{2mm} C \frac{\norm{ \vect{r}_\mathrm{ij} }}{ \norm{\vect{v}_\mathrm{ij}} } < \tau \hspace{10.5mm} \text{distant fly-by.}
\end{split}
\end{equation}
Here $\tau$ is the pivot time-step of the hierarchy level and $C<1$ is a safety buffer factor. The neighbour criteria for bound and unbound particles are illustrated in the top panels of Fig. \ref{fig: ngb}.

After the neighbours of each particle are known we construct the actual subsystems. This is done efficiently in linear time using graph theory methods: the problem corresponds to finding the components of a finite graph $\mathcal{G}$ \citep{Hopcroft1973}. In our case, $\mathcal{G}$ is the disconnected graph represented by the particles (graph vertexes) and their neighbour information (graph edges). Starting from a particle not included in a subsystem we add particles into a new subsystem by traversing the neighbour graph using a depth-first search until no particles remain. The process is repeated for particles not yet visited until each particle has been visited by the search. The algorithm for constructing the subsystems from the neighbour data requires a negligible amount of time and can be performed using serial code.

\subsection{Time-steps}\label{section: dt}

\subsubsection{Time-stepping in collisional N-body simulations}

Compared to collisionless softened gravitational N-body simulations, time-stepping in collisional simulations is a considerably more delicate issue (e.g. \citealt{Dehnen2011}). In our previous N-body code \frost{} the time-steps are calculated from the mutual free-fall $t_\mathrm{ff}$ and fly-by $t_\mathrm{fb}$ time-scales of the simulation particles in a straightforward manner. In \bifrost{} simulations binary and higher multiple systems as well as their encounters change this picture. The free-fall and fly-by time-step assignment is less robust as simulation particles may suddenly gain velocity in few-body interactions. Examples of such situations are strong three and four-body interactions (e.g. \citealt{Valtonen2006}) which commonly occur in simulations with binary systems. Thus, devising a time-step criterion for simulations which is both robust and computationally efficient is challenging. The typical approach is to combine a number of conservative time-step criteria with a possibility to restart the simulation with more accurate time-step parameters if the energy error becomes too large \citep{Aarseth2003, Wang2015}.

The standard time-step criterion for fourth-order order Hermite integrators is the so-called Aarseth criterion computed from the norms of particle accelerations $\norm{\vect{a}_\mathrm{i}}$ and their three first time derivatives $\norm{\vect{j}_\mathrm{i}} = \norm{\dot{\vect{a}}_\mathrm{i}}$, $\norm{\vect{a}_\mathrm{i}^\mathrm{(2)}}$ and $\norm{\vect{a}_\mathrm{i}^\mathrm{(3)}}$ \citep{Aarseth2003}. The acceleration and its first time derivative (jerk) can be directly computed from the particle positions and velocities while the higher derivatives need to be calculated using the force polynomials and the predictor-corrector method. 

Unfortunately, the widely-used time-step recipe for 4th order Hermite integrators cannot be applied for the 4th order forward symplectic integrators in a straightforward manner: the integrator does not use any time derivatives of acceleration. For the current version of the \bifrost{} code we use the time-step assignment of our previous \frost{} code \citep{Rantala2021} based on local fly-by and free-fall timescales of particles supplemented with a novel criterion derived from the gradient force. As usual the final time-step of a particle is the minimum of the set of its different time-steps.

For the \hhsfsi{} integrator of \frost{} the particle time-steps are always pairwise and computed at the beginning of each time-step hierarchy level. Faster hierarchy levels will eventually contain fewer particles than the slower ones so the time-steps are computed from a smaller set of particles as well. This is not a problem for pair-wise particle time-step criteria, but for global criteria depending on e.g. total particle accelerations, pathological situations may occur. One such example is a hard binary system orbiting a low-density system of massive particles. In \bifrost{} we avoid such issues by enforcing that the time-step of a particle cannot increase within a single \hhsfsi{} integration interval: for each hierarchy level we require $\epsilon_\mathrm{i} = \min{\left( \epsilon_\mathrm{i}, \epsilon_\mathrm{i}^\mathrm{level=0} \right)}$. Thus, the time-steps are always based on the entire set of the simulation particles (excluding the particles within the same subsystem, see below) and not just the particles on the current hierarchy level.

\subsubsection{Assigning time-steps for particles in subsystems}

As \bifrost{} simulations contain subsystems the time-step assignment procedure has a number of differences compared to the \frost{} code. Most importantly, when calculating the time-step $\epsilon_\mathrm{i}$ for particle $i$ belonging to subsystem $\mathcal{S}$ the other particles $j$ in the same subsystem are excluded from the time-step assignment. Thus, the time-step of a particle in a subsystem within the \hhsfsi{} hierarchy only depends on the single particles and particles in different subsystems. If all particles on a single hierarchy level end up in the same subsystem, their equal time-step is the pivot time-step of the level.

Each particle in the same subsystem shares the same time-step according to
\begin{equation}
    \epsilon_\mathrm{i} = \min_{ \mathrm{ j \in \mathcal{S} }} \epsilon_\mathrm{j},
\end{equation}
which is enforced after the individual particle time-steps have been assigned using the time-step criteria.

\subsubsection{Time-steps from free-fall and fly-by timescales}\label{section: dt-ff-fb}

We define the free-fall time-steps of the particle as
\begin{equation}\label{eq: dt-freefall}
\epsilon_\mathrm{ff} = \min_{\mathrm{j \neq i}} \epsilon_\mathrm{ff,ij}  = \eta_\mathrm{ff} \min_{\mathrm{j \neq i}} \left( \frac{ \norm{ \vect{r}_\mathrm{ij} }^3 }{ G (m_\mathrm{i}+m_\mathrm{j}) }\right)^{1/2},
\end{equation}
in which the index $j$ runs over all other particles. The fly-by time-step is defined very similarly, namely as
\begin{equation}\label{eq: dt-flyby}
\epsilon_\mathrm{fb} = \min_{\mathrm{j \neq i}} \epsilon_\mathrm{fb,ij}  = \eta_\mathrm{fb} \min_{\mathrm{j \neq i}} \frac{\norm{ \vect{r}_\mathrm{ij}}}{\norm{ \vect{v}_\mathrm{ij} }},
\end{equation}
both definitions following \cite{Pelupessy2012,Janes2014,Rantala2021}.

The two time-step criteria are equal for a bound binary on a circular orbit while for an eccentric binary the fly-by criteria yields smaller time-steps, especially near the periapsis. In collisional N-body systems the fly-by time-step is typically, but not always, smaller than the free-fall time-step. The main benefit of the free-fall time-step is to ensure that the time-step remains small enough even when the relative velocity w.r.t. nearby particles is low, e.g. when a particle is on an eccentric orbit around a massive particle near its apoapsis.

\subsubsection{Time-steps from the FSI gradient acceleration}\label{section: dt-acc}

Next we describe a novel time-step criterion derived from the properties of the fourth-order forward integrator. The change in particle velocities $\Delta \vect{v}_\mathrm{i}$ in the gradient kick operation of \fsi{} is
\begin{equation}
\begin{split}
\Delta \vect{v}_\mathrm{i} &= \frac{2 \epsilon}{3} \tilde{\vect{a}}_\mathrm{i}\\ &= \frac{2 \epsilon}{3} \left( \vect{a}_\mathrm{i} + \frac{\epsilon^2}{24} \sum_\mathrm{j \neq i} \frac{G m_\mathrm{j}}{ \norm{\vect{r}_\mathrm{ji}}^5 } \Big( \norm{\vect{r}_\mathrm{ji}}^2 \vect{a}_\mathrm{ji} - 3 ( \vect{a}_\mathrm{ji} \cdot \vect{r}_\mathrm{ji}) \vect{r}_\mathrm{ji}  \Big) \right)\\
&\equiv \frac{2 \epsilon}{3} \left( \vect{a}_\mathrm{i} + \frac{\epsilon^2}{48} \vect{g}_\mathrm{i} \right) \equiv \frac{2 \epsilon}{3} \left( \vect{a}_\mathrm{i} + \epsilon^2\vect{g}_\mathrm{i}^{\prime} \right),
\end{split}
\end{equation}
using Eq. \eqref{eq: kick-gradient-new} and Eq. \eqref{eq: new-grad-acc}. We also use the definition from Eq. (28) of \cite{Rantala2021} in the last line as a useful shorthand notation. Several time-step criteria can be devised from the expression for the velocity change. One can limit the change in velocity, or limit the ratio of the Newtonian and gradient accelerations as they scale differently proportional to the time-step, $\epsilon$ and $\epsilon^3$.

The first time-step criteria involving accelerations is the commonly used velocity-per-acceleration criterion (e.g. \citealt{Springel2001,Wetzstein2009}) defined as
\begin{equation}\label{eq: vel-acc-crit}
\epsilon = \eta \frac{\norm{ \vect{v}_\mathrm{i} }}{ \norm{ \vect{a}_\mathrm{i} } }.
\end{equation}
The gradient contribution $\tilde{\vect{a}}_\mathrm{i} - \vect{a}_\mathrm{i}$ to the velocity change can be limited as well. Analogously to Eq. \eqref{eq: vel-acc-crit} we can write the ratio of the velocity norm and the gradient contribution norm as
\begin{equation}
\frac{\norm{ \vect{v}_\mathrm{i} }}{ \norm{ \tilde{\vect{a}}_\mathrm{i} - \vect{a}_\mathrm{i} } } = \frac{\norm{ \vect{v}_\mathrm{i} }}{ \epsilon^2 \norm{\vect{g}_\mathrm{i}^{\prime} } }
\end{equation}
from which we can obtain our second time-step criterion as
\begin{equation}\label{eq: dt-nabla1}
\epsilon = \eta_{\nabla} \left( \frac{\norm{ \vect{v}_\mathrm{i} }}{ \norm{\vect{g}_\mathrm{i}^{\prime} } } \right)^{1/3}.
\end{equation}
Here $\eta_{\nabla}$ is the user-defined gradient time-step parameter.
Finally, we can limit the ratio of the Newtonian and gradient accelerations resulting in our final gradient time-step criterion defined as
\begin{equation}\label{eq: dt-nabla2}
\epsilon = \eta_{\nabla} \left( \frac{\norm{ \vect{a}_\mathrm{i} }}{ \norm{\vect{g}_\mathrm{i}^{\prime} } } \right)^{1/2}.
\end{equation}
We refer the time-steps from criteria of Eq. \eqref{eq: dt-nabla1} and Eq. \eqref{eq: dt-nabla2} collectively as the gradient time-steps $\epsilon_\nabla$.
The gradient time-step criteria work well in our test simulations and the energy conservation properties of \bifrost{} are more robust than only using the free-fall and fly-by criteria above. All in all, we argue that the time-step criteria for collisional N-body simulation codes should include at least one criterion which is derived from the properties of the integrator itself, and not just using time-steps from general time-scale arguments alone.

We note a somewhat related time-step criterion for N-body simulations from the literature based on the matrix norm of the tidal tensor $\matr{T} = \nabla \vect{a}$ \citep{Grudic2020}. While tidal tensor being the true spatial gradient of the accelerations we still refer our time-step criterion as gradient time-steps as the relevant accelerations arise from the gradient potential $\nabla \norm{\vect{a}}^2 = 2 \vect{a} \nabla \vect{a}$ \citep{Rantala2021}.

\subsubsection{Time-steps from the time derivative of acceleration}\label{section: dt-jerk}

The final new time-step criterion in \bifrost{} is the so-called jerk criterion which limits the rate of change of acceleration $\vect{j}_\mathrm{i} = \dot{\vect{a}}_\mathrm{i}$ of the simulation particles. The time derivative of acceleration can be computed directly from the position and velocity data of the particles as
\begin{equation}
\vect{j}_\mathrm{i} = \dot{\vect{a}}_\mathrm{i} = \sum_\mathrm{j \neq i} \frac{G m_\mathrm{j} }{\norm{\vect{r}_\mathrm{ji} }^5} \left( {\norm{\vect{r}_\mathrm{ji} }^2} \vect{v}_\mathrm{ji} - 3 (\vect{r}_\mathrm{ji} \cdot \vect{v}_\mathrm{ji}) \vect{r}_\mathrm{ji} \right)
\end{equation}
just as in the Hermite codes. The jerk time-step is defined simply as
\begin{equation}
\epsilon_\mathrm{jerk} = \eta_\mathrm{j} \frac{\norm{\vect{a}_\mathrm{i}}}{\norm{\vect{j}_\mathrm{i}}}
\end{equation}
in which $\eta_\mathrm{j}$ is another user-defined time-step accuracy parameter. In the simulations of this study we set $\eta_\mathrm{j} = \eta_\nabla \sim 0.1$.

\subsubsection{Time-steps for subsystems from external perturbations}
Finally, we ensure that individual subsystems have sufficiently short time-steps that the contribution of the external perturbations on the internal dynamics of the subsystems is adequately taken into account. For systems of two bodies, we use a simple criterion of
\begin{equation}
    \epsilon_\mathrm{pert} = \eta_\mathrm{pert} \left\{ \left|\frac{a}{\dot{a}} \right|, \left| \frac{1}{\dot{e}} \right|\right\}
\end{equation}
based on the changes of the semi-major axis $a$ and eccentricity $e$ due to the external perturbations. The rates of change $\dot{a}$ and $\dot{e}$ can be evaluated using the particle positions, velocities and accelerations in a straightforward manner. For subsystems with more than two particles, we use the total energy and angular momentum and their time derivatives instead of the semi-major axis and eccentricity. In practice, we use typically $\eta_\mathrm{pert} \sim 0.02$.

\subsection{Regularised integration of small subsystems}\label{section: logh}

We include a regularised integrator into the \bifrost{} code to integrate subsystems with high numerical accuracy. Due to the hierarchical nature of the \hhsfsi{} integration the fastest time-step levels of the time-step hierarchy contain subsystems, and depending on the simulated system their number may be high. Most of these are two-body systems. Thus, instead of parallelising the force calculation within a large subsystem \citep{Rantala2020} it is more convenient to integrate a large number of small subsystems simultaneously in parallel. The regularisation algorithm of our choice is the so-called Logarithmic Hamiltonian (\logh{} hereafter) originally discovered by \cite{Mikkola1999} and \cite{Preto1999}, and used subsequently in several other works (\citealt{Mikkola2006,Rantala2017,Rantala2020,Wang2020}, \citealt{Wang_Yi_Han2021}). As most of our subsystems are two-body systems we do not use special chained \citep{Mikkola1993} or minimum spanning tree \citep{Rantala2020} coordinate systems in our \logh{} implementation. The so-called slow-down procedure \citep{Mikkola1996,Wang2020} for perturbed binaries is also unnecessary due to the hierarchical nature of integration in the main \hhsfsi{} code as the expensive evaluation of the perturber forces is not needed. The Gragg-Bulirsch-Stoer (GBS) extrapolation method (e.g. \citealt{Gragg1965,Bulirsch1966,Deuflhard1983,Press2007,Wang_Yi_Han2021}) is used to ensure that the relative error of each dynamical variable remains smaller than the user-defined GBS tolerance $\eta_\mathrm{GBS} \sim 10^{-10}$ during each integration interval. For the end-time iteration tolerance described in Section 2.4 of \cite{Rantala2020} we use a value of $10^{-4}$.

Following the algorithm of \cite{Mikkola2006,Mikkola2008} we time-transform the equations of motion of the N-body subsystem by introducing a new independent variable $s$ with a definition of
\begin{equation}
\derfrac{s}{t} = -U \equiv \bar{U} = \sum_{\mathrm{j>i}} \frac{G m_\mathrm{i} m_\mathrm{j} }{\norm{\vect{r}_\mathrm{j}-\vect{r}_\mathrm{i}}}.
\end{equation}
The previous independent variable, time, becomes a coordinate-like quantity. Using the definition of the binding energy (the canonical conjugate variable of time in the new extended phase-space) we have $\bar{U} = T+B$ the N-body equations of motion become
\begin{equation}\label{eq: logh_coordinates}
\begin{split}
\derfrac{t}{s} &= \frac{1}{T+B}\\
\derfrac{\vect{r}_\mathrm{i}}{s} &= \frac{1}{T+B} \vect{v}_\mathrm{i}
\end{split}
\end{equation}
for the coordinate variables. The velocity equations for the binding energy, actual velocities and particle spins $\vect{S}$ are written as
\begin{equation}\label{eq: logh_velocities}
\begin{split}
\derfrac{B}{s} &= -\frac{1}{\bar{U}} \sum_\mathrm{i} m_\mathrm{i} \vect{v}_\mathrm{i} \cdot \left[ \vect{f}_\mathrm{i} + \vect{g}_\mathrm{i}(\vect{v},\vect{S}) \right]\\
\derfrac{\vect{v_\mathrm{i}}}{s} &= \frac{1}{\bar{U}} \left[ \vect{a}_\mathrm{i} + \vect{f}_\mathrm{i} + \vect{g}_\mathrm{i}(\vect{v},\vect{S}) \right]\\
\derfrac{\vect{S}_\mathrm{i}}{s} &= \frac{1}{\bar{U}} \vect{\Omega}_\mathrm{i} \mathbf{\times} \vect{S}_\mathrm{i}
\end{split}
\end{equation}
in which $\vect{a}_\mathrm{i}$ are Newtonian gravitational accelerations, $\vect{f}_\mathrm{i}$ optional velocity-independent accelerations and $\vect{g}_\mathrm{i}$ are velocity- and or spin-dependent accelerations. The term $\vect{\Omega}_\mathrm{i} \mathbf{\times} \vect{S}_\mathrm{i}$ describes the spin derivatives. In the absence of velocity-dependent accelerations the equations of motion in Eq. \eqref{eq: logh_coordinates} and Eq. \eqref{eq: logh_velocities} can be integrated with a standard leapfrog algorithm. For a Keplerian binary this yields the exact orbit within numerical precision the only error being in the phase of the binary \citep{Mikkola1999,Preto1999}. If velocity-dependent accelerations are present a more complex integration strategy is required which is discussed in the next Section. For more details of the regularisation algorithm and its implementation see e.g. \cite{Rantala2017,Rantala2020}.

\subsection{Velocity- and spin-dependent accelerations in subsystems}\label{section: pnacc}

\subsubsection{Explicit integration strategy with velocity-dependent accelerations}\label{section: auxiliary_integration}

If velocity-dependent terms such as relativistic post-Newtonian corrections (e.g. \citealt{Poisson2014}) or drag forces from e.g. stellar tides \citep{Samsing2018} are present in the equations of motion in Eq. \eqref{eq: logh_velocities} the variables on the left-hand side are also present in the right-hand side and an explicit leapfrog algorithm cannot be constructed. An implicit iterative integration is an option but is very inefficient. A clever trick to make the explicit leapfrog possible is to further extend the phase-space of the system by introducing an auxiliary velocity variable $\vect{w}_\mathrm{i}$ \citep{Hellstrom2010}. Ignoring spins for now the velocity equations become
\begin{equation}
\begin{split}
\derfrac{B}{s} &= -\frac{1}{\bar{U}} \sum_\mathrm{i} m_\mathrm{i} \vect{v}_\mathrm{i} \cdot \left[ \vect{f}_\mathrm{i} + \vect{g}_\mathrm{i}(\vect{w}) \right]\\
\derfrac{\vect{v}_\mathrm{i}}{s} &= \frac{1}{\bar{U}} \left[ \vect{a}_\mathrm{i} + \vect{f}_\mathrm{i} + \vect{g}_\mathrm{i}(\vect{w}) \right]\\
\derfrac{\vect{w}_\mathrm{i}}{s} &= \frac{1}{\bar{U}} \left[ \vect{a}_\mathrm{i} + \vect{f}_\mathrm{i} + \vect{g}_\mathrm{i}(\vect{v}) \right]
\end{split}
\end{equation}
for the original i.e. physical and the auxiliary variables with initial values of $\vect{w}_\mathrm{i}(t_\mathrm{0}) = \vect{v}_\mathrm{i}(t_\mathrm{0})$. Using operator formalism, the kick operator can be written as
\begin{equation}
e^\mathrm{\epsilon \op{U}} = e^\mathrm{\frac{1}{2}\epsilon \op{U}_\mathrm{aux}} e^\mathrm{\epsilon \op{U}_\mathrm{phys}} e^\mathrm{\frac{1}{2}\epsilon \op{U}_\mathrm{aux}}.
\end{equation}
The algorithm can be generalised to the case in which spins are present, only a set of auxiliary spin variables $\vect{Z}_\mathrm{i}$ is required (see \citealt{Rantala2017} for details).

\subsubsection{Spin-independent post-Newtonian accelerations}

We have included an option in \bifrost{} to use post-Newtonian equations of motion for the simulation particles in subsystems instead of the common Newtonian equations of motion. The incorporation of the post-Newtonian terms in the equations of motion enables a number of important relativistic effects including the periapsis advance (leading term PN1.0), and shrinking and circularization of binary orbits due to gravitational wave radiation reaction forces (leading term PN2.5). Our PN implementation for \bifrost{} is close to identical to the one in the \ketju{} code (\citealt{Rantala2017,Mannerkoski2019,Mannerkoski2021,Mannerkoski2022}).

The post-Newtonian terms can be summarised as writing the total expression for the particle accelerations as
\begin{equation}
\derfrac{\vect{v}}{t} = \vect{a} + \vect{a}_\mathrm{PN1.0} + \sum_\mathrm{k=4}^\mathrm{7} \frac{1}{c^k} \vect{a}_\mathrm{PN\hspace{0.35mm}k/2}
\end{equation}
the highest term corresponding to $k=7$ being PN3.5. The PN1.0 term in our implementation originates from \cite{Thorne1985} including terms up to three bodies \citep{Einstein1938,Will2014,Lim2020}. The higher-order terms (PN2.0, PN2.5, PN3.0 and PN3.5) are for binary systems only and are adopted from \cite{Blanchet2014}.

\subsubsection{Spin evolution \& spin-dependent post-Newtonian accelerations}

If spin-dependent post-Newtonian terms are included in the equations of motion of the particles in the subsystems the particle spins will evolve according to PN equations of motion. In addition, the particles orbits themselves will also evolve due to spin-orbit coupling and the conservation of post-Newtonian angular momentum in the used formulation. In typical applications only black holes have spins large enough for the spin-dependent post-Newtonian terms to have any effect on the evolution of the systems. The magnitude of the particle spins is described using the common dimensionless black hole spin parameter $s$ defined as
\begin{equation}
s = \frac{c \norm{\vect{S}} }{G m^2}
\end{equation}
with $0\leq s < 1$ in which $m$ is the mass and $\vect{S}$ is the spin vector of the particle. The spin vectors of the particles evolve as
\begin{equation}
\begin{split}
\derfrac{\vect{S}_\mathrm{i}}{t} &= \vect{\Omega}_\mathrm{PN} \mathbf{\times} \vect{S}_\mathrm{i}\\
\vect{\Omega}_\mathrm{PN} &= \vect{\Omega}_\mathrm{SO} + \vect{\Omega}_\mathrm{SS} + \vect{\Omega}_\mathrm{Q}
\end{split}
\end{equation}
in which the $\Omega_\mathrm{PN}$ term is contributed from the spin-orbit ($\vect{\Omega}_\mathrm{SO}$, i.e. geodetic, mass, or de Sitter precession), spin-spin ($\vect{\Omega}_\mathrm{SS}$, frame dragging or Lense-Thirring effect) and quadrupole ($\vect{\Omega}_\mathrm{Q}$) interactions \citep{Poisson2014}. The formulas for the different spin-dependent post-Newtonian terms are adopted from \cite{Thorne1985} in our current implementation.

The spin PN formulation in our current code version is conservative so the total angular momentum $\vect{J}$ of a few-body system is conserved as
\begin{equation}
\derfrac{\vect{J}}{t} = \derfrac{\vect{L}}{t} + \sum_\mathrm{i} \derfrac{\vect{S}_\mathrm{i}}{t} = 0.
\end{equation}
Thus, the orbits of the bodies will react to the evolution of the particle spins. Especially for a binary system we have
\begin{equation}
\derfrac{\vect{L}}{t} = - \left( \derfrac{\vect{S}_\mathrm{1}}{t} + \derfrac{\vect{S}_\mathrm{2}}{t} \right).
\end{equation}
Both the shape of the orbit and the orientation of the orbital plane change due to the evolving particle spins. Again, the spin effects are small for systems that do not contain a rapidly spinning black hole.

\subsection{Global velocity-dependent accelerations: the PN1.0 term}\label{section: global-pn1}

\begin{figure*}
\includegraphics[width=0.81\textwidth]{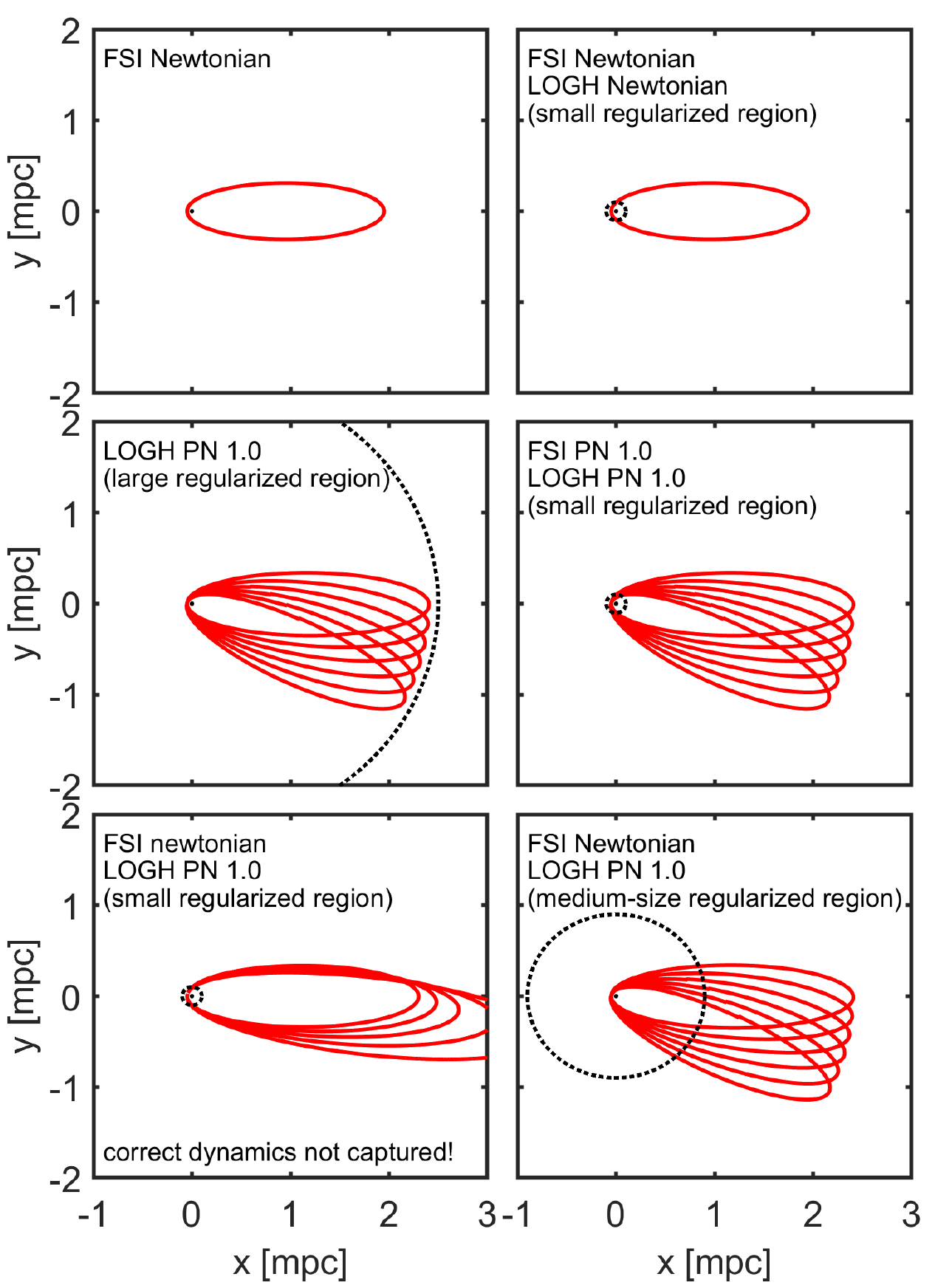}
\caption{An illustration of the importance of the global PN1.0 term. The figure presents a comparison of the effect of including the post-Newtonian PN1.0 term in the equations of motion of a binary system inside the regularised regions (\logh{}, dotted circle) and outside it (FSI). In the two top panels we show that without the PN term the forward integrator alone and with a small regularised region the orbit of the binary is the same. This confirms that the smaller particle crossing in and out of the regularised region does not cause spurious numerical effects in the Newtonian case. The middle left panel shows the fiducial result of the PN1.0 orbital precession when the binary orbit is always within the regularised region. In the middle right panel we show that when the global PN1.0 term applied also outside the regularised region the size of the region can be smaller with dynamics identical to the fully regularised case. The two bottom panels show that without the global PN1.0 term the size of the regularised region compared to the size of the binary orbit determines whether correct dynamics is captured as discussed in the main text. If the regularised region is too small (bottom left panel) spurious effects such as suppression of the orbital precession and increase of the semi-major axis of the binary appear. We argue that the global PN1.0 term is a safe and straightforward solution to avoid the issues with the post-Newtonian terms with finite-radius subsystems in N-body simulations including relativistic binaries.}
\label{fig: PN}
\end{figure*}

In typical N-body simulation codes post-Newtonian terms are only included in the equations of motion of simulation particles in regularised subsystems (e.g. \citealt{Aarseth2012}). There are two main reasons for this. First, the post-Newtonian accelerations are vanishingly small compared to Newtonian accelerations in most stellar systems. The second reason is purely computational as even the simple two-body PN formulation introduces a large number of terms to be evaluated compared to the Newtonian equations of motion. However, due to the cumulative nature of post-Newtonian effects and and finite radius of the subsystems there are physical scenarios in which the PN only in subsystems approach fails.

Writing down the equations of motion of an N-body system with a PN region and a Newtonian region with a radius $r_\mathrm{subsys}$ further qualifies the issue. The equations of motion of such a system are
\begin{equation}\label{eq: subsyspn}
\begin{split}
\derfrac{\vect{r}_\mathrm{i}}{t} &= \vect{v}_\mathrm{i}\\
\derfrac{\vect{v}_\mathrm{i}}{t} &= \vect{a}_\mathrm{Newton} + \sum_\mathrm{j} \left[ 1-H\left( \frac{r_\mathrm{ij}}{r_\mathrm{subsys}} \right) \right]\; \vect{a}_\mathrm{j,PN}
\end{split}
\end{equation}
in which $r_\mathrm{ij}=\norm{\vect{r}_\mathrm{ij}}$ and $H(x)$ is the Heaviside step function. Within the subsystem $(r<r_\mathrm{subsys})$ the equations of motion are PN1.0 while outside the motion is Newtonian. For a bound binary system there are two possibilities how the equations of motion of Eq. \eqref{eq: subsyspn} may lead to unphysical behaviour. First, if the subsystem size is set to a very small value, then we always have $r>r_\mathrm{subsys}$ and the post-Newtonian terms are always ignored. The second option is that the equations of motion are post-Newtonian only on the inner parts of the orbit, and  Newtonian elsewhere. In this case the resulting orbit is not described correctly by the Newtonian or the post-Newtonian orbit. We consider this kind of partially post-Newtonian orbits unphysical. In order to obtain the correct post-Newtonian orbit at all times is to have $r<r_\mathrm{subsys}$.

A straightforward solution is to include the post-Newtonian terms in the equations of motion of the particles outside the subsystems as well. Hereafter we collectively call these terms the global PN1.0 term. The increase of the computational cost by the extra terms can be limited by applying the PN accelerations only for pairwise interactions when either of the particles of is massive enough, e.g. $m_\mathrm{\star}>m_\mathrm{PN} \sim 5 M_\mathrm{\odot}$ which is the approximate lower limit for a mass of a stellar-mass black hole. Now the equations of motion are
\begin{equation}\label{eq: subsyspn-mass}
\begin{split}
\derfrac{\vect{r}_\mathrm{i}}{t} &= \vect{v}_\mathrm{i}\\
\derfrac{\vect{v}_\mathrm{i}}{t} &= \vect{a}_\mathrm{Newton} + \sum_\mathrm{j}  H\left( \frac{m_\mathrm{j}}{m_\mathrm{PN}}\right)\; \vect{a}_\mathrm{j,PN}.
\end{split}
\end{equation}
In practice in the \fsi{} code we perform the post-Newtonian kicks for the simulation particles as 
\begin{equation}
e^\mathrm{ \epsilon \op{H}_\mathrm{FSI+PN} } = e^\mathrm{ \frac{1}{2} \epsilon \op{U}_\mathrm{PN} } e^\mathrm{ \epsilon \op{H}_\mathrm{FSI} } e^\mathrm{ \frac{1}{2} \epsilon \op{U}_\mathrm{PN} }.
\end{equation}
The PN kicks exclude the contribution from the particles in the same subsystem to avoid double-counting of accelerations, just as in the \fsi{} with subsystems in Eq. \eqref{eq: new-integrator-map}.

An illustrative example of a bound two-body system is presented in Fig. \ref{fig: PN}. When the size of the regularised region is large (encompassing the orbit of the binary) the integration is always correctly post-Newtonian. When shrinking the subsystem radius without the global PN1.0 term the orbit of the system will eventually deviate from the correct solution, leading to clearly unphysical behaviour such as decreasing periapsis precession rate or even increase of the semi-major axis of the system. With the global PN1.0 term included the dynamics of the system remains correct even when the size of the subsystem is very small compared to the size of the orbit. It is evident that the accuracy of the post-Newtonian dynamics of the system might still depend on a user-given parameter ($m_\mathrm{PN}$) but we argue that the global PN1.0 with a mass cut has considerably less severe issues than the finite PN radius approach.

\subsection{Regularised integration using \mstar}\label{section: mstar}

\begin{figure}
\includegraphics[width=\columnwidth]{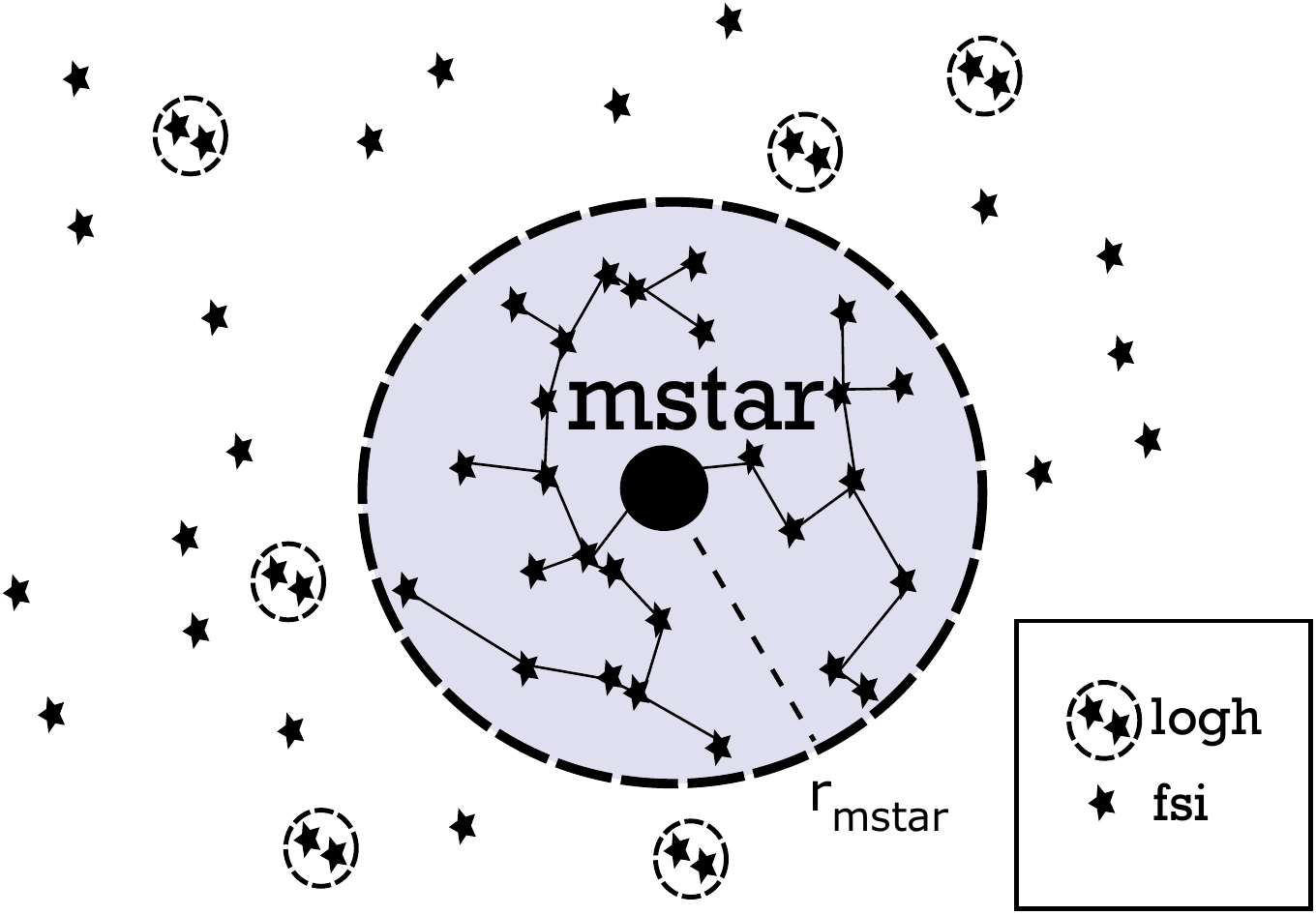}
\caption{An illustration of a regularised \mstar{} region (in light blue) embedded in the simulation domain of the \fsi{} integrator with small regularised subsystems (small dashed circles).}
\label{fig: ngb-mstar}
\end{figure}

We include a possibility to use the regularised \mstar{} integrator \citep{Rantala2020} instead of the forward integrator of the \bifrost{} code. For simulations containing fewer than $\sim$ a few thousand particles it is possible to use \mstar{} instead of the entire \hhsfsi{} integration algorithm. For larger simulations \mstar{} can replace the hierarchy levels with smallest time-steps provided that the particle number of the levels is not too high.

Typical simulations in which the use of \mstar{} is profitable are runs which include one or few SMBHs (or IMBHs) embedded in a tightly bound cluster. A regularised region encompassing a fraction of the influence radius of the massive BH increases the integration accuracy and the running speed of the \bifrost{} code. This technique has been successfully used in a number of simulations using the \ketju{} code (\citealt{Rantala2017,Rantala2018,Mannerkoski2021,Mannerkoski2022}). An illustration of a region integrated by \mstar{} within a \bifrost{} simulation is presented in Fig. \ref{fig: ngb-mstar}.

The regularised integrator \mstar{} and the \logh{} integrator reviewed in Section \ref{section: logh} operate in a closely similar manner both being based in algorithmic regularisation techniques. The main differences are the use of the minimum spanning tree (MST) coordinates and the powerful two-fold parallelization of force loops and GBS sub-step divisions in \mstar{}. These code features make \mstar{} somewhat more accurate and considerably faster than \logh{}, especially for larger simulation systems. The implementation of the post-Newtonian terms in \mstar{} is identical to the one in \bifrost{} presented in Section \ref{section: pnacc}. For additional details of the \mstar{} code see \cite{Rantala2020}.

\subsection{Secular orbit evolution of binary systems}\label{section: secular}

Despite parallelization, a large number of short-period post-Newtonian binary systems may make regularised integration inefficient. Instead of our regularised integrators, we use a secular integration method if the number or binary orbits per the current binary time-step $N_\mathrm{orb} = P_\mathrm{bin}/\epsilon$ is larger than user-given threshold the $N_\mathrm{orb,sec}$, typically of the order of a few. If post-Newtonian terms are not used our secular integrator reduces to a Kepler solver described in Section \ref{section: keplersolver}. The effect of the external perturbations on the dynamics of binary systems are again taken into account in the kick operations of \fsi{}.

The secular equations of motion describe the evolution of the semi-major axis $a$, the orbital eccentricity $e$ and the argument of periapsis $\omega$. We take into account the post-Newtonian evolution originating from the terms PN1.0, PN2.0 and PN2.5. The PN2.5 term responsible for the circularization and shrinking of the orbit changes the semi-major axis and the eccentricity of the binary \citep{Peters1964} as
\begin{equation}\label{eq: secular_a_ecc}
\begin{split}
\left\langle \derfrac{a}{t} \right\rangle_\mathrm{sec} &= -\frac{64}{5} \frac{\beta(m_\mathrm{1},m_\mathrm{2})}{a^3} F(e)\\
\left\langle \derfrac{e}{t} \right\rangle_\mathrm{sec} &= -\frac{304}{15} \frac{\beta(m_\mathrm{1},m_\mathrm{2})}{a^4} e G(e)
\end{split}
\end{equation}
in which the auxiliary functions $\beta(m_\mathrm{1},m_\mathrm{2})$, $F(e)$ and $G(e)$ are defined as
\begin{equation}\label{eq: gw_aux_functions}
\begin{split}
\beta(m_\mathrm{1},m_\mathrm{2}) &= \frac{G^3 m_\mathrm{1} m_\mathrm{2} (m_\mathrm{1}+m_\mathrm{2})}{c^5}\\
F(e) &= \frac{1+\frac{73}{24} e^2 + \frac{37}{96}e^4}{\left( 1-e^2\right)^{7/2}}\\
G(e) &= \frac{1+\frac{121}{304} e^2}{\left( 1-e^2\right)^{5/2}}.
\end{split}
\end{equation}
The periapsis of the orbit advances due to the PN1.0 and PN2.0 terms as
\begin{equation}\label{eq: secular_omega}
\begin{split}
\left\langle \derfrac{\omega}{t} \right\rangle_\mathrm{sec} &= \frac{6 \pi G}{c^2 P}\frac{M}{a(1-e^2)} + \frac{3(18+e^2) \pi}{2c^4 P} \left[ \frac{GM}{a(1-e^2)} \right]^2.
\end{split}
\end{equation}
The second PN2.0 term behaves similarly as the common PN1.0, however causing a weaker effect as it scales proportional to $c^\mathrm{-4}$ instead of $c^\mathrm{-2}$.

We integrate the secular equations of motion of Eq. \eqref{eq: secular_a_ecc} and Eq. \eqref{eq: secular_omega} using a second-order leapfrog integrator. As the time derivatives of the semi-major axis and eccentricity depend on the values of $a$ and $e$ we again use the doubling of phase space method from Section \ref{section: auxiliary_integration} by introducing an auxiliary semi-major axis and an auxiliary eccentricity variable. This yields an explicit leapfrog integration algorithm. The time-steps of the integration are obtained from
\begin{equation}
\epsilon_\mathrm{sec} = \eta_\mathrm{sec} \min{ \left\{ \frac{a}{\left\langle \derfrac{a}{t} \right\rangle}_\mathrm{sec}, \frac{e}{\left\langle \derfrac{e}{t} \right\rangle}_\mathrm{sec}, \frac{\omega}{\left\langle \derfrac{\omega}{t} \right\rangle}_\mathrm{sec} \right\} }    
\end{equation}
with the user-given accuracy parameter $\eta_\mathrm{sec} \sim 0.01$. We note that the secular orbital motion is not exactly the same as the true post-Newtonian motion as e.g. the particle orbital velocities are still Keplerian. However, the computationally efficient secular method captures the long-term post-Newtonian evolution of the binary orbital elements.

The secular integration proceeds in the \bifrost{} code in the following manner. When a binary system is selected for secular integration we first compute the classical Keplerian orbital elements $a$, $e$, $\omega$, inclination $i$, the longitude of the ascending node $\Omega$ and mean anomaly $M$ as
\begin{equation}
\{ \vect{r}_\mathrm{i}, \vect{v}_\mathrm{i}\} \rightarrow \{ a, e, \omega, i, \Omega, M\}.
\end{equation}
After the secular integration we update the positions and velocities of the binary components by performing the transformation 
\begin{equation}
\{ a, e, \omega, i, \Omega, M\} \rightarrow \{ \vect{r}_\mathrm{i}, \vect{v}_\mathrm{i}\}.
\end{equation}
using a Kepler solver discussed in the next Section. We note that the phase of the binary is essentially lost when post-Newtonian orbits are used as we only track secular changes in the orbital elements.

\begin{figure}
\includegraphics[width=\columnwidth]{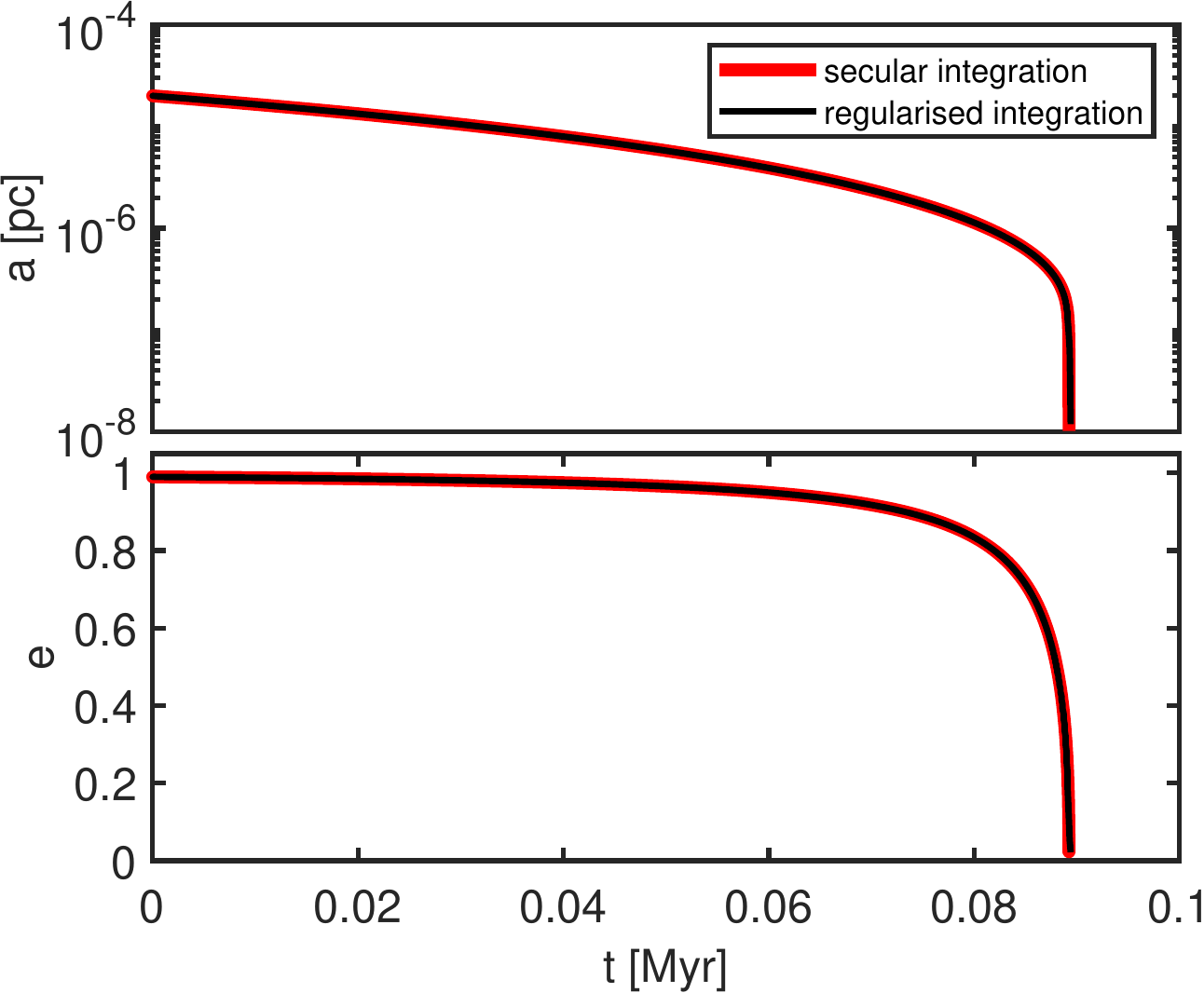}
\caption{A comparison of inspiral and merger of two non-spinning intermediate-mass black holes ($M=1000 M_\mathrm{\odot}$, $q=0.5$) using either regularised (\logh{}) or secular integration (essentially the \citealt{Peters1964} formula) in PN2.5. The results agree very well.}
\label{fig: pn-compare}
\end{figure}

We compare the secular and regularised integration methods for an in-spiraling IMBH binary in Fig. \ref{fig: pn-compare}. The only PN term switched on here is PN2.5. The total mass of the system is $M=1000\;M_\odot$ with a mass ratio $q = m_\mathrm{2}/m_\mathrm{1} = 0.5$. The initial semi-major axis of the binary is set to $a=2\times10^{-5}$ pc and the initial eccentricity is $e=0.99$. The binary shrinks, circularizes and merges rapidly and the evolution of the orbital elements is essentially identical with the two integration methods, with secular integration being orders of magnitude faster.

\subsection{Kepler solver}\label{section: keplersolver}

We implement the solver for Kepler's equation
\begin{equation}
M = E-e \sin{(E)}
\end{equation}
in which $M$ is the mean anomaly and $E$ is the eccentric anomaly, following the approach of \cite{Mikkola2020}. As we use secular integration only for bound binary systems in \bifrost{} we use the standard form of Kepler's equation instead of the elegant but somewhat complex universal variable formulation (see e.g. \citealt{Danby1992,Rein2015, Wisdom2015}).

\subsubsection{Mikkola's cubic approximation}
The first approximation for the eccentric anomaly $E$ is obtained using the cubic approximation of \cite{Mikkola1987}. Defining a new variable $s$ as
\begin{equation}
s = \sin{\left(\frac{E}{3}\right)}     
\end{equation}
the Kepler's equation can be rewritten as
\begin{equation}
\arcsin{(s)} -e\left(s-\frac{4}{3}s^3\right) = \frac{M}{3}.
\end{equation}
Now expanding the $\arcsin{(s)}$ up to third power in $s$ and solving the cubic equation the solution for $s$ (and thus also for $E$) one obtains the result
\begin{equation}
\begin{split}
s &= K \sinh{\left[ \frac{1}{3} \myarsinh{\left( \frac{M}{(1-e)K} \right) } \right]}\\
K &= \left( \frac{1-e}{\frac{1}{a_\mathrm{K}} + \frac{4}{3} e} \right)^{1/2}.
\end{split}
\end{equation}
with the constant $a_\mathrm{K} = 6$ in the standard formulation. An further empirical improved solution can be obtained by setting $a_\mathrm{K} = 6-0.768688675 M$. With this correction the approximate solution for $s$ is exactly correct when $M=\pi$. Overall the cubic approximation with the empirical refinement gives the correct value for the eccentric anomaly within $\sim$ three significant digits \citep{Mikkola1987}. Even though the cubic approximation is typically alone not accurate enough for high-accuracy applications it can be used as a first approximation for $E$ for refinement with subsequent iterative methods, as we do in the code. One should also note that the method cannot be used for parabolic orbits ($e=1$). In this case one can use a slightly perturbed $\tilde{e} = e-10^{-6}$ instead of the original eccentricity for the cubic formulas.

\subsubsection{Further iterative methods}

After the first approximation for the eccentric anomaly $E$ has been obtained using the cubic approximation we proceed with using iterative root-finding methods until the user-given desired accuracy is reached. In the current version of the code we first try the standard Newton-Raphson method. This method is typically sufficient to find the solution for Kepler's equation with desired accuracy but is not always guaranteed to do so. Thus, we supplement our Kepler solver with the common bisection method which is guaranteed to find the solution in the cases when the Newton-Raphson iteration method fails. We accept the solution for the eccentric anomaly when $|\Delta E|<\eta_\mathrm{Kepler}$ in which $|\Delta E|$ is the difference of the values for $E$ in two consecutive iteration rounds. We typically set $10^{-10} \lesssim \eta_\mathrm{Kepler} \lesssim 10^{-6}$ in our simulations.

\subsection{Secular orbit evolution of triple systems}\label{section: secular-triple}

Relatively isolated hierarchical triple systems can be computationally extremely expensive even for regularised integrators. Thus, an approximate but efficient integration method is again desirable. Our integration technique of choice for such triple systems is again secular integration, widely used in the literature (e.g. \citealt{Marchal1990,Correia2016,Naoz2013,Naoz2016,Toonen2016,Hamers2016,Hamers2021}).

The secular three-body integration is limited to stable systems while the unstable i.e. soon dissolving systems are integrated using regularisation methods. We assess the stability of a three-body system before subsystem integration using the stability criterion of \cite{Mardling2001} (see also \citealt{Eggleton1995,Vynatheya2022}). A three-body system is assumed stable if $R_\mathrm{p} = a_\mathrm{out} (1.0-e_\mathrm{out}) > R_\mathrm{p,crit}$ with the critical outer pericenter distance $R_\mathrm{p,crit}$ defined as
\begin{equation}
R_\mathrm{p,crit} = 2.8 \left[ (1+q_\mathrm{out} \frac{1+e_\mathrm{out}}{\left( 1-e_\mathrm{out}\right)^{1/2}}) \right]^{2/5}\left( 1 - \frac{3 i}{10 \pi}\right) a_\mathrm{in}.
\end{equation}
Here $q_\mathrm{out} = m_\mathrm{2}/(m_\mathrm{0}+m_\mathrm{1})$ is the so-called outer mass ratio, $a_\mathrm{in}$ and $a_\mathrm{out}$ the inner and outer semi-major axis, $q_\mathrm{out}$ is the eccentricity of the outer orbit and $i$ is the mutual inclination of the orbits.

\begin{figure}
\includegraphics[width=\columnwidth]{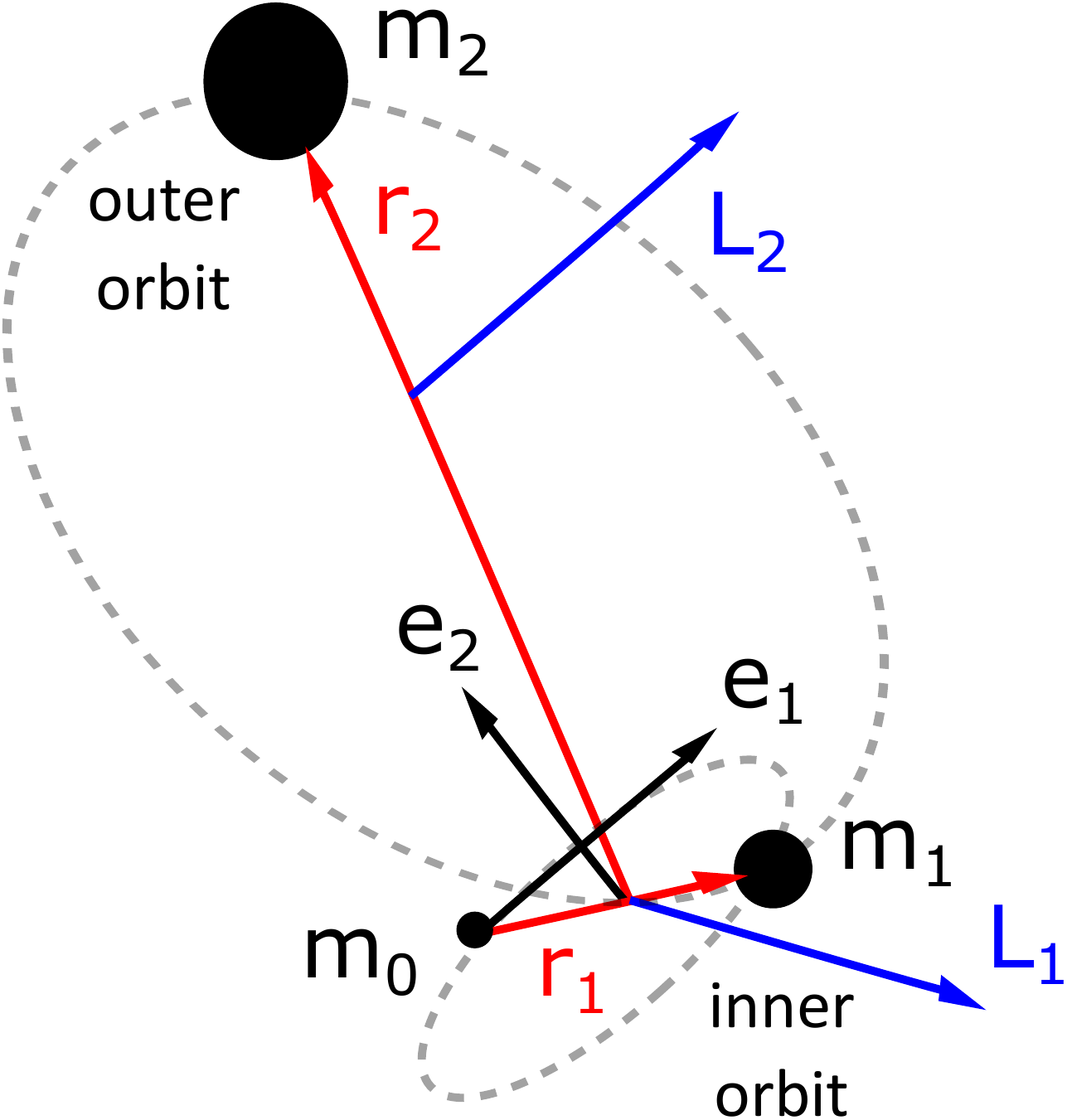}
\caption{An illustration of a hierarchical triple system with position ($\vect{r}_\mathrm{1}$, $\vect{r}_\mathrm{2}$ in red), orbital angular momentum ($\vect{L}_\mathrm{1}$, $\vect{L}_\mathrm{2}$ in blue) and eccentricity vectors ($\vect{e}_\mathrm{1}$, $\vect{e}_\mathrm{2}$ in black) shown. Note that the sizes of the inner and outer orbits (dashed grey ellipses) are not to scale.}
\label{fig: triple}
\end{figure}

The three-body secular integration is most conveniently performed using the angular momentum and eccentricity vectors as the dynamical variables of the system. Following the notation of \cite{Correia2016} (with slight modifications to be consistent with the notation of this study) the inner (subscript $i=1$) and outer (subscript $i=2$) angular momentum and eccentricity vectors are defined as
\begin{equation}
\begin{split}
\vect{L}_\mathrm{i} &= \beta_\mathrm{i} \vect{r}_\mathrm{i} \times \vect{v}_\mathrm{i} = L_\mathrm{i} \hat{\mathbf{k}}_\mathrm{i} = \beta_\mathrm{i} \sqrt{\mu_\mathrm{i} a_\mathrm{i} (1-e_\mathrm{i}^2)}\; \hat{\mathbf{k}}_\mathrm{i}\\
\vect{e}_\mathrm{i} &= \frac{\dot{\vect{r}}_\mathrm{i} \times \vect{L}_\mathrm{i} }{\mu_\mathrm{i} \beta_\mathrm{i}} - \frac{\vect{r}_\mathrm{i}}{ \norm{\vect{r}_\mathrm{i}} }.
\end{split}
\end{equation}
The mass constants in the definition equations are defined as $\beta_\mathrm{01} = m_\mathrm{0} m_\mathrm{1} / m_\mathrm{01}$, $\beta_\mathrm{2} = m_\mathrm{2} m_\mathrm{01} / (m_\mathrm{2} + m_\mathrm{01})$, $\mu_\mathrm{1} = G m_\mathrm{01}$ and $\mu_\mathrm{2} = G(m_\mathrm{2}+m_\mathrm{01})$ with $m_\mathrm{01} = (m0 + m1)$. An example illustration of a hierarchical triple systems with relevant vectors is shown in Fig. \ref{fig: triple}.

The secular equations of motion of a hierarchical three-body system up to octupole order using the angular momentum and eccentricity vectors can be formally written as
\begin{equation}\label{eq: eom-triple}
\begin{split}
\left\langle \derfrac{\vect{L}_\mathrm{1}}{t} \right\rangle_\mathrm{sec} &= \dot{\vect{L}}_\mathrm{1,quad}(\vect{L}_\mathrm{1},\vect{L}_\mathrm{2},\vect{e}_\mathrm{1}) + \dot{\vect{L}}_\mathrm{1,oct}(\vect{L}_\mathrm{1},\vect{L}_\mathrm{2},\vect{e}_\mathrm{1},\vect{e}_\mathrm{2})\\
\left\langle \derfrac{\vect{L}_\mathrm{2}}{t} \right\rangle_\mathrm{sec} &= - \left\langle \derfrac{\vect{L}_\mathrm{1}}{t} \right\rangle_\mathrm{sec}\\
\left\langle \derfrac{\vect{e}_\mathrm{1}}{t} \right\rangle_\mathrm{sec} &= \dot{\vect{e}}_\mathrm{1,quad}(\vect{L}_\mathrm{1},\vect{L}_\mathrm{2},\vect{e}_\mathrm{1}) + \dot{\vect{e}}_\mathrm{1,oct}(\vect{L}_\mathrm{1},\vect{L}_\mathrm{2},\vect{e}_\mathrm{1},\vect{e}_\mathrm{2})\\
\left\langle \derfrac{\vect{e}_\mathrm{2}}{t} \right\rangle_\mathrm{sec} &= \dot{\vect{e}}_\mathrm{2,quad}(\vect{L}_\mathrm{1},\vect{L}_\mathrm{2},\vect{e}_\mathrm{1},\vect{e}_\mathrm{2}) + \dot{\vect{e}}_\mathrm{2,oct}(\vect{L}_\mathrm{1},\vect{L}_\mathrm{2},\vect{e}_\mathrm{1},\vect{e}_\mathrm{2}).
\end{split}
\end{equation}
The expressions for the derivatives $\dot{\vect{L}}_\mathrm{1,quad}$, $\dot{\vect{L}}_\mathrm{1,oct}$, $\dot{\vect{e}}_\mathrm{1,quad}$, $\dot{\vect{e}}_\mathrm{1,oct}$, $\dot{\vect{e}}_\mathrm{2,quad}$ and $\dot{\vect{e}}_\mathrm{2,oct}$  can be found e.g. from Eq. (9) - (20) of \cite{Correia2016}. Note than the equations of motion are both coupled and implicit so an integration technique resembling the auxiliary variable method described in Section \ref{section: auxiliary_integration} is required to integrate the equations of motion beyond first order. 

The equations of motion of the secular triple system above in Eq. \eqref{eq: eom-triple} are purely Newtonian. It is straightforward to include post-Newtonian correction terms into the equations of motion of the inner binary \citep{Kidder1995,Correia2016} causing the relativistic advance of the inner periapsis with the PN1.0 term
\begin{equation}
\left \langle \derfrac{\vect{e}_\mathrm{1}}{t} \right \rangle_\mathrm{sec, PN1.0} = \frac{3 \mu_\mathrm{1} \eta_\mathrm{1}}{c^2 a_\mathrm{1} (1-e_\mathrm{1}^2)} \hat{\mathbf{k}}_\mathrm{1} \times \vect{e}_\mathrm{1}
\end{equation}
in which $\eta_\mathrm{1} = \left( \mu_\mathrm{1}/a_\mathrm{1}^3\right)^{1/2}$ is the mean motion of the inner binary. The circularization and shrinking of the inner orbit caused by the inclusion of the PN2.5 gravitational wave radiation reaction term changes the magnitude of the inner angular momentum vector as
\begin{equation}
\left \langle \derfrac{L_\mathrm{1}}{t} \right \rangle_\mathrm{sec, PN2.5} = \frac{1}{2} \left[ \frac{\dot{a}_\mathrm{1}}{a_\mathrm{1}} - 2\frac{e_\mathrm{1} \dot{e}_\mathrm{1}}{1-e_\mathrm{1}^2}\right] L_\mathrm{1}.
 \end{equation}
The evolution of the inner semi-major axis $\dot{a}_\mathrm{1}$ and the norm of the inner eccentricity vector $\dot{e}_\mathrm{1}$ can be obtained directly from the \cite{Peters1964} formula in Eq. \eqref{eq: secular_a_ecc}. In this study we do not include relativistic effects on the outer orbits of hierarchical triple systems. The external perturbation of the hierarchical triple systems are taken into account in kick operations on the \fsi{} code side.

For the time-step of the secular triple integration we use a fraction of the outer orbital period, i.e. $\epsilon_\mathrm{X} = \eta_\mathrm{X} P_\mathrm{2}$ with $\eta_\mathrm{X} \sim 0.1$. After the integration we perform coordinate transformations from the angular momentum and eccentricity vectors to obtain the positions and velocities of the three bodies on their orbits as 
\begin{equation}
\{ \vect{L}_\mathrm{1},\vect{L}_\mathrm{2},\vect{e}_\mathrm{1},\vect{e}_\mathrm{2}, M_\mathrm{1}, M_\mathrm{2}\} \rightarrow \{ \vect{r}_\mathrm{i}, \vect{v}_\mathrm{i}\}
\end{equation}
in which the mean anomalies $M_\mathrm{1}$, $M_\mathrm{2}$ of the inner and outer orbits are random. As in Section \ref{section: secular} we need to use our Kepler solver described in Section \ref{section: keplersolver}.

\subsection{Stellar evolution}\label{section: sse}

The dynamics of real star clusters can be strongly affected by the evolution of individual stars in several ways. First, mass loss from stellar winds and supernovae explosions influence the early evolution of stellar systems. It is therefore essential to include these processes in N-body simulations and track the mass of each individual star particle. 
In addition to tracking the particle masses, incorporating detailed stellar prescriptions beyond point-mass particles (radius, luminosity) enables comparing simulated clusters with observations. Numerical experiments that have combined accurate descriptions of both gravitational and stellar evolution have revealed the formation channels of several observed exotic objects. For instance, blue stragglers, observed in the inner region of many star clusters \citep{Sandage1953, Piotto2004}, form through collisions or mass transfer between main-sequence stars \citep{Chatterjee2013, Kremer2020}. Such results could only be obtained with the inclusion of comprehensive and precise stellar evolution prescriptions.
Similarly,  N-body models,  combined with the most updated stellar evolution treatments, are proving to be valuable computational tools to gain new insights concerning gravitational wave phenomenology as shown by a recent set of numerical simulations (e.g. \citealt{Rizzuto2022}).

We incorporate single stellar evolution effects linking the synthetic package \sse{} \citep{Hurley2000} with \bifrost. 
This package consists of a large set of analytical functions to approximate the main properties of a star (radius, mass, luminosity, spin, core mass and core radius) as a function of time from the main-sequence phase to the remnant stage. The analytical functions have been calibrated to reproduce the observed stellar evolution tracks and they provide reliable estimates for a wide range of mass ($0.08 \Msun \lesssim M \lesssim  100 \Msun$) and metallicity ( $0.0001 \lesssim Z \lesssim 0.01$). The original prescriptions presented in \cite{Hurley2000} have been enriched with several new treatments whose implementation is described in detail in \cite{Banerjee2020}. 
First of all, stellar winds treatments for light blue variables stars have been included following \cite{Belczynski2010}. Secondly, recipes for pair-instability and pulsation pair-instability supernova models have been incorporated in the remnant formation prescriptions following \cite{Fryer2012} and \cite{Belczynski2016}. Also, the analytical expression for the natal kick velocities of black holes and neutron stars now depend explicitly on the fallback fraction \citep{Banerjee2020}.
In addition, electron capture supernovae following \cite{Podsiadlowski2004} and \cite{Gessner2018} have been included. With such prescriptions, simulations produce neutron stars with low-velocity kicks that are therefore likely retained in medium-size star clusters.

\subsection{Mergers of stars and compact objects}\label{section: mergers}

\subsubsection{Merger criteria}

We allow our particles to merge during a simulation run if any of the several merger criteria are fulfilled for a pair of particles. The merger criteria are based on physical characterisations of compact object (white dwarf, neutron star, BH) mergers, tidal disruption of stars, and stellar mergers. The merger criteria are checked in the beginning of the \fsi{} integration.

The first merger criterion is the gravitational wave driven coalescence timescale $\tau_\mathrm{gw}$ of a compact bound binary system. We merge two particles of the binary if their mutual coalescence timescale is shorter than their current time-step in the time-step hierarchy. For a binary system with initial semi-major axis $a_\mathrm{0}$ and eccentricity $e_\mathrm{0}$ the merger timescale can be evaluated from the integral expression
\begin{equation}
\tau_\mathrm{gw} = \frac{15}{304} \frac{a_\mathrm{0}^4}{\beta(m1,m2)} \frac{1}{g^4(e_\mathrm{0})} \int_\mathrm{0}^\mathrm{e_\mathrm{0}} \frac{g^4(e)(1-e^2)^{5/2}}{e \left( 1+\frac{121}{304} e^2\right)} de
\end{equation}
in with the auxiliary function $\beta(m_\mathrm{1},m_\mathrm{2})$ was defined in Eq. \eqref{eq: gw_aux_functions} and $g(e)$ is defined as
\begin{equation}
g(e) = \frac{e^{12/19}}{1-e^2} \left(1 + \frac{121}{304} e^2 \right)^{870/2299}
\end{equation}
following \cite{Maggiore2007}. For circular orbits the expression for $\tau_\mathrm{gw}$ becomes
\begin{equation}
\tau_\mathrm{gw} = \frac{5}{256} \frac{a_\mathrm{0}^4}{\beta(m_\mathrm{1},m_\mathrm{2})}.
\end{equation}
Instead of calculating the integral at every time-step we use an estimate for $\tau_\mathrm{gw}$. We have performed the integration beforehand for a large number of different $e_\mathrm{0}$ and tabulated the results. The value of the integral and thus $\tau_\mathrm{gw}$ is obtained in our code by interpolating the table values.

The next compact object merger criterion is based on the innermost stable circular orbit (ISCO) around a Schwarzschild black hole. The radius of this orbit is
\begin{equation}
r_\mathrm{isco} = \frac{6GM_\bullet}{c^2},
\end{equation}
which corresponds to three times the Schwarzschild radius $R_\mathrm{sch}$ of the black hole. A somewhat more conservative option is to perform the merger at $10\:R_\mathrm{sch}$. At this separation the post-Newtonian equations of motion are still well-behaved and the energy radiated in gravitational waves agrees with the energy lost by the binary reasonably well \citep{Mannerkoski2019}. 

Compact objects may tidally disrupt stars in the code following a simple prescription. A star in a bound binary is tidally disrupted by the compact companion mass $M_\bullet$ if the pericenter distance falls below the tidal disruption distance, i.e. $r_\mathrm{peri} < r_\mathrm{tde}$ \citep{Kochanek1992} in which 
\begin{equation}
r_\mathrm{tde} = 1.3 \left( \frac{m_\star + M_\bullet}{m_\star} \right)^{1/3} R_\star.
\end{equation}
Here $m_\star$ is the mass of the non-compact star and $R_\star$ is its radius. We also ensure that the pericenter is reached within the next time-step. If the star is unbound to the compact object we instead check whether the star is currently close enough to the compact object, namely if  $r<3\: r_\mathrm{tde}$.
Finally, we merge two stars in the code if they overlap, that is $r<r_\mathrm{overlap}$ in which the overlap radius is simply defined as
\begin{equation}
r_\mathrm{overlap} = R_{\star,\mathrm{1}} + R_{\star,\mathrm{2}}.
\end{equation}

\subsubsection{Merger remnant properties - Newtonian}\label{section: merger-remnant}

We assume that in a particle merger linear and angular momentum are conserved. The position and velocity of the merger remnant are those of the center-of-mass of the two particles defined as
\begin{equation}
\begin{split}
\vect{r}_\mathrm{remnant} &= \frac{m_\mathrm{1} \vect{r}_\mathrm{1} + m_\mathrm{2} \vect{r}_\mathrm{2} }{m_\mathrm{1}+m_\mathrm{2}}\\
\vect{v}_\mathrm{remnant} &= \frac{m_\mathrm{1} \vect{v}_\mathrm{1} + m_\mathrm{2} \vect{v}_\mathrm{2} }{m_\mathrm{1}+m_\mathrm{2}}.
\end{split}
\end{equation}
The spin of the merger remnant also inherits the remaining orbital angular momentum as 
\begin{equation}
\vect{S}_\mathrm{remnant} = \vect{L} + \vect{S}_\mathrm{1} + \vect{S}_\mathrm{2} = \frac{m_\mathrm{1} m_\mathrm{2}}{m_\mathrm{1}+m_\mathrm{2}} \vect{r}_\mathrm{21} \times \vect{v}_\mathrm{21} + \vect{S}_\mathrm{1} + \vect{S}_\mathrm{2}.
\end{equation}
To infer the stellar properties (mass, radius, luminosity, mass, etc.) and the correct stellar type of a merger remnant we utilise the routine mix.f of the binary stellar evolution package \bse{} (for more details see \citealt{Hurley2002}).

\subsubsection{Merger remnant properties - relativistic}\label{section: gwrecoil}

When two black holes merge we include an option to take the relativistic mass loss $\Delta m_\mathrm{gw}$ and the relativistic recoil kick velocity $\vect{v}_\mathrm{kick}$ of the merger remnant into account. The kick velocity and direction, final black hole mass and spin are obtained using the fitting formulas of \cite{Zlochower2015} which are fits to numerical simulations performed in full general relativity. As noted in \cite{Mannerkoski2022} the model is still approximate due to the inherent limitations of the fitting functions.

\subsection{Escapers}\label{section: escapers}
We remove gravitationally unbound particles from the simulation if they are sufficiently far away from the center-of-mass of the simulated star cluster and move outwards. For escaping binary or multiple systems the distance and boundness criteria are checked using the center-of-mass of the escaping system. The escape distance from the center-of-mass of the simulated system is a user-defined free input parameter. For typical star clusters an escape distance of $r_\mathrm{esc} \sim 100\; r_\mathrm{h}$ is a reasonable choice.

\subsection{Energy book-keeping}\label{section: bookkeeping}

The simulated N-body system may lose (or gain) energy if processes such as particle mergers, gravitational-wave recoil kicks, removal of escapers and stellar evolution are enabled. The full expression for the total energy of an N-body system in the current version of \bifrost{} is thus
\begin{equation}
H = E_\mathrm{Newton} + E_\mathrm{esc} + E_\mathrm{gw} + E_\mathrm{merg} + E_\mathrm{sse}
\end{equation}
in which the terms account for losses (gains) to the escapers, gravitational waves and recoil kicks, mergers and single stellar evolution, reading from left to right after the Newtonian energy.

For very close binaries for which relativistic effects are important we have included an option in the code to use post-Newtonian expressions for energy (and angular momentum) in orders PN1.0 and PN2.0 instead of the basic Newtonian formulas (e.g. \citealt{Blanchet2003,Memmesheimer2004,Blanchet2014,Poisson2014,Avramov2021}). The expressions are somewhat lengthy and cumbersome so we will not repeat them here. 

\subsection{Adaptive energy error restarts}\label{section: error-restart}

We include a procedure in \bifrost{} to re-run an integration interval if too much energy error (compared to a user-given tolerance $\eta_\mathrm{restart}$) accumulated during the interval. The procedure is similar as in the \nbody{} series of direct summation codes \citep{Aarseth1966,Spurzem1999,Aarseth2003,Wang2015}.

We save the dynamical state of the simulation before each integration interval. If the relative energy error compared to the beginning of the interval is too large, i.e. $|\Delta E/E|>\eta_\mathrm{restart}$ we restore the previous saved physical state of the simulation. We typically use values $10^{-8} \lesssim \eta_\mathrm{restart} \lesssim 10^{-5}$ for the error restart tolerance parameter. 

The time-step accuracy parameters are lowered by 50\% each consecutive with each consecutive restart for each accuracy parameter. We also increase the neighbour radius $r_\mathrm{ngb}$ of the subsystem neighbour search by 15\% with each restart. After the sufficient accuracy is reached the original time-step accuracy parameters and subsystem sizes are restored. If the energy error starts to increase again we accept the current result and continue the simulation.

\section{Scaling and timing with binary systems}\label{section: 4}

\subsection{The scaling of \frost{} versus the scaling of \bifrost{}}

The computation of gravitational accelerations in \bifrost{} remains largely unchanged from our previous code version, \frost{} \citep{Rantala2021}. The direct-summation $\bigO{N^2}$ acceleration calculation loops are the computationally most expensive tasks performed by both codes. We expect that the parallel subsystem integration, especially secular, is in most cases efficient enough to make its cost a subdominant component of the wall-clock time budget of \bifrost. We perform several scaling and timing tests to confirm this expectation.

\subsection{Strong scaling of star cluster simulations with different binary fractions}\label{section: scaling-strong}

\begin{figure}
\includegraphics[width=\columnwidth]{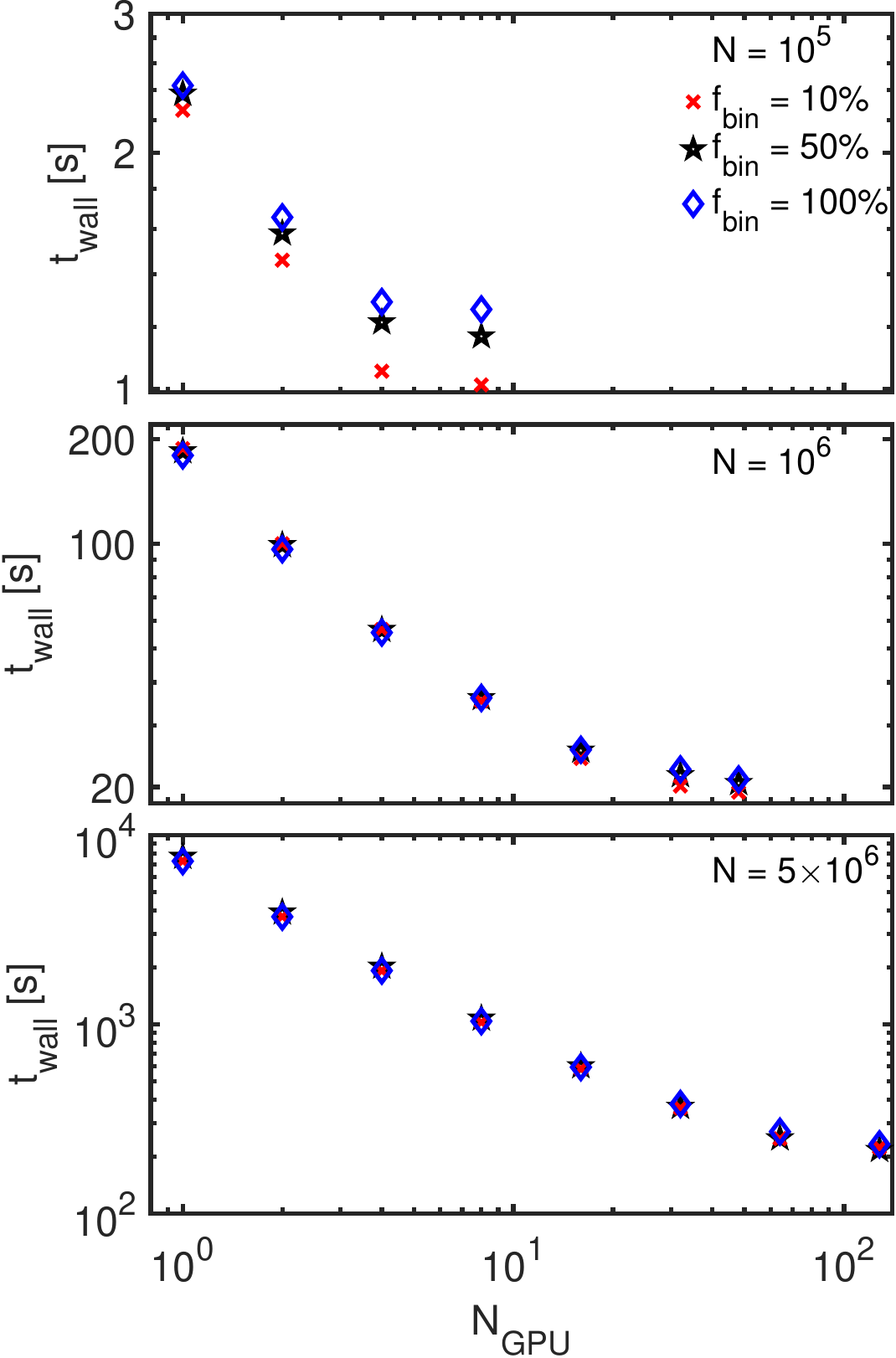}
\caption{Strong scaling of \bifrost{} with $N=10^5$ (top panel), $N=10^6$ (middle panel) and $N=5\times10^6$ (bottom panel) simulation particles and three different binary fractions of $f_\mathrm{bin} = 10\%, 50\%$ and $100\%$. The star cluster initial conditions and the hardware used are described in detail in the text. In the smaller $N=10^5$ runs the code scaling stalls at $N_\mathrm{GPU} \sim 8$. Runs with smaller binary fractions scale better as the computational cost of the subsystem integration is non-negligible as discussed in the text. With $N=10^6$ the code scales up to $N_\mathrm{GPU} = 48$ GPUs but scaling is not ideal with higher GPU numbers. The trend is similar with the high particle number $N=5\times10^6$ tested up to $N_\mathrm{GPU} = 128$. The binary fraction has a negligible effect on the timing and scaling of the two larger runs as the direct summation acceleration calculations dominate the wall-clock time budget. The results are consistent with the scaling tests of the \frost{} code, as expected.
}
\label{fig: timing-ngpu}
\end{figure}

The total wall-clock time budget of \bifrost{} is mostly elapsed in all-particle pair-wise $\bigO{N^2}$ operations and in linear $\bigO{N}$ operations. The $\bigO{N^2}$ loops are required for time-step and subsystem assignment and direct-summation acceleration calculations. The subsystem integration is by far the most expensive linear operation of the code so the miscellaneous contribution from other linear operations of the code can be ignored here. As the subsystem integrations are independent of each other their cost grows linearly with increasing number of binary systems. The number of binary systems can be expressed \citep{Kupper2011} using the binary fraction $f_\mathrm{bin}$ as
\begin{equation}
N_\mathrm{bin} = \frac{1}{2} f_\mathrm{bin} N.
\end{equation}
The total wall-clock time can now be formally estimated as
\begin{equation}
T_\mathrm{wall} \approx T_\mathrm{N^2} N^2 + T_\mathrm{bin} f_\mathrm{bin} N
\end{equation}
in which the two time constants $T_\mathrm{N^2}$ and $T_\mathrm{bin}$ depend on the simulated system, user-given code parameters and the hardware configuration. In typical \bifrost{} applications the first term dominates. For initial conditions smaller than approximately $N=10^5$ simulation particles the subsystem integration may have a non-negligible contribution to the total wall-clock time budget, depending on the binary and multiple system population. For the scaling tests in this study we use star clusters with different binary fractions but do not include primordial triple systems or even more complex higher multiple systems.

We setup nine star cluster initial conditions (ICs) using our novel IC generator described in Appendix \ref{section: SCIC}. We use three different total particle numbers ($N=10^5$, $N=10^6$, $N=5\times10^6$) and three different binary fractions ($f_\mathrm{bin} = 10\%$, $50\%$, $100\%$). We evolve the star cluster models using \bifrost{} for a single integration interval of $T=10^{-3}$ Myr on the supercomputer Raven of MPCDF\footnote{Max Planck Computing and Data Facility, \url{www.mpcdf.mpg.de}} in Garching, Germany. The supercomputer nodes used for the timing and scaling tests of this study contain each 72 CPU cores from two Intel Xeon IceLake-SP 8360Y processors and 4 Nvidia A100-SXM4 GPUs. We use up to 32 supercomputer nodes totalling $128$ GPUs and $2304$ CPUs in the strong scaling tests.

The relevant user-given accuracy parameters for the strong scaling test runs are the following. For the time-steps we use a value of $\eta = 0.2$ for the fly-by and free-fall time-step criteria and $\eta = 0.1$ for the jerk and gradient time-steps. Most of the subsystems are integrated using secular methods. For fly-bys and multiplets the regularised integrator \logh{} has a GBS tolerance of $\eta_\mathrm{GBS} = 10^{-10}$ and a relative end-time tolerance of $10^{-4}$. We allow the particles to merge and escape during the brief scaling test runs but as expected the number of these events is small.

The results of the strong scaling test runs are presented in Fig. \ref{fig: timing-ngpu}. The smaller $N=10^5$ initial conditions scale up to $N_\mathrm{GPU} = 8$ GPUs with the scaling becoming somewhat inefficient for more than four GPUs. Initial conditions with smaller binary fractions scale better with the maximum wall-clock time difference being of the order of $25\%$. For the medium $N=10^6$ and large $N=5\times10^6$ initial conditions the binary fraction has a negligible effect on the total wall-clock time or scaling of the code as the $\bigO{N^2}$ are computationally far more expensive than the $\bigO{N_\mathrm{bin}}$ subsystem integration. This is despite the fact that the number of binary systems is very large, up to $N_\mathrm{bin} = 2.5\times10^6$ in the run with $N=5\times10^6$ and $f_\mathrm{bin} = 100\%$. The simulations with million-body initial conditions scale up to $N_\mathrm{GPU} = 48$ while runs with $N=5\times10^6$ scale up to the largest tested number of GPUs, i.e. $N_\mathrm{GPU} = 128$. As with the smaller initial conditions the scaling becomes less ideal when the number of GPUs increases. The scaling behaviour of \bifrost{} in the performed test runs is consistent with the scaling of the previous code version \frost{} which has been shown to scale up to $N_\mathrm{GPU} = 40 N/10^6$ GPUs \citep{Rantala2021}. Thus, we conclude that despite the increased complexity of the new simulation code \bifrost{} its scaling remains at the level of its predecessor code \frost{}. This was our original aim for the scaling performance of the new code.

\subsection{Wall-clock timing of star cluster simulations with different binary fractions}

\begin{figure}
\includegraphics[width=\columnwidth]{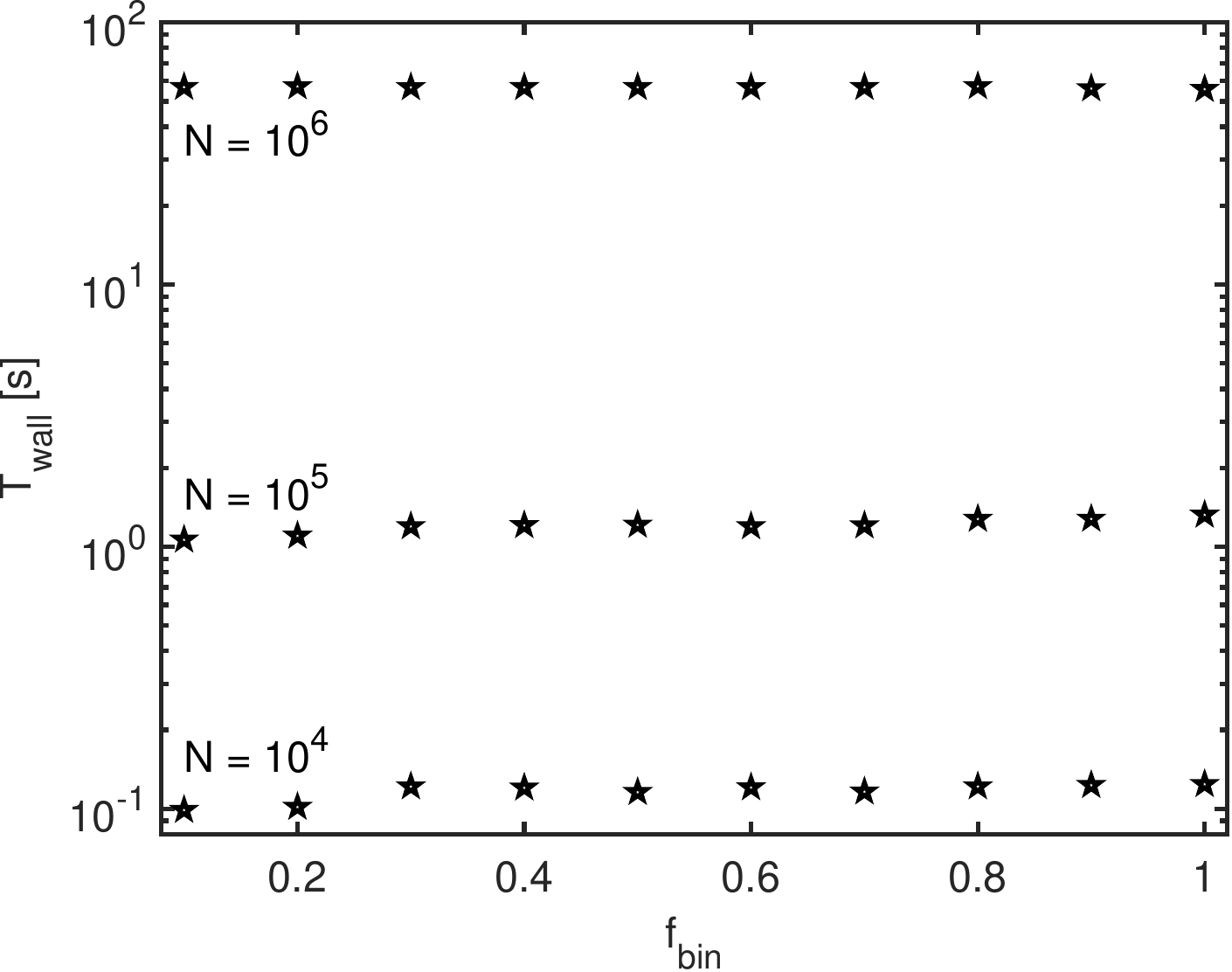}
\caption{Timing of simulations of 30 star cluster models with three different particle numbers from $N=10^4$ to $N=10^6$ and ten different binary fractions ranging from $f_\mathrm{bin} = 10\%$ to $100\%$ simulated with the \bifrost{} code. The timing simulations were run on a single supercomputer node with $N_\mathrm{GPU}=4$ and $N_\mathrm{CPU} = 72$ for $T=10^{-3}$ Myr. The elapsed wall-clock time only has a weak dependence on the binary fraction in the test runs with the maximum increase from the $10\%$ to $100\%$ binary fraction is only $25\%$.}
\label{fig: timing-fbin}
\end{figure}

Next we focus further on the effect of binary fraction on the wall-clock running speed of \bifrost. For these tests we only use a single hardware configuration, namely a single Raven supercomputer node with $N_\mathrm{CPU} = 72$ and $N_\mathrm{GPU} = 4$. For more details about the supercomputer and the hardware see the previous Section \ref{section: scaling-strong}.

We generate additional 30 star cluster initial conditions with three different particle numbers ($N=10^4$, $N=10^5$ and $10^6$) and ten different binary fractions ranging from $f_\mathrm{bin} = 10\%$ to $f_\mathrm{bin} = 100\%$ with increments of $10\%$. Again, the initial conditions are run for $T=10^{-3}$ Myr using accuracy parameters from the strong scaling tests in the previous section.

The timing tests with the varying binary fraction are presented in Fig. \ref{fig: timing-fbin}. The wall-clock time has only a very weak dependence on the binary fraction $f_\mathrm{bin}$ for the tested initial conditions. Increasing the binary fraction from $f_\mathrm{bin} = 10\%$ to $100\%$ only increases the elapsed wall-clock time by $25\%$, $24\%$ and $3\%$ for the initial conditions with $N=10^4$, $N=10^5$ and $10^6$ particles, respectively. The running speed of the simulations with the largest initial condition is least affected by the increased binary fraction as explained in the previous section. 
The elapsed wall-clock times of the runs correspond to $720$ Myr, $67$ Myr and $1.5$ Myr of simulation time per 24 hours of wall-clock time for the initial conditions of three different particle numbers on a single supercomputer node. The running times of simulations between different particle numbers do not exactly follow the $\bigO{N^2}$ scaling as the half-mass radii of the star clusters are obtained from cluster mass-size relations as described in Appendix \ref{section: SCIC}. As the more massive star clusters are on average less dense, the average time-steps are longer and the tenfold increase in particle number does not result in a hundredfold increase in the elapsed wall-clock time in this code performance test.

We emphasise that the presented \bifrost{} running speed estimates are the upper limit for the performance of the code. This is due to the fact that for the timing tests for the 30 different initial conditions the chosen simulation time was relatively short. In simulations with lengths comparable to the mass segregation or core-collapse time-scales of the initial conditions more frequent close encounters between single and binary systems, and formation and evolution of subsystems with more than two components would slow down the code from its ideal running speed.

\section{Code applications and performance}\label{section: 5}

\subsection{Types and nature of energy error in \bifrost}

We use relative total energy error compared to the simulation start defined as 
\begin{equation}
\left| \frac{\Delta E_\mathrm{tot}}{E} \right| \equiv \left| \frac{E(t)-E(t_\mathrm{0})}{E(t_\mathrm{0})} \right| \hspace{2.5cm} \text{total}
\end{equation}
to characterise the accuracy of our \bifrost{} simulations. Here the total energy is defined as in Section \ref{section: bookkeeping}. Another useful quantity is the relative energy error compared to the previous interval which we define as
\begin{equation}
\left| \frac{\Delta E_\mathrm{s}}{E} \right| \equiv \left| \frac{E(t)-E(t-\tau)}{E(t_\mathrm{0})} \right| \hspace{2cm} \text{single interval}
\end{equation}
in which $\tau$ is the duration of a single integration interval in \bifrost.

The sources of energy error in \bifrost{} simulations can be informally divided into two categories. We label the two types of error sources as well-behaving errors and stochastic errors. Given a general time-step accuracy parameter $\eta_\mathrm{0} \ll 1$ the relative energy error after a single integration interval will be $|\Delta E/E|_\mathrm{0}$. The exact value of the relative energy error accumulated during a single integration interval depends on the dynamical state of the N-body system. The error Hamiltonians of symplectic integrators are proper Hamiltonians themselves even though they are complex and in most cases unknown. The dynamics generated by the terms of the error Hamiltonians perturb the integrated solution from the true exact solution, causing energy error \citep{Chin2007a}. If the time-step accuracy parameter is adjusted to another reasonable value $\eta_\mathrm{1}$, the new relative energy error per integration interval will be
\begin{equation}\label{eq: fourthorder}
\left| \frac{\Delta E}{E} \right|_\mathrm{1} = \left( \frac{\eta_\mathrm{1}}{\eta_\mathrm{0}} \right)^4 \left| \frac{\Delta E}{E} \right|_\mathrm{0}.
\end{equation}
This is due to the fact that \hhsfsi{} is a fourth-order integrator \citep{Rantala2021}. When decreasing the time-step accuracy parameter $\eta$ the energy error will decrease until at some point the increasing floating-point round-off error begins to dominate the error. Due to these facts, well-behaving error is unavoidable in our simulations though it can be controlled in a robust and systematic manner. Finally, the relative total energy error will grow even if the relative energy errors of individual integration intervals are constant. The total energy error grows linearly in time since the individual errors are not unbiased and thus accumulate faster than what is expected from a simple random walk (e.g. \citealt{Rein2015a}).

Stochastic errors in \bifrost{} have two sources. First, even though subsystem integration is in most situations very accurate, occasionally the energy error from integrating especially difficult particle configurations may exceed the level of error from the \hhsfsi{} integration. Second, particles may end up being in situations in which their assigned time-step is too long to capture the relevant dynamics. This can occur e.g. if a particle suddenly gains velocity in a strong few-body interaction. When exactly such events occur is essentially random, thus the label stochastic error. When visualising the total relative energy error as a function of time this causes a discontinuity, a sudden increase in the error, as opposed to smooth error growth. Increasing the accuracy of the simulation, i.e. making the time-step parameters $\eta$ smaller, decreases the frequency of how often stochastic error events occur. At the same time also there will be less well-behaving error, according to Eq. \ref{eq: fourthorder}. However, decreasing the time-step parameters also increases the required computational time. Thus in typical \bifrost{} simulations some amount of stochastic error must be tolerated in order to perform simulations efficiently. 

\subsection{A star cluster with equal-mass particles}\label{section: gc-equalmass}

\begin{figure}
\includegraphics[width=\columnwidth]{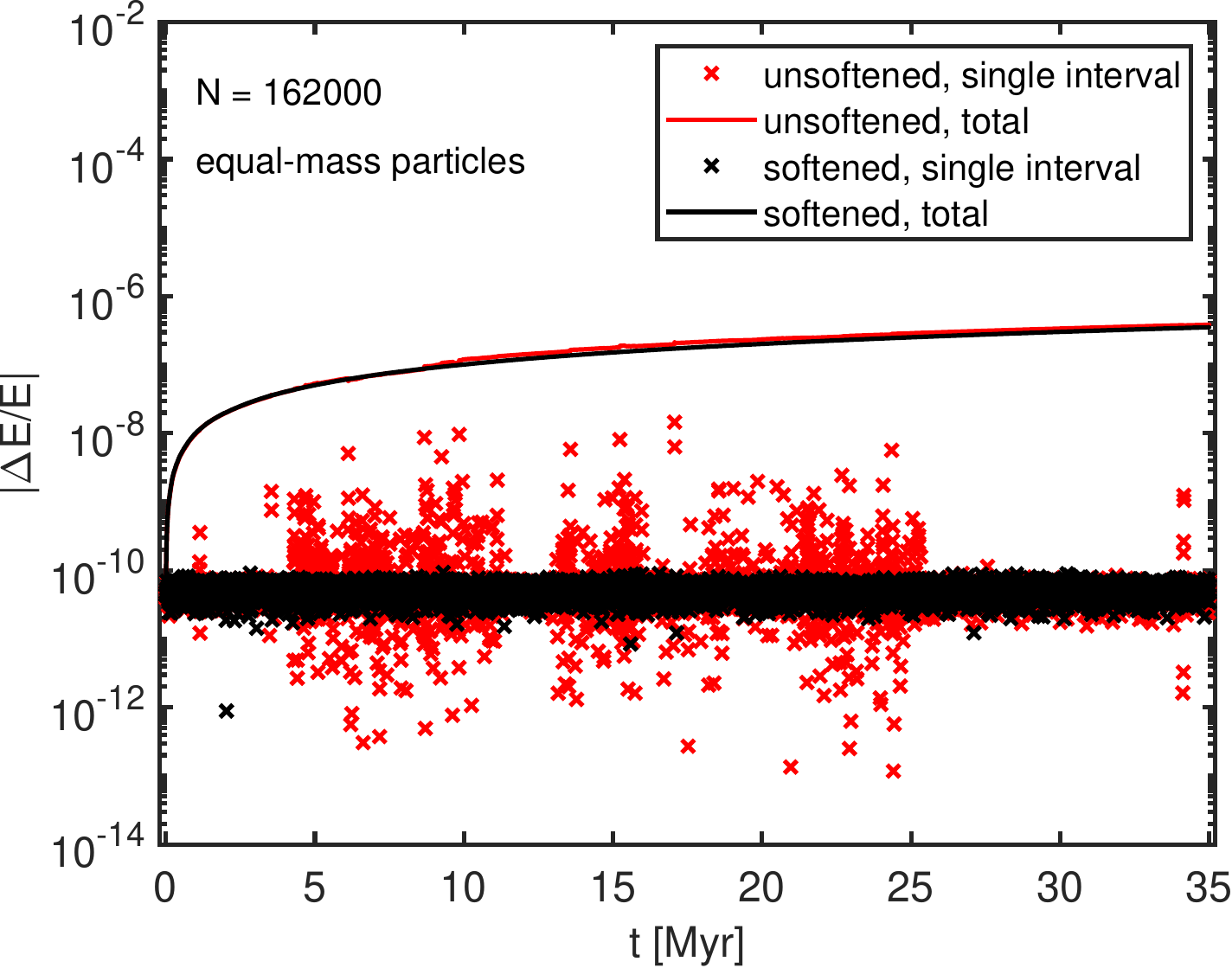}
\caption{Comparing the \bifrost{} code (unsoftened, in red) with a \frost-like (softened, in black) setup in a star cluster simulation of $N=162000$ equal-mass particles. The unsoftened \bifrost{} run occasionally has a larger energy error per integration interval when there are strong close encounters in the simulation. The total relative energy error behaves very similarly in the two simulations.}
\label{fig: energy-frost-bifrost}
\end{figure}

We begin to assess the accuracy of the new \bifrost{} code by running two star cluster simulations with identical initial conditions, one with gravitational softening and one without. The softened run closely corresponds to the simulations performed with our earlier code version \frost{}, which did not include specialised subsystem integration methods. The gravitational softening for this comparison test is implemented as in \cite{Rantala2021} using the standard \cite{Plummer1911} softening approach. Note that the gravitational softening is implemented in \bifrost{} only for the purposes of this comparison run and the standard code version uses subsystem integration instead of softening.

We setup the star cluster model initial conditions as described in Appendix \ref{section: SCIC}. The Plummer model consists of $N=161000$ equal-mass particles of $m_\mathrm{\star} = 0.5 M_\odot$ each and has a half-mass radius of $r_\mathrm{h}=3.1$ pc. The mass and half-mass radius of the cluster are chosen to resemble a median Milky Way globular cluster \citep{Heggie2003}.

The user-given accuracy parameters are set as follows. For the time-steps we have $\eta = 0.2$ and for the fly-by and free-fall time-steps and $\eta = 0.1$ for the jerk and gradient time-steps. In the softened run the gravitational softening length is set to $\epsilon_\mathrm{soft} = 10^{-3}$ pc, and in the unsoftened run we use the same value for the subsystem neighbour radius, i.e. $r_\mathrm{ngb} = 10^{-3}$. In the softened runs only the standard fourth-order forward integrator is used. Most of the subsystems in the unsoftened simulation are close fly-bys which are integrated using the regularised \logh{} integrator of \bifrost. The GBS tolerance parameter of \logh{} is set to $\eta_\mathrm{GBS} = 10^{-10}$ and the relative end-time tolerance parameter to $10^{-4}$. Both simulations are evolved for $T=35$ Myr with an integration interval duration (corresponding to the maximum time-step) of $10^{-3}$ Myr.

We compare the energy conservation in the \bifrost{} and the softened \frost-like simulation in Fig. \ref{fig: energy-frost-bifrost}. In the softened run the mean relative energy error per integration interval is only $|\Delta E_\mathrm{s}/E| \sim 5\times10^{-11}$. The total energy error grows linearly in time, as expected. In the unsoftened \bifrost{} run the mean relative energy error is very close to the value of the softened run, $|\Delta E_\mathrm{s}/E| \sim 8\times10^{-11}$. The energy error in intervals during which extremely close hyperbolic encounters occur is occasionally larger up to $|\Delta E_\mathrm{s}/E| \sim \times10^{-8}$. However, the contribution of these intervals to the total relative energy error is small and the energy error behaviour of the \bifrost{} run is very close to the softened run. We conclude that \bifrost{} is able to match the accuracy of a softened simulation code despite the zero gravitational softening in the code.

\subsection{Star clusters with a realistic IMF}\label{section: gc-mw}

\begin{figure}
\includegraphics[width=\columnwidth]{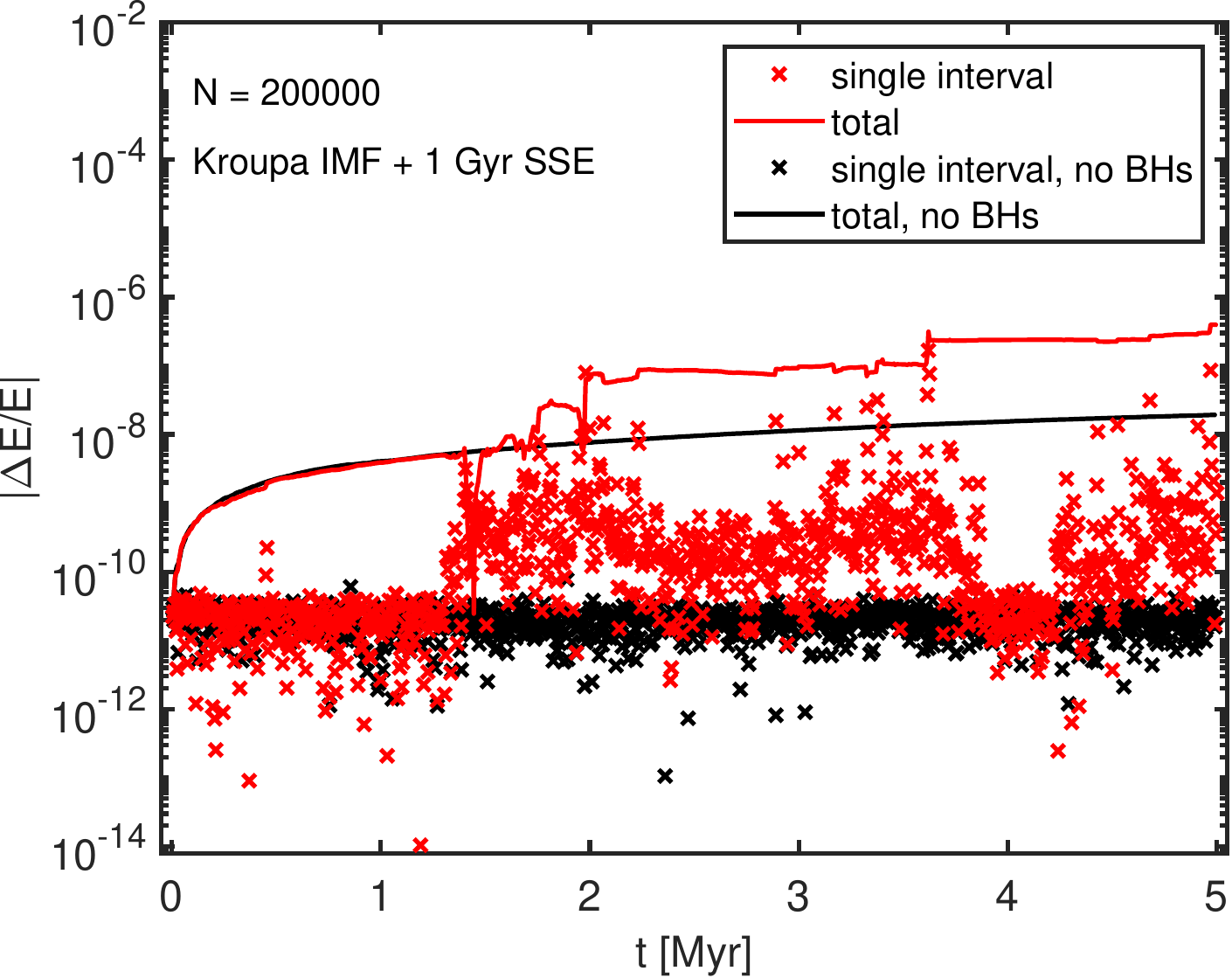}
\caption{Relative energy error in two star cluster simulation with $N=200000$ particles. In the first simulation the stellar mass function is an evolved Kroupa IMF (in red) while the ICs of the other run are identical but the black holes have been removed (in black). The energy conservation of the run without black holes closely resembles the simulation results with equal-mass particles in Fig. \ref{fig: energy-frost-bifrost} the final relative energy error being $|\Delta E_\mathrm{tot}/E| ~ \sim 2\times10^{-8}$. With black holes included the simulation is more demanding and relative energy error is somewhat larger, $|\Delta E_\mathrm{tot}/E| ~ \sim 3\times10^{-7}$ after $5$ Myr.}
\label{fig: energy-median-bh-nobh}
\end{figure}

We next setup more realistic initial conditions. Keeping the total mass of the cluster ($M=8.1\times10^4 M_\odot$) and its half-mass radius ($r_\mathrm{h} = 3.1$ pc) unchanged we now populate the cluster using a stellar population ($Z=0.01 Z_\mathrm{\odot}$) evolved to an age of $1$ Gyr from the \cite{Kroupa2001} initial mass function (IMF). This totals in $N=2\times10^5$ simulation particles of which $N_\bullet = 357$ are black holes. The masses of the least and the most massive black holes are $M_\bullet^\mathrm{min} = 5.5 M_\odot$ and $M_\bullet^\mathrm{max} = 41.6 M_\odot$, respectively. All the other simulation particles are less massive than $m_\mathrm{\star} = 1.9 M_\odot$.

We simulate the evolution of the star cluster model for $T=5$ Myr using \bifrost{}. The energy conservation in the simulation is presented in Fig. \ref{fig: energy-median-bh-nobh}. Initially the relative energy error of the simulation behaves as in the equal-mass particle runs in Section \ref{section: gc-equalmass} as the relative error per integration interval is less than $|\Delta E_\mathrm{s}/E| \sim 10^{-10}$. This behaviour continues until approximately $T=1.3$ Myr after which the mean relative energy error per integration interval increases by an order of magnitude. The final total relative energy error at the end of the simulation is approximately $|\Delta E_\mathrm{tot}/E| \sim 3\times 10^{-7}$.

In order to trace the origin of the energy error we repeat the simulation with a version of the initial conditions without black holes. While the mass and density profile of the star cluster model remains relatively unchanged, the mass function of the simulation particles now lacks objects more massive than $m_\mathrm{\star} = 1.9 M_\odot$. The results of this run are shown Fig. \ref{fig: energy-median-bh-nobh}. In contrast to the simulation run with all particles, without the black holes the energy conservation is almost identical to the run with equal-mass simulation particles in Fig. \ref{fig: energy-frost-bifrost}.

\begin{figure}
\includegraphics[width=\columnwidth]{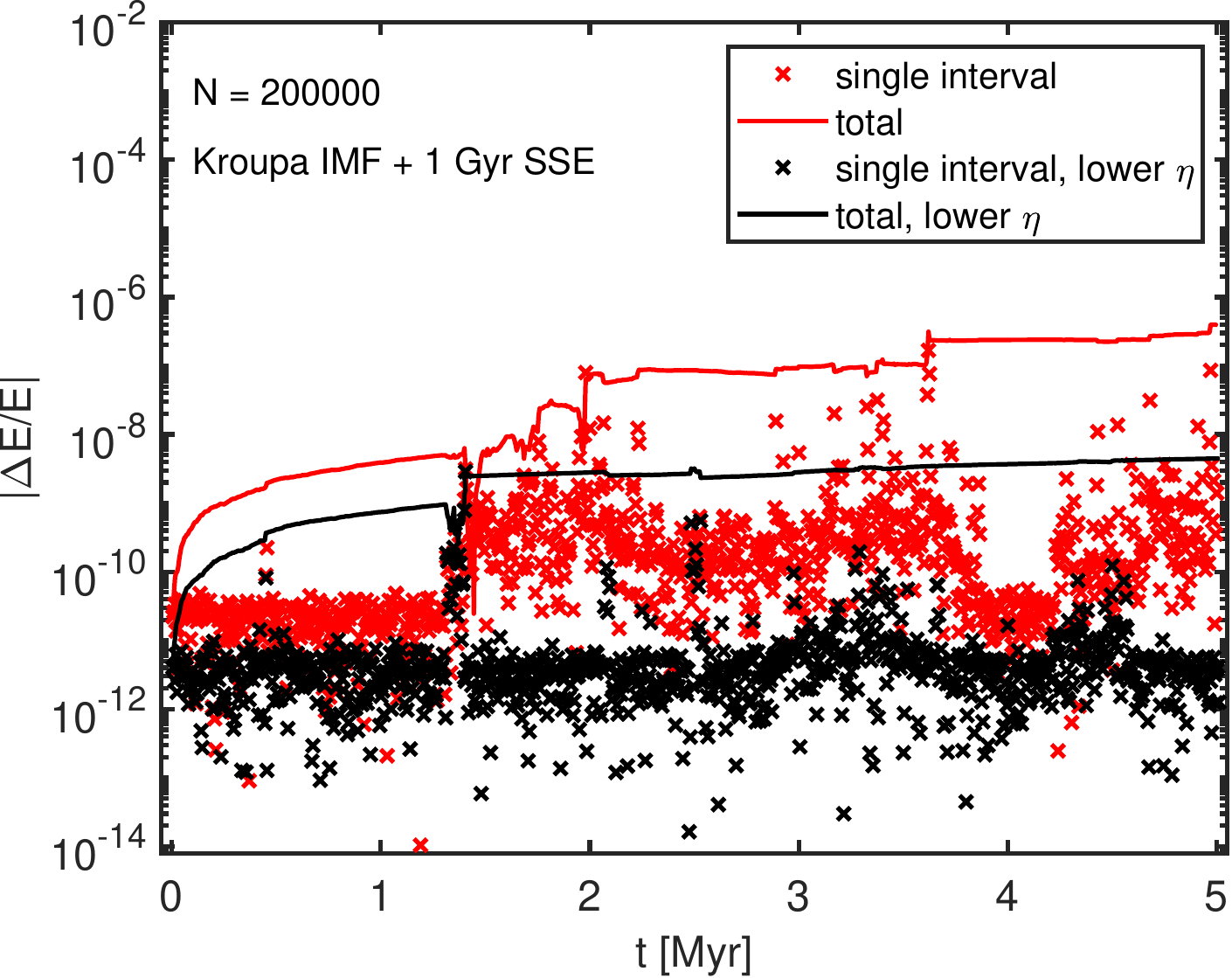}
\centering
\caption{The energy conservation of the star cluster simulation with black holes (in red) from Fig. \ref{fig: energy-median-bh-nobh} and its more accurate counterpart (in black). Lowering the time-step accuracy parameters $\eta$ by a factor of $2$ has two effects. First, the well-behaving relative energy error per integration interval (black crosses) decreases to less than $|\Delta E_\mathrm{s}/E| < 10^{-11}$. In addition, the intervals with larger stochastic error become less frequent. The combined effect is that now the total relative energy error (solid black line) shows almost no discontinuities, accumulating linearly as function of time. After $T=5$ Myr of evolution the total relative energy error is $|\Delta E_\mathrm{tot}/E| \sim 5\times10^{-9}$.}
\label{fig: energy-eta}
\end{figure}

We repeat the simulation with the black holes with increased code accuracy by lowering all the time-step accuracy parameters $\eta$ by a factor of $2$. The results of this run are presented in Fig. \ref{fig: energy-eta} alongside the original simulation. In the new more accurate simulation the relative energy error per integration interval falls below $|\Delta E_\mathrm{s}/E| \sim 5\times10^{-11}$ during most intervals. In addition the amount of stochastic error is smaller with the total relative energy error closely following the linear error growth trend. The final total energy error after $T=5$ Myr is only $|\Delta E_\mathrm{tot}/E| \sim 5\times10^{-9}$. Thus, the stochastic energy error originating from the simulated dynamics of the most massive particles in the runs can be reduced by tuning the user-given accuracy parameters.

\subsection{A star cluster with a 50\% binary fraction}

\begin{figure}
\includegraphics[width=\columnwidth]{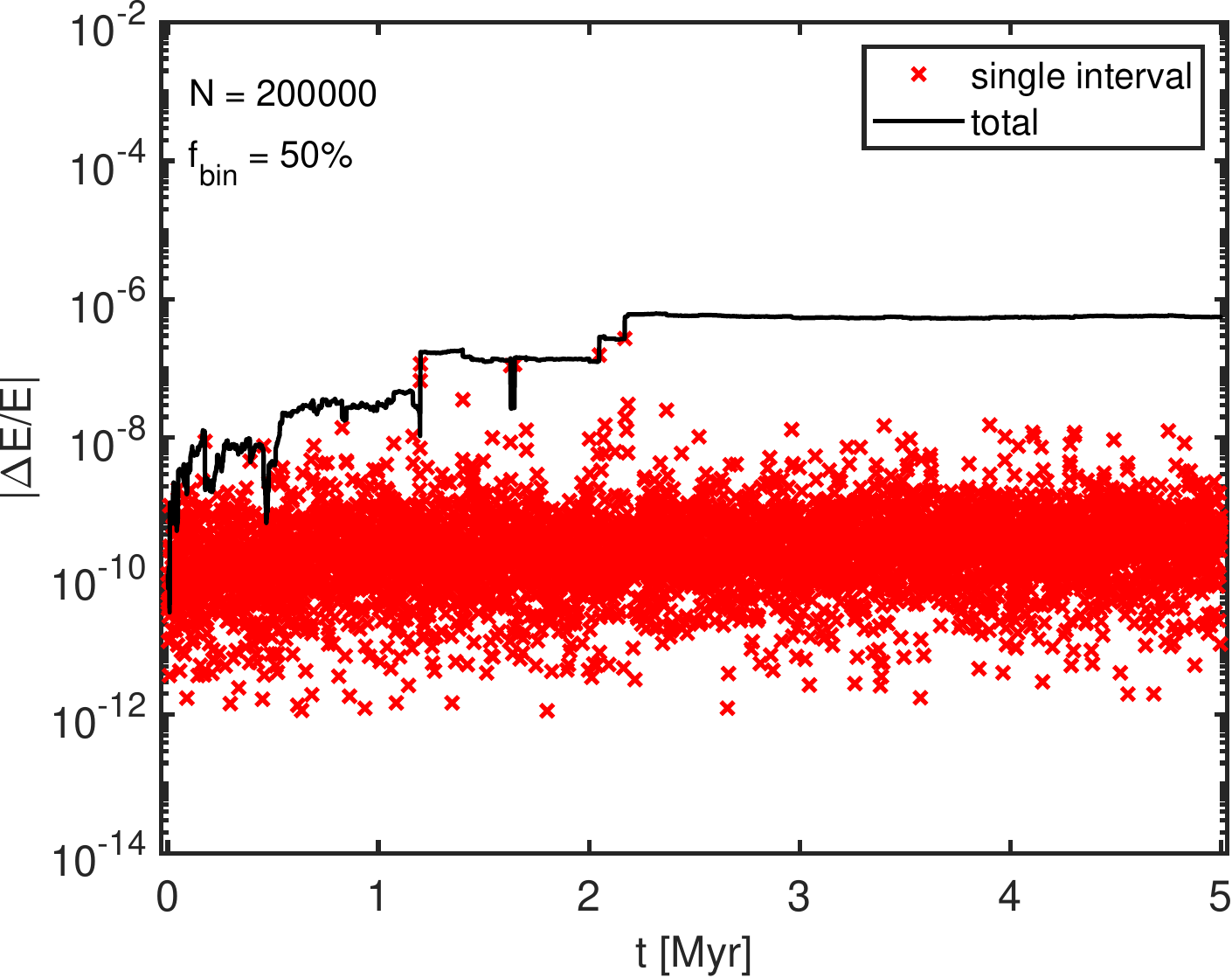}
\centering
\caption{Results of a star cluster simulation with $N=200000$ and a large binary fraction of $f_\mathrm{bin} = 50\%$. The mean relative energy error per integration interval is $|\Delta E_\mathrm{s}/E| \sim 7\times10^{-10}$ which is comparable to the results of the runs with no binary stars. However, the scatter of $|\Delta E_\mathrm{s}/E|$ per integration interval is now larger. The stochastic error originating especially from strong binary-binary interactions dominates the total energy error after $T=2.2$ Myr. This can be seen as the flat behaviour of the total relative energy error (black solid line) during in the latter half of the simulation. The final total relative energy error is $|\Delta E_\mathrm{tot}/E| \sim 6\times10^{-7}$ for the run.}
\label{fig: 050}
\end{figure}

Next we construct a star cluster with a high $f_\mathrm{bin} = 50\%$ binary fraction. The global structural properties of the star cluster model such as total mass and radial density profile (and hence the half-mass radius $r_\mathrm{h}$) remain unchanged from the previous sections. The initial semi-major axis distribution of the binaries before $1$ Gyr of binary stellar evolution is reciprocal or log-uniform while the eccentricity distribution is thermal. The masses of the secondary stars for primaries more massive than $m_\mathrm{1} = 5 M_\odot$ are drawn from a constant mass ratio distribution. The details of the binary star population are explained in more detail in Appendix \ref{section: SCIC}. We note most of the total energy of the star cluster is stored in the binaries. The energy of the binary population exceeds the energy of a $f_\mathrm{bin} = 0\%$ stellar cluster model by a factor of $\sim 7$.

We run the $f_\mathrm{bin} = 50\%$ star cluster model for $T=5$ Myr with \bifrost. The energy conservation results of the run are presented in Fig. \ref{fig: 050}. The mean relative energy error per time interval is $|\Delta E_\mathrm{s}/E| \sim 7\times10^{-10}$, a value comparable to the accuracy of the runs without binary systems in the previous sections, although with a larger scatter. With a high binary fraction in the star cluster, strong binary-single and binary-binary interactions are now common. A number of integration intervals have a higher stochastic error due to especially strong interactions up to $|\Delta E_\mathrm{s}/E| \sim 3\times10^{-7}$. This dominates the total relative energy error after $T=2.2$ Myr in the run. The total relative energy error at the end of the simulation is $|\Delta E_\mathrm{tot}/E| \sim 6\times10^{-7}$.

\subsection{An in-spiraling and merging IMBH binary with relativistic recoil kick in a star cluster}\label{section: gc-mw-gwrecoil}

We further test the \bifrost{} code by running a simulation with the star cluster initially including an IMBH binary close to merging. The binary has a mass ratio of $q=m_\mathrm{2}/m_\mathrm{1} = 0.4$ and a total mass of $M=4\times10^3M_\odot$, and randomly oriented spins with the dimensionless spin parameter being $s=0.1$ for both the black holes. The orbital elements of the in-spiraling binary are $a = 10^{-4}$ pc and $e = 0.995$ to ensure a rapid merger. The initial orbital period of the binary is approximately $P=1.5$ years. We do not include other binary systems in the run, i.e. $f_\mathrm{bin} = 0.0$.

We run the system with identical accuracy parameters as in the previous Sections. As expected the IMBH binary circularises and shrinks rapidly and merges already at $t=2.8\times10^{-3}$ Myr, corresponding to $\sim 1850$ initial orbital periods of the binary. The resulting merger remnant IMBH receives a relativistic recoil kick of $v_\mathrm{kick} = 211$ km/s which is more than enough to unbind it from the globular cluster ($v_\mathrm{esc} = 15$ km/s) it resides in. The merger remnant loses $M_\mathrm{gw} = 137 M_\odot$ of mass-energy in gravitational wave emission and has a spin parameter of $s=0.61$. The effective inspiral spin parameter of this particular merger event was $\chi_\mathrm{eff} = 0.041$.

\begin{figure}
\includegraphics[width=\columnwidth]{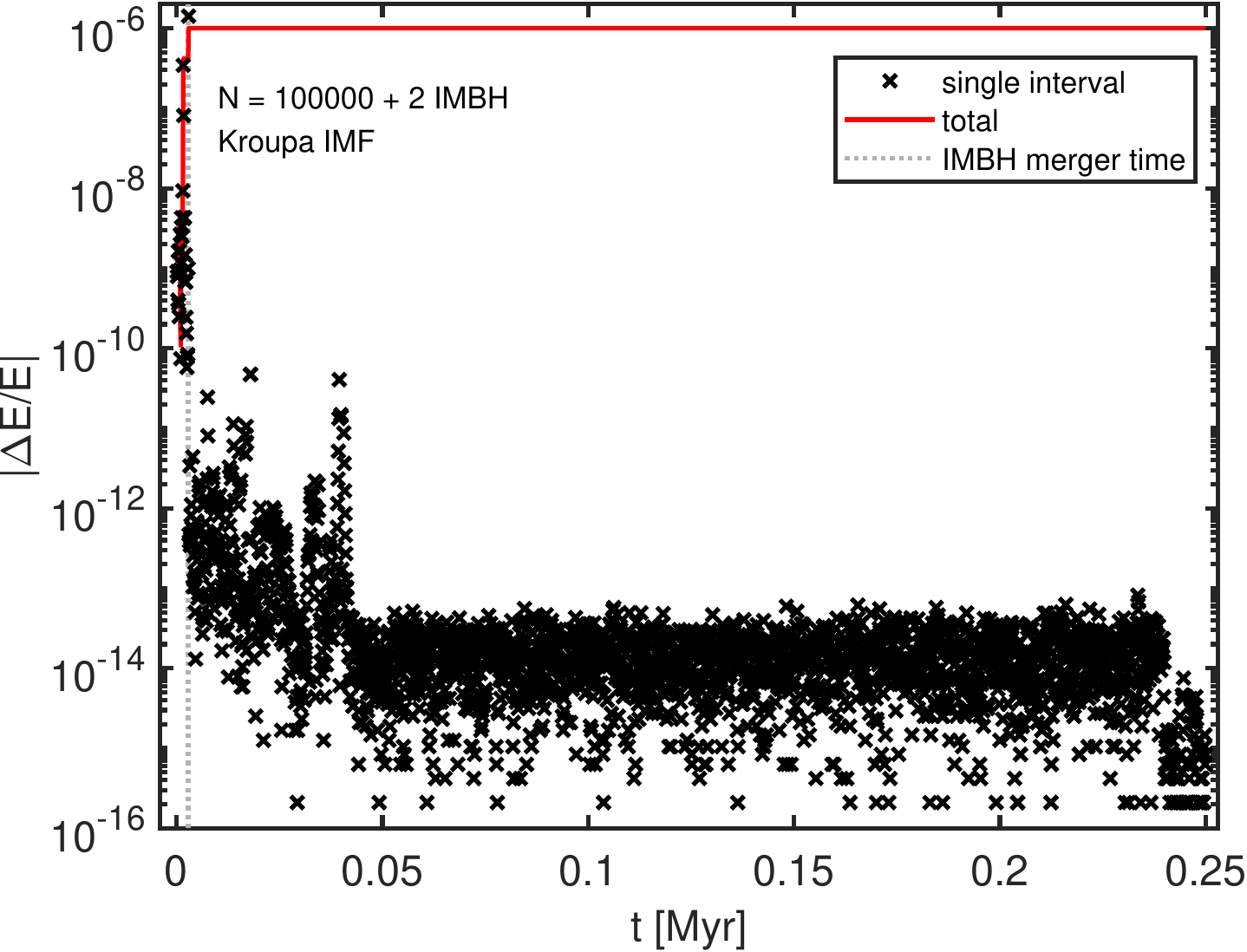}
\caption{Total (red) and per single interval (black) relative energy error in a simulation run in which an IMBH binary inspirals, merges, receives a relativistic recoil kick and escapes its host star cluster. The initial conditions are constructed so that the binary merges almost immediately, at $T=2.8\times10^{-3}$ Myr. The relative energy error per integration interval is the largest at merger time, $|\Delta E_\mathrm{s}/E|\sim 10^{-6}$. The merger remnant receives a recoil kick of $211$ km/s causing it to escape the cluster at $T=0.24$ Myr. The relative energy error per integration interval rapidly decreases from $|\Delta E_\mathrm{s}/E|\sim 10^{-6}$ to $|\Delta E_\mathrm{s}/E|\sim 10^{-13}$ as the IMBH moves through the lower-density outer parts of the star cluster.}
\label{fig: energy-recoil}
\end{figure}

We run the simulation until the now gravitationally unbound IMBH reaches the escaper radius of $r_\mathrm{esc}$ and is removed from the simulation. This occurs at $t=0.23$ Myr and the final simulation time is $t=0.24$ Myr. The code performance remains good during the inspiral, merger and the subsequent escape of the IMBH through the star clusters into its outskirts. The final relative energy error compared to the simulation start is $|\Delta E_\mathrm{tot}/E| \sim 10^{-6}$. The relative energy error of the simulation is presented in Fig. \ref{fig: energy-recoil}. Most of the energy error is accumulated during the inspiral of the IMBH binary. After the IMBH merger the energy conservation is again excellent with the mean relative energy error during an integration interval being less than $|\Delta E_\mathrm{s}/E_\mathrm{0}| \sim 10^{-13}$.

\subsection{A core-collapsing star cluster}

\begin{figure}
\includegraphics[width=\columnwidth]{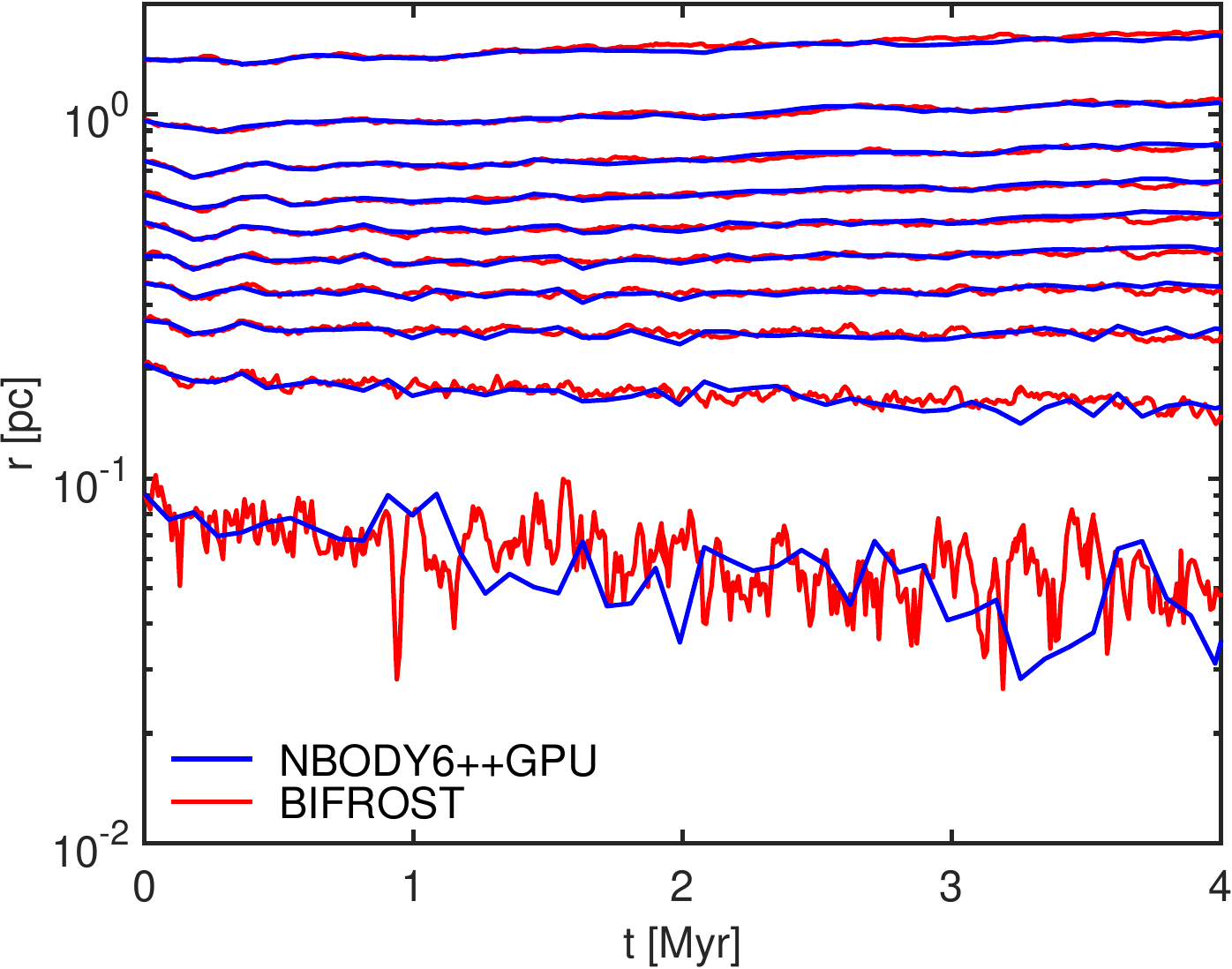}
\centering
\caption{The Lagrangian radii of a core-collapsing star cluster simulated both with \nbodysix{} (blue) and \bifrost{} (red). From bottom to up the Lagrangian radii enclose $1\%$ and $10\%$ to $90\%$ of the total stellar mass of the cluster. The minimum central density $\sim 30$ times the original central density is reached around $T=4$ Myr after which it begins to decrease again. The results of the two simulations codes agree very well.}
\label{fig: lagrange}
\end{figure}

\begin{figure}
\includegraphics[width=\columnwidth]{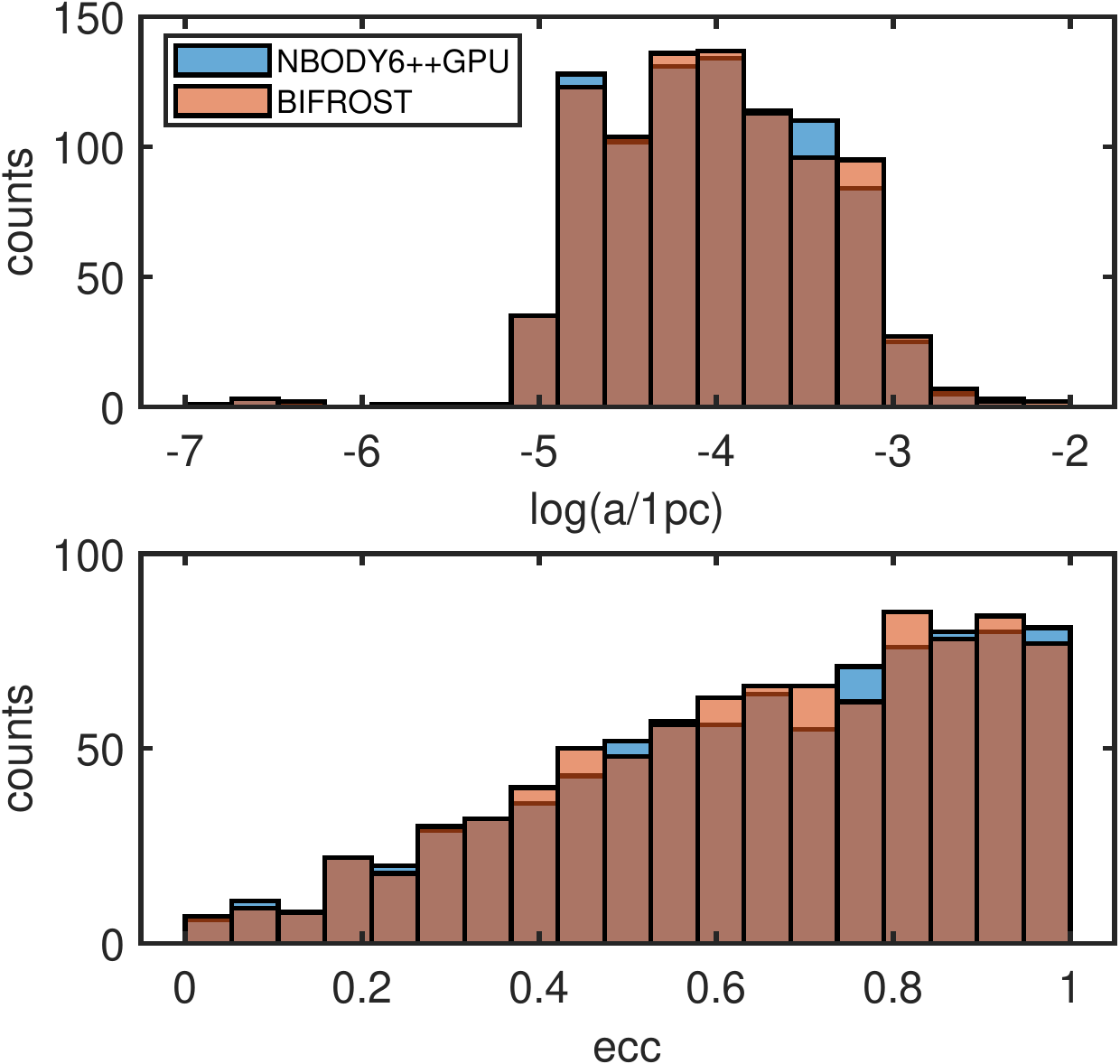}
\centering
\caption{The semi-major axis (top panel) and the eccentricity distributions of binary stars at the end of the core-collapse simulation both with \nbodysix{} (in blue) and \bifrost{} (in orange). The distributions do not evolve very much during the $4$ Myr simulation time are almost identical with the two codes.}
\label{fig: binpop}
\end{figure}

\begin{figure}
\includegraphics[width=\columnwidth]{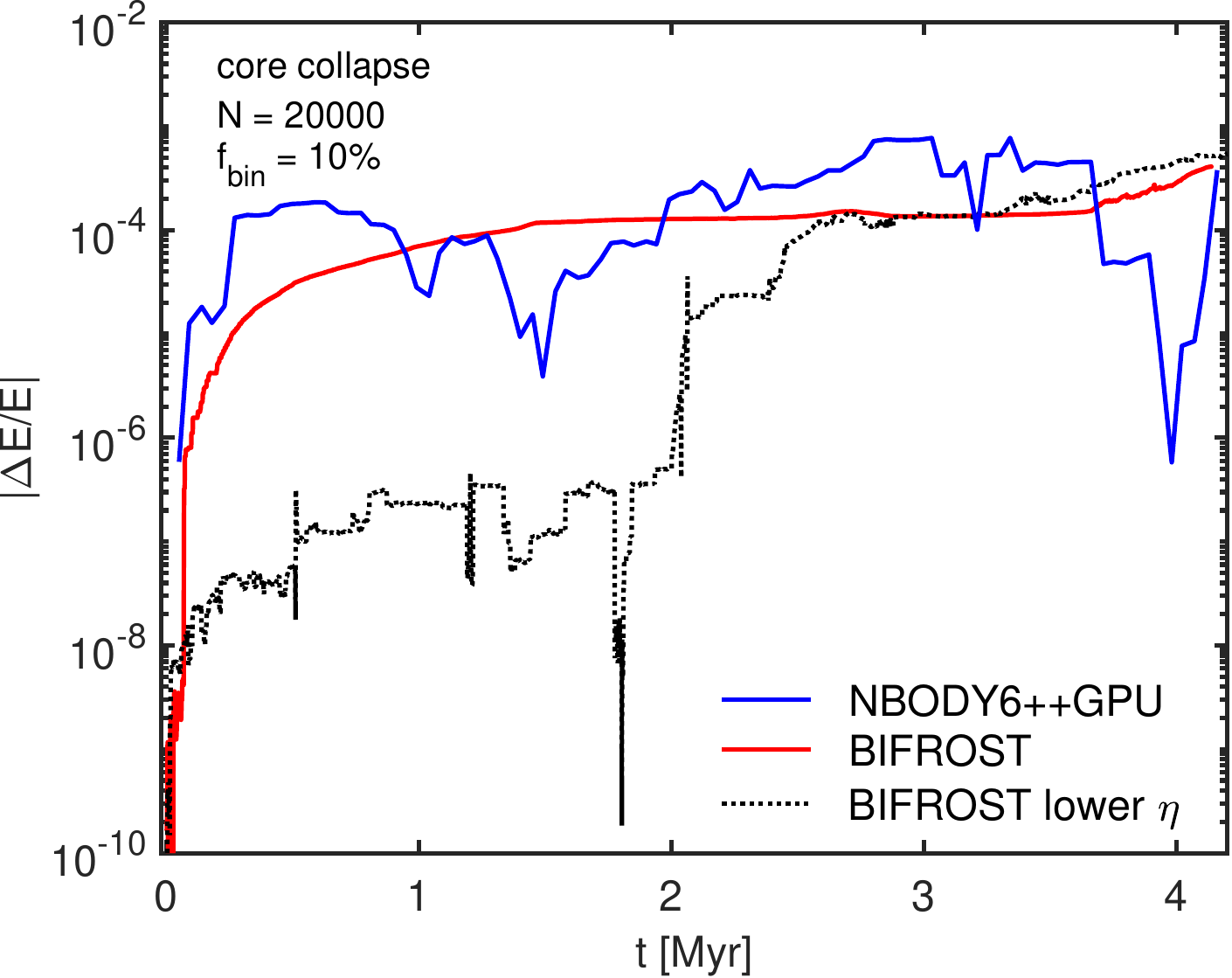}
\centering
\caption{The total relative energy error in a simulation of a core-collapsing star cluster with $N=20000$ and $f_\mathrm{bin} = 10\%$. The results \bifrost{} simulation with $\eta=0.2$ (solid red line) agree very well with the \nbodysix{} (solid blue line) comparison run the final energy error being $|\Delta E_\mathrm{tot}/E| \sim 3\times 10^{-4}$ at $T=4$ Myr. These results are also in agreement with the literature results of core collapse simulations as discussed in the text. A slower and initially more accurate \bifrost{} run (dashed black line) with $\eta = 0.025$ reaches similar final energy error as the two other runs as stochastic energy error from strong few-body encounters in the collapsing core dominates the total energy error budget after $T=2$ Myr.}
\label{fig: nbody6}
\end{figure}

Finally we test the \bifrost{} in the numerically extremely challenging simulation setup of a core-collapsing star cluster. We setup a $N=20000$ Plummer model with a binary fraction of $f_\mathrm{bin} = 10\%$ a half-mass radius of $r_\mathrm{h} = 0.5$ pc. With a Kroupa IMF the expected core-collapse timescale is only a few Myr. We run the star cluster setup for $T=4 $ Myr after which the central density reaches its maximum approximately $30$ times the original density, after which the core density begins to decrease again. We include a run simulated with \nbodysix{} for comparison purposes.

The Lagrangian radii enclosing $1\%$ and $10\%$ to $90\%$ of the total stellar mass are presented in Fig. \ref{fig: lagrange}. In addition, we show the semi-major axis and eccentricity distributions of binary stars at the end of the simulations in Fig. \ref{fig: binpop}. The simulation results of the two codes agree very well both for the global structure of the star cluster and its binary population properties.

The total relative energy errors of the core collapse run in two \bifrost{} runs and the \nbodysix{} comparison run are presented in Fig. \ref{fig: nbody6}. We use accuracy parameters \texttt{ETAI}$=0.01$ and \texttt{ETAR}$=0.01$ for \nbodysix. This choice of parameters yields very similar results as the \bifrost{} run with the standard time-step accuracy parameter of $\eta = 0.2$. The total energy error rapidly increases above $|\Delta E_\mathrm{tot}/E| = 10^{-6}$ in the runs the final error being $|\Delta E_\mathrm{tot}/E| \sim 3\times 10^{-4}$ after $T=4$ Myr of evolution. The results is very well in line with the collisional code accuracies in core collapse simulations in the literature (e.g. \citealt{Konstantinidis2010,Pelupessy2012,Wang2020b}). Another \bifrost{} run with a higher accuracy of $\eta = 0.025$ has energy error less than $|\Delta E_\mathrm{tot}/E| \sim 5\times 10^{-7}$ until $T=2$ Myr until the stochastic error from strong encounters in the collapsing core begins to dominate. The final energy error after the core collapse is essentially independent of the chosen $\eta$ while it changes the elapsed wall-clock time linearly. With the same numerical accuracy \nbodysix{} is marginally faster than \bifrost{} for the initial conditions presented here. We note that for more massive systems $N>10^5$ \nbodysix{} cannot run systems with high binary fractions which \bifrost{} is able to do very well as shown in Section \ref{section: 4}.

\section{Future work}\label{section: 6}

While our \bifrost{} code contains a large number of novel features compared to the earlier \frost{} code version a number of updates remain to be implemented in future work. A number of most important numerical methods and astrophysical models are discussed below with their recent implementations in the literature. 

\begin{itemize}

    \item \textit{Binary and multiple stellar evolution}. The current \bifrost{} version only includes single stellar evolution from the \sse{} package \citep{Hurley2000}. In order to accurately simulate interacting and rapidly evolving binary populations a module for binary stellar evolution is evidently needed. The main challenge to be solved in future work is the coupling of a chosen binary evolution package to the integration of subsystems, especially for systems with more than two bodies. While the binary stellar evolution code \bse{} of \cite{Hurley2002} is a widely-used package in N-body simulation codes other options such as \texttt{binary\_c} (\citealt{Izzard2018} and references therein) or \texttt{SEVN} \citep{Spera2015} remain. An interesting possibility is to use the recent secular multiple stellar evolution code \mse{} \citep{Hamers2021} as it can handle arbitrary dynamical configurations of evolving stars.

    \item \textit{Tidal drag and capture of stars near black holes.} Orbital energy can be transferred into internal degrees of freedom when a star encounters a black hole, up to the point that an initially parabolic or hyperbolic encounter leads to a capture of the star. Furthermore, tidal interactions can cause the orbit of the star rapidly decay until the star is disrupted and accreted \citep{Stone2017}. The two processes are of paramount importance considering the growth of intermediate-mass black holes in dense star clusters (e.g. \citealt{Rizzuto2021,Rizzuto2022}). A tidal frag force implementation resembling the PN2.5 formulation by \cite{Samsing2018} is especially suitable for regularised integrators in \bifrost.

    \item \textit{Orbit-average-accurate treatment of perturbed binaries.} The treatment of perturbed binary systems in the current code version is not orbit-average-accurate, although the evolution of binary systems is very close to the results of the \nbodysix{} code as demonstrated in Fig. \ref{fig: binpop}. In order to always ensure the correct orbit-average-accurate evolution of binary systems in regularised integration, a slow-down treatment of perturbed binaries (e.g. \citealt{Mikkola1996,Wang2020}) will be added in a future version of our code.

    \item \textit{Tree or fast multipole method (FMM) gravity solver.} Even though simulations of million-body star clusters using GPU-accelerated direct-summation $\bigO{N^2}$ codes have been performed \citep{Wang2015,Wang2016}, they remain extremely expensive. In addition, studying the evolution and dynamics of nuclear star clusters (e.g. \citealt{Neumayer2020}) while resolving individual stars requires simulation codes capable of simultaneously integrating $N=10^7$--$10^8$ simulation particles. The two widely-used options for gravitational force calculations beyond a million bodies are tree \citep{Barnes1986} and FMM solvers (e.g. \citealt{Dehnen2014}). Tree and FMM solvers are typically employed in galaxy- or cosmological-scale simulations \citep{Springel2021} allowing for very large particle numbers, but also recently in collisional star cluster simulations as well (e.g. \texttt{PeTar}, \citealt{Wang2020b} and \texttt{Taichi}, \citealt{Mukherjee2021,Mukherjee2022}). We plan to include either tree or FMM solver for \bifrost{} as well in further studies.

    \item \textit{Coupling to a hydrodynamics solver.} Star clusters form embedded in giant molecular clouds and gas plays an important role in their early gravitational dynamics (\citealt{Lada2003,PortegiesZwart2010,Longmore2014,Lahen2020,Li2022}). A number of codes both with collisional gravitational dynamics and gas hydrodynamics have been recently developed (e.g. \texttt{TORCH} \citealt{Wall2020}, \texttt{STARFORGE} \citealt{Grudic2021}, \texttt{SIRIUS} \citealt{Hirai2021}, \texttt{EKSTER} \citealt{Rieder2022}). Our goal is to include a modern smoothed particle hydrodynamics (SPH) module employing the improvements of standard SPH as e.g. presented by \cite{Hu2014} into \bifrost{}. The algorithm will be capable of using GPUs for the required tree-based particle neighbour searches and SPH kernel operations. Simulation codes using both SPH and hierarchical integration (with CPUs) already exist second-order codes, e.g. in the \texttt{GADGET-4} code \citep{Springel2021}.
\end{itemize}

\section{Conclusions}\label{section: 7}

We have developed and tested the novel N-body simulation code \bifrost{} specialising in star cluster simulations with a large fraction of binary and multiple star systems. Based on our earlier GPU-accelerated hierarchical fourth-order forward integrator code \frost{} the new code includes both regularised (\texttt{LOGH}, \texttt{MSTAR}) and secular integration (binary, triple) methods for compact subsystems. 

In the regularised subsystems, the equations of motion of the simulation particles contain post-Newtonian terms up to PN3.5 order including three-body Einstein–Infeld–Hoffmann terms in PN1.0. Spin-dependent PN terms can be enabled by the user as well. In the secular integration the highest orbit-averaged post-Newtonian term is PN2.5. For the \fsi{} integration we have the possibility to include the so-called global PN1.0 term which enables post-Newtonian dynamics also outside subsystems.

In addition to the new subsystem integrator, \bifrost{} now includes single stellar evolution using the \sse{} tracks and prescriptions of various particle merger processes. We include gravitational-wave inspiral mergers for compact binary systems, tidal disruption events between compact objects and stars, and finally a prescription for stellar mergers. For binary black hole mergers we include relativistic mass loss and gravitational-wave recoil kicks using fitting functions from numerical relativity simulations.

We have presented the timing and scaling tests for the \bifrost{} code simulating star cluster models between $N=10^4$ and $N=5 \times 10^6$ particles. The binary fractions of the star clusters range from $f_\mathrm{bin} = 10\%$ to $f_\mathrm{bin} = 100\%$ with the highest number of binary systems being $N_\mathrm{bin} = 2.5\times10^6$. For particle numbers above $N \gtrsim 10^5$ the binary fraction $f_\mathrm{bin}$ has only a small effect on the scaling and running times of the simulations. We confirm that the strong scaling of the novel \bifrost{} code is very similar as the scaling of its precursor code \frost. The maximum number of GPUs to be used in a simulation with $N$ particles is still approximately $N_\mathrm{GPU} \sim 40\times N / 10^6$. The maximum number of simulation particles for \bifrost{} is approximately a few million, depending on the density and compactness of the system as well as the desired simulation time.

We have tested the accuracy of our new code in a series of star cluster simulations. The various star cluster simulations examine the effect of the stellar IMF and the binary fraction on the amount of accumulated energy error. In general, simulations with more massive particles and higher binary fractions are less accurate due to more frequent close and strong encounters. By carefully choosing the user-given code accuracy parameters the total relative energy error was always less than $|\Delta E/E_\mathrm{tot}|\sim 6 \times 10^{-7}$ in the runs. We note that the energy error can be further decreased by decreasing the time-step accuracy parameters $\eta$ at the cost of increased code running times. 

Finally, we have performed more simulations of more extreme stellar-dynamical scenarios. We modelled the inspiral and rapid merger of an IMBH binary embedded in a star cluster. The merger remnant black hole received a strong gravitational-wave recoil kick of over $v=200$ km/s and soon escaped the star cluster. The total relative energy error of the simulation was approximately $|\Delta E_\mathrm{tot}/E| \sim 10^{-6}$. In order to further test of our code we simulated the core collapse of a star cluster with $N=20000$, $r_\mathrm{h} = 0.5$ pc and $f_\mathrm{bin} = 10\%$. The total relative energy error of the core collapse simulation was $|\Delta E_\mathrm{tot}/E| \sim 3 \times 10^{-4}$, a result comparable to the accuracy of other collisional simulation codes in the literature (e.g. \citealt{Konstantinidis2010,Pelupessy2012,Wang2020b}. We also performed a run of the same initial conditions using \nbodysix{} in order to directly compare the global properties of the star cluster simulated with \bifrost{} with another code. The running times and energy errors of the particular star cluster model with the two codes are very similar. Most importantly, star cluster properties including the Lagrangian radii and the binary star population evolve in an analogous manner.

Hierarchical fourth-order forward symplectic integrators are a promising alternative for both Hermite integrators in traditional direct-summation N-body simulations and second-order leapfrog-type symplectic integrators, especially when supplemented with accurate subsystem integrators. Efficient parallelization of our \bifrost{} code allows for simulating stellar systems with arbitrary binary fractions up to $f_\mathrm{bin} = 100\%$, a feature which only few other simulation codes include \citep{Wang2020b}. In addition to the arbitrary binary fractions we highlight the capability of \bifrost{} to simulate relativistic binary systems. This is possible as both the secular and regularised subsystem integrators in \bifrost{} contain post-Newtonian terms in their equations of motion.

The version of \bifrost{} presented in this study is an efficient and versatile tool for simulating stellar systems from few-body systems to massive star clusters and low-mass nuclear star clusters including the inner parsecs of the Milky Way centre. The immediate near-future code updates including binary stellar evolution and hydrodynamics will upgrade \bifrost{} to an even more powerful tool for studying the formation, evolution and possible disruption of star clusters in a wide range of mass scales. In addition, incorporating a tree or FMM gravity solver will eliminate the current wall-clock time bottleneck placed by the direct summation approach in the near future. This will enable simulating the evolution of the most massive star clusters in the Universe, and even full low-mass isolated dwarf galaxies and their mergers.

\section*{Data availability statement}
The data relevant to this article will be shared on reasonable request to the corresponding author.

\section*{Acknowledgments}
The numerical simulations were performed using facilities hosted by the Max Planck Computing and Data Facility (MPCDF) and the Leibniz Supercomputing Centre (LRZ), Germany. TN acknowledges support from the Deutsche Forschungsgemeinschaft (DFG, German Research Foundation) under Germany's Excellence Strategy - EXC-2094 - 390783311 from the DFG Cluster of Excellence "ORIGINS". F.P.R. acknowledges the support by the European Research Council via ERC Consolidator Grant KETJU (no. 818930).

%%%%%%%%%%%%%%%%%%%%%%%%%%%%%%%%%%%%%%%%%%%%%%%%%%
%%%%%%%%%%%%%%%%%%%% REFERENCES %%%%%%%%%%%%%%%%%%
\bibliographystyle{mnras}
\interlinepenalty=10000

%\bibliography{references}
%%%%%%%%%%%%%%%%%%%%%%%%%%%%%%%%%%%%%%%%%%%%%%%%%%
%%%%%%%%%%%%%%%%% APPENDICES %%%%%%%%%%%%%%%%%%%%%

\appendix

\section{Star cluster initial conditions}\label{section: SCIC}
\subsection{A novel fast initial conditions generator}

We setup the star cluster initial conditions (IC) for the purposes of this study using our new initial conditions generator we term \scic{} (for star cluster initial conditions). Our IC generator code is functionally very similar to the widely used \mcluster{} code \citep{Kupper2011}. The main difference is the speed of the two codes: generating a million-body star cluster IC takes only a few seconds using \scic. The reason for the speed of our new code is the efficient MPI parallelization allowing the use of several supercomputer nodes, a feature which \mcluster{} as a shared-memory single-node code lacks. Besides the standard $\bigO{N^2}$ loop parallelization also time-consuming $\bigO{N}$ loops such as single and binary stellar evolution are parallelised in \scic. The code is written in standard C language with a number of Fortran libraries from the literature included for stellar evolution. In order to summarised the basic features of the \scic{} code we have collected the main fixed and user-given parameters into Table \ref{tab: scic}. Further details of the features of the code are given below.

\subsection{Star cluster models}\label{section: clustermodels}

We sample the masses of individual stars using the common \cite{Kroupa2001} initial mass function. In the code we typically use the number of stars as the user-given input parameter instead of the total cluster mass. This choice is made to avoid the iterative search for the correct particle number to match the given cluster mass when stellar evolution is used.

The spherically symmetric mass density profile of the star clusters is set according to the \cite{Plummer1911} profile
\begin{equation}
\rho(r) = \frac{3 M_\mathrm{p}}{4 \pi a_\mathrm{p}} \left( 1+\frac{r^2}{a_\mathrm{p}^2} \right)^{-5/2}
\end{equation}
in which $M_\mathrm{p}$ and $a_\mathrm{p}$ are the mass and scale radius of the model, respectively. The positions of single stars and binary center-of-masses are sampled from the corresponding cumulative mass profile using the standard inversion sampling technique as the inverse function $r(M)$ is easy to calculate. Two random angles from a unit sphere are used to get the Cartesian components of the position vector. The scale radius $a_\mathrm{p}$ is related to the half-mass radius of the model as $r_\mathrm{h} = \left( 2^{2/3}-1 \right)^{-1/2}a_\mathrm{p}$. In \texttt{SCIC} the half-mass radius is either user-given or obtained from the observed mass-radius relations of young star clusters and nuclear star clusters. For lower-mass star clusters we use the relation
\begin{equation}
\frac{r_\mathrm{h}}{1 \mathrm{pc}} = 2.55 \left( \frac{M}{10^4 M_\mathrm{\odot}} \right)^{0.24}    
\end{equation}
of \cite{Brown2021}. For more massive nuclear star clusters we devise a simple ad-hoc relation
\begin{equation}
\begin{split}
\log_\mathrm{10}\left(\frac{R_\mathrm{e}}{1 \mathrm{pc}}\right) &= \frac{1}{2}, & M<10^{6.5} M_\mathrm{\odot}\\
\log_\mathrm{10}\left(\frac{R_\mathrm{e}}{1 \mathrm{pc}}\right) &= \frac{1}{2} \log_\mathrm{10}\left( \frac{M}{1 M_\mathrm{\odot}} \right) - \frac{11}{4}, & M\geq10^{6.5} M_\mathrm{\odot}
\end{split}    
\end{equation}
with $r_\mathrm{h} = ((1/2)^{-2/3}-1)^{-1} R_\mathrm{e}$ for the \cite{Plummer1911} model. The formula phenomenologically captures the observed flat radius distribution at smaller masses and the power-law behaviour above the threshold mass of $M \sim 10^{6.5} M_\mathrm{\odot}$ (see e.g. Fig. 7 of \citealt{Neumayer2020}).

The space velocities of single stars and binary center-of-masses are obtained using the \cite{Plummer1911} potential 
\begin{equation}
\phi(r) = -\frac{G M}{a_\mathrm{p}} \left( 1 + \frac{r^2}{a_\mathrm{p}^2} \right)^{-1/2}
\end{equation}
and the energy distribution function
\begin{equation}
f(\mathcal{E}) = \begin{cases}
\frac{24 \sqrt{2} a_\mathrm{p}^2}{7 \pi^3 G^5 M_\mathrm{p}^5} (\mathcal{E})^{7/2} &\hspace{1cm} \mathcal{E}\geq0\\
0 & \hspace{1cm}\mathcal{E}<0
\end{cases}
\end{equation}
using $\mathcal{E} = -E=-\frac{1}{2}m\norm{\vect{v}}^2-\phi(r)$ together with the von Neumann rejection sampling method (see e.g. the Art of Computational Science online material\footnote{www.artcompsci.org by P. Hut \& J. Makino} and \citealt{Heggie2003} for extensive practical details of the implementation). As with the positions the isotropic components of the velocity vectors are obtained by sampling two angles uniformly from the unit sphere.

\begin{table}
	\caption{The main initial conditions generator \scic{} parameters, their symbols and their typical values as described in the text.}
	\label{tab: scic}
	\begin{tabular}{lll}
		\hline
		Parameter & Symbol & Details\\
		\hline
		particle number & $N$ & $\lesssim 10^8$\\
		IMF & $\xi(m)$ & \cite{Kroupa2001}\\
		half-mass radius & $r_\mathrm{h}$ & user-given or observed\\
		binary fraction & $f_\mathrm{bin}$ & 0 $\leq f_\mathrm{bin} \leq 1.0$\\
		semi-major axis distribution & $f(a)$ & flat in log space\\  
        min, max semi-major axis & $a_\mathrm{min}$, $a_\mathrm{max}$ & $a_\mathrm{min} \leq a_\mathrm{max}$\\ 
        eccentricity distribution & $f(e)$ & thermal\\
        binary pairing mass limit & $m_\mathrm{pair}$ & $5 M_\mathrm{\odot}$\\
        mass ratio distribution & $f(q)$ & const., $0.1 \leq q \leq 1.0$\\
        stellar population age & $t_\star$ & $<$ age of the Universe\\
        stellar population metallicity & $Z_\star$ & $10^{-4} < Z\star < 0.03$\\
        artificial binary breaking limit & $r_\mathrm{break}$ & $\sim$ a few times $a_\mathrm{max}$\\
		\hline
	\end{tabular}
\end{table}

\subsection{Binary stars}

Defining the number of single stars and binary systems in a star cluster as $N_\mathrm{s}$ and $N_\mathrm{b}$ respectively the total number of stars is $N = N_\mathrm{s} + 2 N_\mathrm{b}$. The number of binary systems can also be expressed using the binary fraction $f_\mathrm{bin}$ of the cluster as $N_\mathrm{b} = 1/2 f_\mathrm{bin} N$. We use the binary fraction as an user-given input parameter defining the binary content of the initial conditions.

The initial binary component masses are paired above $m_\mathrm{pair} = 5 M_\mathrm{\odot}$ using a flat mass ratio distribution $q = m_\mathrm{2}/m_\mathrm{1} \propto \text{const}$ with $0.1 \leq q < 1.0$ (\citealt{Kiminki2012,Sana2012,Kobulnicky2014}). In the practical implementation beginning from the most massive star we assign secondary star masses by finding the actual closest mass to the sampled random secondary mass from the remaining stars until no primary stars above $5 M_\mathrm{\odot}$ remain.

For binary stars the initial semi-major axis distribution is flat in logarithmic space between the user given minimum and maximum values of $a_\mathrm{min}$ and $a_\mathrm{max}$. For reasonable cluster models $a_\mathrm{max} \ll r_\mathrm{h}$. The binary eccentricity distribution is assumed to be thermal initially, i.e $f(e) = 2e$.

The age and metallicity of the stellar population are user-given initial parameters. At the moment our code supports a single stellar population per star cluster. In the case of an evolved stellar population the properties of the single (e.g. type, mass, radius) and binary stars (e.g. types, masses, radii, semi-major axes, eccentricities) of the given age and metallicity are computed using the widely-used \sse{} \citep{Hurley2000} and \bse{} \citep{Hurley2002} stellar evolution libraries. The \scic{} calls the libraries in parallel (a single task per binary) through a simple C-to-Fortran code interface. Due to the parallel approach \scic{} is faster than \mcluster{} in stellar evolution as the latter code only has a serial library interface.

We allow the binary stars to merge during binary stellar evolution using the \bse{} library merging criteria \citep{Hurley2002}. In addition to the \bse{} merger criteria we merge all systems which orbit closer than the sum of the radii (or sum of ISCO radii) of the components of the binary. We also check the systems with a single compact object after stellar evolution for tidal disruption events just as in the \bifrost{} code described in Section \ref{section: mergers}. The merger remnant properties are obtained using the \bse{} package. Depending on the types and masses of the merging objects there may or may not to be a remnant. Finally we check the binary systems containing a single or two compact objects for gravitational-wave driven inspirals within the next $0.01$ Myr again as in Section \ref{section: mergers}. In our test simulations with \bifrost{} initial conditions generated using \mcluster{} usually lead in up to a few tens of tidal disruption mergers within the first few time-steps immediately in the beginning of the \bifrost{} simulation. For initial conditions generated with \scic{} this somewhat spurious burst of merger events does not occur due to a more detailed set of merger criteria in the IC generator code.

After stellar evolution the relative positions and velocities of the binary star components are added to the position and velocity of the center-of-mass of the system. While the binary semi-major axes and eccentricities are given by the stellar evolution libraries we assign random values to the other orbital elements (inclination $i$, longitude of the ascending node $\Omega$, argument of periapsis $\omega$ and mean anomaly $M$). For obtaining the eccentric anomaly required for the positions and velocities we use the same Kepler solver as in the \bifrost{} code described in Section \ref{section: keplersolver}.

\subsection{Massive black holes}

We have included an option to add a central black hole in the star cluster initial conditions in our code. The user-given input parameters for the massive black hole particle are the mass and the magnitude of spin vector $s$. Typically we conservatively set $s$ to 10\% of the maximal black hole spin. For a single central black hole the initial orientation of the spin vector does not matter as the cluster model is spherically symmetric anyway so we decide to align it with the z-axis of the coordinate system.

\subsection{Avoiding artificial binaries and multiplets}

Even when setting the binary fraction $f_\mathrm{bin} = 0\%$ a star cluster initial condition typically contains a few binary systems formed by random chance which we refer to here as artificial binaries. The number $N_\mathrm{art}$ of these systems increases with increasing cluster particle number $N$ and decreasing half-mass radius $r_\mathrm{h}$ which both increase the stellar number density of the cluster.

A binary system which survives in a star cluster for a long time needs to be sufficiently tight (e.g. \citealt{Heggie2003}), typically on milliparsec-scale separations or less. This separation scale is considerably smaller than the half-mass radius of typical star clusters (of the order of a few pc) or mean inter-stellar distance in the cluster centers (of the order of $0.01$ pc to $0.1$ pc). Still, the sampling of the initial star particle positions independently of each other will inevitable create close-by particle pairs. An artificial binary is formed if the relative velocity of the two particles is small enough for the binary to be bound. The artificial binaries typically have high eccentricities originating from their required small apocenter velocities. 

Artificial binaries formed in the initial conditions generator are not a problem in itself but rather complicate the comparison of simulation runs using same star cluster model properties with different random seeds, especially when $f_\mathrm{bin}$ is low and $N$ is small. As artificial binaries are most very eccentric they are thus computationally expensive to integrate compared to typical binary systems. This may have an effect on for example timing and scaling tests of a simulation code with a set of initial conditions with a low or zero $f_\mathrm{bin}$ as initial conditions with tight eccentric binaries are the slowest to integrate in wall-clock time. We emphasised that for fair and controlled timing and scaling tests initial conditions containing artificial binaries should be avoided. Thus, a robust way beyond simply removing artificial binaries by hand should be devised.

We also note that in the case of $f_\mathrm{bin}>0$ (typically short-lived) artificial multiplets of three or more stars will also occur. As these artificial multiple systems are rarely stable they will quickly dissolve in a strong few-body interaction (e.g. \citep{Valtonen2006}). This may lead to spurious formation of high-velocity stars or artificially enhanced merger rates. We however note in addition that including physical multiple systems (as motivated by observations) is an interesting prospect for future work. 

A simple method to avoid artificial binaries and multiplets in the initial conditions generation is to impose a minimum distance $r_\mathrm{break} \sim$ a few times $a_\mathrm{max}$ between particles which are not members of the same binary system. We search for the shortest inter-particle distances between the $N_\mathrm{s}+N_\mathrm{b}$ single stars and binary systems. If an artificial binary is detected (i.e. $a<r_\mathrm{break}$) the positions and velocities of the particles in question are re-sampled just as they were sampled for the first time: the resulting new realisation of the cluster model is as realistic a cluster model as the original one, only the random numbers differ. The process may need to be repeated up to $\sim 10$ times to get rid of all the artificial binaries. The recipe also holds for the binary center-of-masses as for single stars as the binaries treated as point particles before assigning the positions and velocities of the binary components in the final phase of the IC generation.

However, differences in the partially re-sampled random realisations begin to unfortunately arise when either $a_\mathrm{max}$ is large or the cluster is very dense: re-sampled particles from the centre of the cluster will end up more probably to larger radii and the density profile of the cluster starts to deviate from the desired one and the $f_\mathrm{bin}$ begins to develop a radial gradient. A solution for this issue is to only re-sample the angular coordinates (both for positions and velocities) leaving the original radial position $\norm{\vect{r}_\mathrm{i}}$ and space velocity $\norm{\vect{r}_\mathrm{i}}$ unchanged. A practical algorithm in our \scic{} code first performs the angular re-sampling method for the artificial binaries using up to $10$ iterations before switching to the original re-sampling if artificial systems remain. After another $5$-$10$ iteration rounds typically only few or zero artificial binaries remain in the cluster model.

%%%%%%%%%%%%%%%%%%%%%%%%%%%%%%%%%%%%%%%%%%%%%%%%%%

% Don't change these lines
\bsp	% typesetting comment
\label{lastpage}
\end{document}